\documentclass{emulateapj}
\usepackage{apjfonts}

\newcommand{\scaledown}{\epsscale{0.95}}
\newcommand{\scaleup}{\epsscale{1.1}}
\newcommand{\scaleupp}{\epsscale{1.15}}
\newcommand{\plotter}{\plotone}
\newcommand{\plotterr}{\plotone}

\newcommand{\breaker}{}

\newcommand{\tableclear}{}

\newcommand{\etal}{et al.}

\newcommand{\msun}{M_{\sun}}

\newcommand{\fgas}{f_{\rm gas}}

\newcommand{\mergercalcurl}{\scriptsize{\path{http://www.cfa.harvard.edu/~phopkins/Site/mergercalc.html}}}

\shorttitle{Which Mergers Matter?}
\shortauthors{Hopkins \etal}
\slugcomment{Submitted to ApJ, August 20, 2009}
\begin{document}

\title{Mergers and Bulge Formation in $\Lambda$CDM: 
Which Mergers Matter?}
\author{ 
Philip F.\ Hopkins\altaffilmark{1}, 
Kevin Bundy\altaffilmark{1},
Darren Croton\altaffilmark{2},
Lars Hernquist\altaffilmark{3},
Dusan Keres\altaffilmark{3,4},
Sadegh Khochfar\altaffilmark{5},
Kyle Stewart\altaffilmark{6},
Andrew Wetzel\altaffilmark{1},
Joshua D.\ Younger\altaffilmark{3,7}
}
\altaffiltext{1}{Department of Astronomy, University of California 
Berkeley, Berkeley, CA 94720, USA} 
\altaffiltext{2}{Centre for Astrophysics \&\ Supercomputing, Swinburne University 
of Technology, P.O.\ Box 218, Hawthorn, VIC 3122, Australia} 
\altaffiltext{3}{Harvard-Smithsonian Center for Astrophysics, 60
  Garden Street, Cambridge, MA 02138, USA} 
\altaffiltext{4}{W.~M.\ Keck
  Postdoctoral Fellow at the Harvard-Smithsonian Center for
  Astrophysics} 
\altaffiltext{5}{Max Planck Institut 
f{\"u}r Extraterrestrische Physik, Giessenbachstr., D-85748, Garching, Germany}
\altaffiltext{6}{Center for Cosmology, Department of Physics and
  Astronomy, The University of California at Irvine, Irvine, CA 92697, USA}
\altaffiltext{7}{Hubble Fellow, 
Institute for Advanced Study, Einstein Drive, Princeton, 
NJ 08540, USA}

\begin{abstract}

We use a suite of semi-empirical models to predict the galaxy-galaxy merger 
rate and relative contributions to bulge growth as a function of mass 
(both halo and stellar), redshift, and mass ratio. 
The models use empirical constraints on the halo occupation distribution, 
evolved forward in time, to robustly identify where and when galaxy mergers 
occur. Together with the results of 
high-resolution merger simulations, this allows 
us to quantify the relative contributions of mergers with different 
properties (e.g.\ mass ratios, gas fractions, redshifts) to the bulge population. 
We compare with observational constraints, and find good agreement. 
We also provide useful 
fitting functions and make public a code\footnote{\mergercalcurl} to 
reproduce the predicted merger rates and contributions 
to bulge mass growth.
We identify several robust conclusions. 
(1) Major mergers dominate 
the formation and assembly of $\sim L_{\ast}$ bulges and  
the total spheroid mass density, but minor mergers 
contribute a non-negligible $\sim30\%$. 
(2) This is mass-dependent: 
bulge formation and assembly is dominated by more minor mergers 
in lower-mass systems. In higher-mass systems, most bulges originally 
form in major mergers near $\sim L_{\ast}$, but assemble in increasingly 
minor mergers. 
(3) The minor/major contribution is also morphology-dependent: 
higher $B/T$ systems preferentially form in more major mergers, 
with $B/T$ roughly tracing the mass ratio of the largest recent merger; 
lower $B/T$ systems preferentially form in situ from minor minors. 
(4) Low-mass galaxies, being gas-rich, require more  
mergers to reach the same $B/T$ as high-mass systems. 
Gas-richness dramatically suppresses the absolute efficiency of 
bulge formation, but does not strongly influence the relative 
contribution of major versus minor mergers. 
(5) Absolute merger rates at fixed mass ratio increase 
with galaxy mass. 
(6) Predicted merger 
rates agree well with those observed in pair and morphology-selected 
samples, but there is evidence that some morphology-selected samples 
include contamination from minor mergers. 
(7) Predicted rates also agree with the integrated growth in bulge mass density 
with cosmic time, but with factor $\sim2$ uncertainty in both -- 
up to half the bulge mass density could come from non-merger processes. 
We systematically vary the model assumptions, 
totaling $\sim10^{3}$ model permutations, 
and quantify the resulting uncertainties. 
Our conclusions regarding the importance of different 
mergers for bulge formation are very robust to these changes. 
The absolute predicted merger rates are systematically uncertain 
at the factor $\sim2$ level; uncertainties grow at the lowest masses and 
high redshifts.

\end{abstract}


\section{Introduction}
\label{sec:intro}

In the now established $\Lambda$CDM 
cosmology, structure grows hierarchically \citep[e.g.][]{whiterees78}, making 
mergers an inescapable element in galaxy formation. 
Thirty years ago, \citet{toomre77} proposed the ``merger hypothesis,''
that major mergers between spirals could result in elliptical galaxies, 
and the combination of detailed observations of 
recent merger remnants \citep{schweizer82,LakeDressler86,Doyon94,ShierFischer98,James99,
Genzel01,tacconi:ulirgs.sb.profiles,dasyra:mass.ratio.conditions,dasyra:pg.qso.dynamics,
rj:profiles,rothberg.joseph:kinematics,
vandokkum:dry.mergers} and e.g.\ faint shells and tidal 
features around ellipticals \citep{malin80,malin83,schweizer80,
schweizerseitzer92,schweizer96} have lent considerable 
support to this picture \citep[e.g.][]{barneshernquist92}.

Mergers are also linked to starburst galaxies and luminous 
quasars. By exciting tidal torques that lead to rapid inflows of gas into 
the centers of galaxies \citep{barnes.hernquist.91,barneshernquist96}, 
mergers provide the fuel to power intense starbursts \citep{mihos:starbursts.94,
mihos:starbursts.96}, to
feed rapid black hole growth \citep{dimatteo:msigma,hopkins:lifetimes.letter,
hopkins:lifetimes.methods}, and, through various associated feedback
channels, convert blue, star-forming galaxies into quiescent, red
ones \citep[e.g.][]{springel:red.galaxies,springel:models}.
Mergers are inevitably associated with the most luminous 
star-forming systems, from ULIRGs in the local Universe 
\citep{soifer84a,soifer84b,joseph85,sanders96:ulirgs.mergers} 
to bright sub-millimeter galaxies at high redshifts 
\citep{alexander:bh.growth,younger:smg.sizes,shapiro:highz.kinematics,
tacconi:smg.mgr.lifetime.to.quiescent}, and properties 
ranging from their observed kinematics, 
structural correlations, and clustering link these populations 
to massive ellipticals today 
\citep[][]{LakeDressler86,Doyon94,ShierFischer98,James99,
Genzel01,rothberg.joseph:kinematics,rothberg.joseph:rotation,
hopkins:clustering}. 

Observations have similarly linked mergers to 
at least some of the quasar population 
\citep{sanders88:quasars,canalizostockton01:postsb.qso.mergers,
guyon:qso.hosts.ir,dasyra:pg.qso.dynamics,bennert:qso.hosts}. 
Although the precise role of mergers is debated, the 
existence of tight correlations between black hole mass 
and spheroid properties such as stellar mass 
\citep{magorrian}, velocity dispersion \citep{FM00,Gebhardt00}, 
and binding energy \citep{hopkins:bhfp.obs,hopkins:bhfp.theory,
aller:mbh.esph,younger:minor.mergers} imply 
that the growth of black holes, dominated by bright quasar 
phases \citep{soltan82,salucci:bhmf,yutremaine:bhmf,
hopkins:bol.qlf,shankar:bol.qlf}, is 
fundamentally linked to the growth of spheroids.

Despite the importance of galaxy mergers, the galaxy-galaxy merger 
rate and its consequences remain the subject of considerable 
theoretical and observational debate. 
Halo-halo merger rates have been increasingly well-determined with 
improvements to high-resolution dark matter only simulations, 
with different groups and simulations 
yielding increasingly consistent results \citep[see e.g.][]{gottlober:merger.rate.vs.env,
stewart:mw.minor.accretion,stewart:merger.rates,
fakhouri:halo.merger.rates,wetzel:mgr.rate.subhalos,genel:merger.rates.perprogenitor}. 
But mapping halo-halo mergers to galaxy-galaxy mergers is non-trivial, 
and there are a number of apparent disagreements in the literature 
(both theoretical and observational) 
over the absolute rate of galaxy mergers as a function of 
galaxy mass and merger mass ratio. Moreover, although most of the 
literature has focused on the most violent events (major mergers), 
various high-resolution 
simulations have shown that a sufficiently large 
number of minor mergers (in a sufficiently 
short period of time) can do as much to build bulge\footnote{In 
this paper, we take ``bulge'' to refer to classical spheroids 
(Sersic index $\gtrsim2$, somewhat dispersion-supported spheroids), 
which are believed to primarily form in mergers, unless otherwise 
specified. We also use the term ``bulge'' to refer to any classical 
spheroid -- small bulges in disks through S0 and elliptical galaxies. 
} mass 
as a smaller number of major mergers \citep{naab:minor.mergers,
bournaud:minor.mergers,younger:minor.mergers,
cox:massratio.starbursts,hopkins:disk.survival}. So it remains a subject of 
debate whether or not minor mergers are important for (or may even dominate) 
bulge formation. Indeed, several questions arise. 
What is the galaxy-galaxy merger rate as a function of mass ratio, galaxy 
mass, and redshift? Which mergers -- major or minor -- are most important 
to bulge formation? Is this a function of galaxy mass and/or bulge-to-disk ratio? 
What is the typical merger history through which most of the bulge mass 
(and, by implication, black hole mass) in the Universe was assembled? 
What room does this leave for secular or non-merger related processes?

The relative importance of mergers of different mass 
ratios is critical for understanding the structure of spheroids and the 
related processes above. Although, in
principle, a sufficiently large number of 
minor mergers could build the same absolute bulge mass as a 
couple of major mergers, various simulations have shown that the two 
scenarios produce very different structural properties in the remnants, 
including rotation and higher-order kinematics, 
flattening and isophotal shapes, profile shapes, central densities, 
effective radii, and triaxialities \citep{weil94:multiple.merger.kinematics,
weil96:multiple.merger.scalings,burkert:anisotropy,
naab:size.evol.from.minor.mergers,
hoffman:dissipation.and.gal.kinematics}. 
Sufficiently minor mergers may not even preferentially form 
elliptical-like ``classical'' bulges, but rather disk-like 
``pseudo-bulges,'' which have typically been associated with 
secular (non-merger) processes \citep{younger:minor.mergers,
elichemoral:pseudo.from.minor}.  
These different channels clearly imply dramatically different formation timescales 
and histories, also important for understanding the star formation 
histories, stellar populations, colors, 
abundances and $\alpha$-enrichment of spheroids, and their gradients 
\citep[see e.g.][]{mihos:gradients,hopkins:cusps.ell,
ruhland:dry.mergers.and.color.mag.relation,
foster:metallicity.gradients}. 

Basic questions such as whether or not 
individual galaxies move continuously in small 
increments in the Hubble diagram, or can change types 
significantly, depend on the kind of mergers in which they form. 
Clearly, if e.g.\ small bulges in late-type disks 
are preferentially formed in early major mergers that destroy 
the disk (a new disk being later accreted), or if they form in situ 
in minor mergers that only partially affect the disk, the 
implications for their evolution and the demographics of bulges 
and disks are substantial. Moreover, to the extent that 
starburst and/or AGN activity are coupled to bulge formation in mergers, 
the magnitude, duty cycles, and cosmological evolution of 
these events is clearly directly linked to the kinds of mergers triggering 
activity -- one would expect continuous, high-duty cycle but low-level 
activity from sufficiently minor mergers, with more dramatic, dusty, 
shorter duty-cycle activity in major mergers \citep{hopkins:seyferts,
hopkins:seyfert.limits,hopkins:seyfert.bimodality}. 
Whether or not observations could, in principle, see evidence of 
merger-induced activity in these systems also depends on 
the kinds of mergers that dominate bulge formation, 
as does the question of whether or not every bulge/massive elliptical 
passed through a ULIRG/quasar phase in its formation. 

Conditions are ripe to address these questions. In addition to the convergence 
in theoretical predictions of the halo-halo merger rate, observations have 
greatly improved constraints on halo occupation statistics: namely, the stellar 
mass distributions of galaxies hosted by a halo/subhalo of a given 
mass. Observational constraints from various methods yield consistent results 
with remarkably small scatter (discussed below), and have been applied 
from redshifts $z=0-4$ (albeit with increasing uncertainties at higher 
redshifts). Furthermore, to lowest order, many of the 
most salient properties for this application appear to depend primarily on 
halo mass. That is, at fixed $M_{\rm halo}$ other properties such 
as redshift, environment, color, satellite/central galaxy status, or morphology 
have a minor impact 
\citep[see e.g.][]{yan:clf.evolution,
zentner:substructure.sam.hod,tinker:hod,cooray:highz,weinmann:obs.hod,
vandenbosch:concordance.hod,
conroy:mhalo-mgal.evol,zheng:hod.evolution}, 
the dependence of these properties on mass 
and one another all being expected consequences of a simple 
halo occupation distribution. Populating well-determined 
halo mergers with well-constrained galaxy properties 
can yield predictions for the galaxy-galaxy merger rate without reference to any 
(still uncertain) models of galaxy formation and with small uncertainties 
\citep[as demonstrated in][]{stewart:merger.rates}. 
Meanwhile, numerical simulations are beginning to converge 
in predicting how the efficiency of bulge formation 
scales with merger mass ratio and other basic parameters (orbital 
parameters, gas fractions, etc.), making it possible to robustly 
predict how much bulge should be formed by 
each event in a given merger history. 

In this paper, we present a simple, empirical model using this approach to 
predict galaxy-galaxy merger rates, and the relative contribution of 
mergers as a function of mass ratio, each as a function of galaxy mass, 
redshift, and other properties. We compare variations to the modeling 
within the range permitted by theory and observations, and show that 
the questions above can be answered in a robust, largely empirical fashion, and without 
reference to specific models of galaxy formation. 

In \S~\ref{sec:model} \&\ \S~\ref{sec:hod}, we outline the semi-empirical 
model adopted and show how observed halo occupation constraints lead to 
galaxy-galaxy merger rates. 
In \S~\ref{sec:mgr.rates} we use this semi-empirical model to predict galaxy-galaxy 
merger rates as a function of mass ratio, galaxy stellar mass, and redshift. 
In \S~\ref{sec:compare} we use these predicted merger rates together with the 
results of high-resolution simulations to identify the relative importance of 
different mergers as a function of mass ratio, mass, and redshift. 
In \S~\ref{sec:robustness}, we extensively vary the model parameters 
to test the robustness of these conclusions.
In \S~\ref{sec:discuss}, we summarize our conclusions.

Except where otherwise specified, we adopt a WMAP5 cosmology with 
$(\Omega_{\rm M},\,\Omega_{\Lambda},\,h,\,\sigma_{8},\,n_{s})$=
$(0.274,\,0.726,\,0.705,\,0.812,\,0.96)$ \citep{komatsu:wmap5} 
and a \citet{chabrier:imf} IMF, and appropriately
normalize all observations and models shown.
The choice of IMF
systematically shifts the normalization of stellar masses herein, but
does not otherwise change our comparisons. 

Throughout, we use the notation $M_{\rm gal}$ to denote the baryonic 
(stellar+cold gas) mass of galaxies; the stellar, cold gas, and dark 
matter halo masses 
are denoted $M_{\ast}$, $M_{\rm gas}$, and $M_{\rm halo}$, respectively. 
When we refer to merger mass ratios, we use the same subscripts 
to denote the relevant masses used to define a mass ratio 
(e.g.\ $\mu_{\rm gal} = M_{\rm gal,\, 2}/M_{\rm gal,\,1}$), 
always defined such that $0<\mu<1$ ($M_{\rm gal,\,1}>M_{\rm gal,\,2}$).

\breaker
\section{The Semi-Empirical Model}
\label{sec:model}

In order to track merger histories with as few assumptions as possible, 
we construct the following semi-empirical model, motivated by 
the halo occupation framework. Essentially, we assume galaxies obey 
observational constraints on disk masses and gas fractions, and then 
predict the properties of merger remnants. The model is described 
in greater detail in \citet{hopkins:disk.survival.cosmo}, and a similar 
variant based on subhalo mergers 
presented in detail in \citet{hopkins:groups.qso,hopkins:groups.ell}, 
but we summarize the key properties here. 

Note that what we describe here is our ``default'' model. In \S~\ref{sec:robustness}, 
we will systematically vary {\em every} part of this model, and show that 
our conclusions are robust.

\subsection{Step 1: The Halo+Subhalo Mass Function}

At a given redshift we initialize the halo+subhalo population, onto which 
we will ``paint'' galaxies. We do this following the standard 
\citet{shethtormen} mass function, calibrated to match the 
output of high-resolution N-body simulations.
Note that the {\em total} halo mass function (which is what 
we use to map central galaxy stellar mass to primary halo mass) 
is nearly identical to the halo+subhalo mass function -- as such, 
including the subhalos explicitly at this stage 
\citep[which we can do following the fits or 
methodology in e.g.][]{valeostriker:monotonic.hod,
gao:subhalo.mf,nurmi:subhalo.mf,vandenbosch:subhalo.mf} 
makes almost no difference to any of our results. 
This choice, and the known $\sim5\%$ uncertainties in halo mass functions are negligible 
compared to other uncertainties in our model. 
We will also, in \S~\ref{sec:robustness}, 
consider how systematic changes in the halo MF owing to different cosmological 
parameters affect our results, and show that the differences are minimal. 

This defines a primary halo mass function. There are of course subhalos in 
each halo -- defining the systems that will ultimately merge and therefore be of interest here -- 
but these come in the first place from mergers of other halos, which 
we will need to model. We describe this below.

\subsection{Step 2: Painting Central Galaxies Onto Halos}

At a given redshift, we use the halo occupation formalism to 
construct a mock sample of galaxies, each with a parent halo or subhalo. 
Specifically, we begin with the observed galaxy stellar mass function (MF), 
which we take as given. 
Generally, we will be interested in all galaxies (i.e.\ the total galaxy stellar mass 
function). But if we wish to specifically identify only gas-rich or star-forming galaxies, 
we can just use observations that separate the mass function of those galaxies alone 
(i.e.\ that of star-forming or ``blue'' galaxies).\footnote{
At $z<2$, we adopt the measurements of the type-separated 
galaxy stellar mass functions from \citet{ilbert:cosmos.morph.mfs}, 
if such separation is desired. At 
redshifts $z>2$, type-separated MFs are no longer available, 
so we simply assume all systems 
are star-forming; however, the fraction of massive 
galaxies that are ``quenched'' and red has become sufficiently low 
by $z=2$ (and is rapidly falling) that it makes little difference 
(e.g.\ adopting the upper limit -- that the red fraction at all 
masses at $z>2$ is equal to that at $z=2$ -- makes no difference to 
our predictions).}

We then assign each galaxy to a halo or subhalo in a simple manner following 
the standard halo occupation methodology described in 
\citet{conroy:hod.vs.z}; ensuring, by construction, that 
the galaxy mass function and galaxy clustering 
(as a function of stellar mass, galaxy color, and physical scale) is exactly reproduced. 
This essentially amounts to a simple ``rank ordering'' procedure, 
whereby both galaxies and halos+subhalos are rank-ordered in 
mass\footnote{In what follows, the term ``mass'' of a subhalo will always, 
unless otherwise specified, refer to the ``infall'' or maximum mass of the 
subhalo, pre-accretion (i.e.\ the maximum mass the system had while it 
was still a primary halo, before being accreted and becoming a subhalo). 
This is what is important for the galaxy properties, and what enters into 
the HOD rank-ordering methodology. Obviously, at a later time 
after accretion, tidal stripping can arbitrarily reduce the subhalo mass, 
but this should not change the original accreted galaxy mass.}, 
and then assigned to one another in a one-to-one fashion. 

\subsubsection{Uncertainties in Galaxy Masses and the Mapping}

The observational uncertainties in the galaxy abundance are a potential 
source of uncertainty in the model, especially at high redshifts. 
We will therefore consider three different determinations of the total 
galaxy stellar MF as a function of redshift, in our ``default'' model 
(the model is otherwise identical, but will use one or another determination 
of the MF). 
Further variations are explored in \S~\ref{sec:robustness}, as well. 
These are taken from 
(1) \citet{conroy:hod.vs.z} (our ``default'' choice, when 
no other is specified),
(2) \citet{perezgonzalez:mf.compilation}, 
and (3) \citet{fontana:mfs}. 

Each of these stellar MFs is directly constrained as a 
function of redshift out to at least $z>4$, more than sufficient for our (generally 
lower-redshift) predictions here. 
The specific choices are not important -- we could just as well use a 
number of other observations in the literature 
\citep[e.g.][]{bundy:mfs,pannella:mfs,franceschini:mfs,
borch:mfs,fontana:highz.mfs,brown:mf.evolution,
marchesini:highz.stellar.mfs,kajisawa:stellar.mf.to.z3}. 
We choose these three because they bracket the extremes in all of these different 
observations. Specifically, the differences between mass functions (1)-(3) 
represent the range across all these different samples. Those, in turn, 
reflect uncertainties owing to a combination of cosmic variance, 
selection/completeness, and different methods used to determine 
galaxy stellar masses \citep[which can easily contribute $\sim0.3-0.5\,$dex 
uncertainties in especially the 
high-redshift stellar mass measurements; see][]{moster:stellar.vs.halo.mass.to.z1,
marchesini:highz.stellar.mfs,behroozi:mgal.mhalo.uncertainties}. 
The compilation in \citet{perezgonzalez:mf.compilation} draws from a large 
number of different observations, and thus is a representative 
``average.'' The mass function choice in \citet{conroy:hod.vs.z} 
is interesting because the authors do not directly present 
a stellar MF measurement. Rather, they present a fitted stellar MF 
versus redshift that is fitted to both various observations of 
stellar masses and to the history of galaxy SFR versus mass and redshift. 
As such, the well-known discrepancy between high-redshift 
SFRs and stellar mass buildup is apparent \citep[see][]{hopkinsbeacom:sfh}. 
The authors address this by adjusting the stellar MF at high redshifts 
to fit the SFR measurements. As such, it represents a completely 
independent constraint on the stellar MF evolution, and 
is representative of the stellar MF that would be obtained with 
fairly radically different assumptions (for example, a stellar 
IMF that evolves with redshift). To this extent, it provides something of 
an upper limit to the uncertainties in the galaxy mass-halo mass mapping. 

Note that other methods of HOD fitting (other than 
our rank-ordering approach) yield very similar results. 
Allowing e.g.\ for scatter in the $M_{\rm gal}-M_{\rm halo}$ relation 
or fitting some prior assumed functional form (say e.g.\ a double-power 
law relation between these quantities, separately for central and satellite 
galaxies, with assumed lognormal scatter), yields little difference 
from the rank-ordering method and very small ($<0.1\,$dex) scatter 
\citep[see e.g.][]{conroy:hod.vs.z,wang:sdss.hod,
perezgonzalez:hod.ell.evol,brown:hod.evol,behroozi:mgal.mhalo.uncertainties}. 
We consider such methodological distinctions in \S~\ref{sec:robustness}, 
and find that they make little difference to our conclusions.

\subsubsection{Attaching Other Galaxy Properties}

If and when other galaxy properties are required as input for the model, 
these too are assigned according to observations. For example, 
in \S~\ref{sec:compare.fgas} we consider galaxy gas fractions. These 
are assigned to each galaxy according the observed correlations between galaxy 
gas mass and stellar mass, which have been quantified at a range of 
redshifts from $z=0-3$. As discussed there, we have compiled 
observations from the available sources, 
spanning this redshift range and a stellar mass range from 
$M_{\ast}\sim 10^{10}-10^{12}\,\msun$ (more than sufficient dynamic 
range for the predictions of interest here), 
specifically from \citet{belldejong:disk.sfh,mcgaugh:tf,
calura:sdss.gas.fracs,shapley:z1.abundances,erb:lbg.gasmasses,
puech:tf.evol,mannucci:z3.gal.gfs.tf,cresci:dynamics.highz.disks,
forsterschreiber:z2.sf.gal.spectroscopy,erb:outflow.inflow.masses,
mannucci:z3.gal.gfs.tf}. 

We find that these observations can be well-approximated as a function of 
galaxy stellar mass and redshift with the fitting functions presented in 
\citet{stewart:disk.survival.vs.mergerrates}, and therefore simply adopt those, 
with a simple lognormal scatter term of $\sim0.2\,$dex representative 
of the observed scatter at a given stellar mass. 
But adopting any individual measurement of these quantities instead gives 
very similar results.\footnote{At $z=0$, the gas fractions measured are based
  on measured atomic HI gas fractions; \citet{belldejong:tf} correct
  this to include both He and molecular H$_{2}$; \citet{mcgaugh:tf}
  correct for He but not H$_{2}$; \citet{kannappan:gfs} gives just the
  atomic HI gas fractions \citep[this leads to slightly lower
    estimates, but still within the range of uncertainty plotted;
    H$_{2}$ may account for $\sim20-30\%$ of the dynamical mass, per
    the measurements in][]{jogee:H2.masses}.  We emphasize that these
  gas fractions are lower limits (based on observed HI flux in some
  annulus).  At $z=2$,
  direct measurements are not available for most samples; the gas masses from
  \citet{erb:lbg.gasmasses} are estimated indirectly based on the
  observed surface densities of star formation and assuming that the
  $z=0$ Kennicutt law holds. However other observations suggest that it 
  does indeed hold \citep{bouche:z2.kennicutt}, and  
   other indirect estimates 
  yield similar results \citep{cresci:dynamics.highz.disks,mannucci:z3.gal.gfs.tf}. 
  Moreover recent observations have been able to directly measure the molecular 
  gas content of galaxies at $z\sim2-4$, and where available these 
  measurements agree very well with the assumed $f_{\rm gas}(M_{\ast},\,z)$ 
  relations used here \citep[see e.g.][]{tacconi:high.molecular.gf.highz}. We therefore 
  conclude that although appropriate caution is due at high 
  redshift, radical departures from our assumptions are unlikely. 
  }
Of course, the uncertainties in gas fractions and other properties increase at higher 
redshifts, but these are still sub-dominant in their ultimate effects compared to the 
growth in uncertainty in the stellar mass function itself.

\subsection{Step 3: Halo-Halo Mergers}

The next step towards identifying merging galaxies is to identify 
merging halos. The halo-halo merger rate  
(i.e.\ the rate at which formerly primary halos are accreted onto other primary halos 
and thus become subhalos) is well-measured in 
cosmological simulations and defined in the extended Press-Schechter formalism. 
In our default model, we determine this rate, as a function of halo 
mass, merging halo mass ratio, and redshift, from the fits 
presented in \citet{fakhouri:halo.merger.rates}, determined 
from high-resolution N-body simulations \citep{springel:millenium}. 
We can include scatter in this as well; \citet{fakhouri:merger.rate.env} quantify 
the scatter in halo-halo merger rates across populations. 
For a given halo population and arbitrary time interval, we can then 
statistically assign all halo-halo mergers that have occurred in that interval.
We vary the determination of the halo-halo merger rate in 
\S~\ref{sec:robustness} (using e.g.\ the results of different 
dark matter simulations, simulations with 
gas included, and selecting halos differently), and find that 
it makes little difference. 

The galaxy properties of the secondary are defined at the time 
when it first becomes a subhalo -- i.e.\ when the secondary halo mass is 
its infall/maximum pre-accretion mass (when the halo-halo 
merger occurs). The primary galaxy continues to obey the 
normal $M_{\rm gal}-M_{\rm halo}$ relation for its total halo mass. 
So if the $M_{\rm gal}-M_{\rm halo}$ relation evolves with redshift, 
the primary galaxy will lie on the relation defined at the moment the 
actual merger occurs. The secondary will lie on the relation defined 
at the moment of the halo-halo merger (the subhalo instantaneous 
mass by the time of merger being much smaller). 
This is the basis for the rank-ordering method of assigning 
subhalo populations, and is key to the agreement 
between the observed small-scale clustering of galaxies 
and that produced by the models here.

\subsection{Step 4: From Halo-Halo to Galaxy-Galaxy Merger}

Of course, a halo-halo merger is not a galaxy-galaxy merger. 
We need to follow the recent subhalos until the time when the 
galaxies themselves will actually merge. 
There are two general methods to do this. The first, and our choice in 
this ``default'' model, is to assign a ``merger time'' -- essentially, a 
delay timescale, usually calibrated from the results of high-resolution 
simulations, that represents the time for orbital decay from the initial 
halo-halo merger. 
This is the standard approach adopted in many semi-analytic models and 
models based on the extended Press-Schechter formalism. 
There are different possible choices for this merger time, 
and we will consider several of them in \S~\ref{sec:robustness}. 
For our default model, we adopt the most common: the dynamical 
friction time. Specifically, we use the dynamical friction timescale for 
galaxy-galaxy merger with respect to the initial halo-halo merger, 
calibrated as a function of galaxy mass, mass ratio, redshift, 
and orbital parameters in high-resolution merger simulations 
in \citet{boylankolchin:merger.time.calibration}. 
Specifically the formula they present is:
\begin{equation}
t_{\rm df} = 0.0216\,H^{-1}\,\frac{(M_{1}/M_{2})^{1.3}}{\ln{(1+M_{1}/M_{2})}}\,
\exp{\left[1.9\,\eta \right]}\,\left[ \frac{r_{c}(E)}{r_{\rm vir}} \right]
\end{equation}
where $H$ is the Hubble constant at $z$, 
$M_{1}$ and $M_{2}$ the halo masses at the moment of the halo-halo merger, 
$\eta=j/j_{c}(E)$ is the standard ``circularity parameter'' ($\langle\eta\rangle\approx0.5$), 
and $r_{c}(E)$ is the circular radius for an orbit with energy 
$E$ (trivially related to the pericentric passage distance $r_{\rm peri}/r_{\rm vir}$, 
more usually quoted). 
Note that this explicitly depends on the merger orbital parameters. 
This allows us to incorporate the scatter in merger times seen 
in full cosmological simulations. 
Specifically, the cosmological distribution of the orbital parameters 
$\eta$ and $r_{\rm peri}/r_{\rm vir}$ 
are presented in \citet{benson:cosmo.orbits,khochfar:cosmo.orbits}. 
Drawing randomly from these distributions, we can 
thus determine some Monte Carlo merger population.\footnote{Note 
that recently, \citet{wetzel:sat.orbit.vs.halomz} has shown that these distributions depend 
non-trivially on mass and redshift. However, the sense of that dependence 
is such that, if anything, the average merger timescales adopted here are 
always upper limits. As such, incorporating this more detailed dependence 
only strengthens our conclusions (see our discussion in \S~\ref{sec:robustness}).}

In general, varying the prescription for the merger ``delay,'' 
across the entire physically plausible range, as we do 
in \S~\ref{sec:robustness}, leads to 
factor $\sim2$ systematic differences in the merger rate. 

An alternative approach 
to following galaxies to their galaxy-galaxy merger from the halo-halo 
merger is to track the subhalos directly in cosmological simulations. 
In other words, 
given some cosmological population of halos, we can follow the bound substructure of the 
subhalo after an initial halo-halo merger, until the subhalo reaches some sufficiently small 
radius or is completely disrupted, at which point we define the ``merger'' to have occurred. 
Given this methodology, we can define a ``subhalo'' merger rate -- i.e.\ a merger rate of 
subhalos being destroyed in primary halos (as opposed to a rate of those subhalos simply 
``falling into'' those halos). If the simulation is sufficiently high resolution, the 
subhalo final merger/destruction should correspond closely to the 
actual galaxy-galaxy merger between the primary halo central galaxy and the 
satellite galaxy hosted by the subhalo (actually, so long as the subhalos are tracked 
long enough that the ``remaining'' merger time, after subhalo destruction, is small 
relative to the Hubble time, this is sufficient). 

With such a measured subhalo destruction 
rate from simulations, we can convolve with the determined 
$M_{\rm gal}-M_{\rm halo}$ distribution and define directly the galaxy-galaxy merger rate. 
Again we stress that $M_{\rm halo}$ for the subhalos, as enters into this 
calculation and determines $M_{\rm gal}$ is the infall or maximum pre-accretion 
mass (as it generally should be, since the galaxy, pre-accretion, would lie 
on the $M_{\rm gal}-M_{\rm halo}$ relation for a normal central galaxy, not ``knowing'' 
it would be accreted at some time in the future), as the subhalo instantaneous (post-accretion, 
stripped) mass goes to zero at the time of destruction/merger. 
Such an approach 
automatically accounts for e.g.\ the orbit distribution of subhalos; however, it has 
its own uncertainties -- subhalos are not baryonic galaxies, and the subhalo-subhalo 
merger time can differ from the galaxy-galaxy merger time by factors of up to several 
(owing to the resonant effects of baryons, and finite resolution limits). 
In any case, in \S~\ref{sec:robustness}, we will consider several such determinations of the 
subhalo merger rates instead of our merger ``delay'' 
approach, with each using slightly different methodologies 
\citep[see e.g.][]{stewart:merger.rates,kravtsov:subhalo.mfs,zentner:substructure.sam.hod,
vandenbosch:subhalo.mf}, and show that they 
yield similar results to our merger ``delay'' approach. We should 
note that the formula from \citet{boylankolchin:merger.time.calibration} is calibrated as the total 
time from halo-halo merger to galaxy-galaxy merger in live simulations; thus, it 
implicitly includes all the effects seen in a full simulation (e.g.\ continuous mass 
loss/stripping of the satellites, resonant effects on near-passage of the galaxies, 
and baryonic effects on the halos). 

Thus, after the convolution with the merger timescale, we obtain 
the rate of galaxy-galaxy mergers 
as a function of galaxy mass $M_{\ast}$, redshift $z$, and galaxy-galaxy 
baryonic mass ratio $\mu_{\rm gal}$.

\subsection{Step 5 (Optional): Linking Populations Across Different Epochs}

If we desire only instantaneous quantities (e.g.\ the merger rate), then 
what we have already described is entirely sufficient, and we can simply re-initialize the model 
at any redshift where we want to make predictions.  

However, we will occasionally desire integral quantities (for example, how many mergers 
a given galaxy is likely to have experienced in its history). 
There are several choices of method to calculate these. 

The simplest is our choice in the ``default'' model. 
Recall, we only need to integrate quantities such as the merger history 
{\em statistically}. And we will quote quantities such as the number of mergers 
for the primary branch of the halo merger tree. 
As such, we can use the fact that for a halo 
of mass $M_{0}$ at $z=0$, the median 
primary progenitor mass at higher redshift $\langle M_{h}(M_{0},\,z)\rangle$ 
(and its scatter) is a 
well-measured function from cosmological simulations 
(in other words, for a halo of mass $M_{0}$, we know the distribution of progenitor 
masses $M_{h}(M_{0},\,z)$ at some higher $z$). 
We can then integrate over average histories, for example for the merger rate: 
\begin{equation}
\langle N_{\rm merger}(M_{0},\,\mu) \rangle 
= \int \frac{dN_{\rm merger}}{{\rm d}z}(M_{h}[M_{0},\,z],\,z,\,\mu)\,{\rm d}z\ .
\end{equation}
In our default model, we adopt the fits to the average halo 
growth tracks/progenitor mass distribution 
given as a function of halo mass and redshift in \citet{neistein:natural.downsizing}, 
calibrated to match the results of high-resolution $N$-body simulations. 
Of course, at each redshift $z$ in the integral above, our steps $1-4$ are 
implicit in obtaining the relevant galaxy properties. 

Note that this is effectively the same as beginning with some population 
of halos and evolving them forward along the average growth tracks 
defined by $M_{h}$ above, 
where we integrate the halo mass forward using the merger rates 
and assign any shortfall to ``unresolved'' or ``diffuse'' accretion 
(since only mergers with mass ratios $\mu_{\rm gal}\gtrsim10$ are 
of real significance here, it makes no difference if this mass technically 
comes from very small halos or truly diffuse material). This 
guarantees that the halo+subhalo population matches the 
total halo mass function at all times. 
Re-assigning quantities such as $M_{\rm gal}(M_{\rm halo})$ 
at each time, according to the observational constraints at each redshift, 
is equivalent to assigning the galaxy some net growth in 
stellar mass (star formation rate) and gas mass (inflow minus outflow), 
but without a prior on how much of each is in a bulge or disk component.\footnote{
If, for example via some rapid mergers, a galaxy exceeds the 
$M_{\rm gal}(M_{\rm halo})$ relation, 
our model implicitly assumes its growth stalls until 
the halo ``catches up''; if it falls below the relation, it experiences 
the necessary gas accretion and star formation for itself to 
catch up.
}
We can therefore compare the implied growth rate to observations, 
such as the observed stellar mass-star formation rate relation 
\citep{noeske:sfh}.
For example, the interpolation between HODs at different redshifts 
implies that the SFR in star-forming galaxies scales very crudely 
as $\propto M_{\ast}^{(0.3-0.6)}\,(1+z)^{(1.5-2.5)}$ at $z\lesssim2$ 
(with a sharp decline above the turnover mass $\sim L_{\ast}$), 
which agrees well with a number of different observational 
estimates \citep[see e.g.][]{blain:ir.lf.synthesis.model,
noeske:2007.sfh.part1,noeske:sfh,papovich:ssfr,martin:uv.lum.history,
bell:morphology.vs.sfr,damen:ssfr.highz.massive.gals}. 
In fact, a more rigorous comparison shows quite good agreement -- 
this exercise, for a halo occupation model that is effectively 
equivalent to the one used here, is presented in \citet{conroy:hod.vs.z},
who find very good agreement both with observations of the 
integrated SFR versus redshift and the specific SFR in galaxies 
at different masses. 
Also, see \citet{lee:2009.uv.lum.vs.mhalo} and 
\citet{zheng:hod.evolution}, 
who perform a corresponding analysis
and obtain similar conclusions. 
The variation in predictions between our different models 
is comparable to that in the different models considered in these papers. 
(And recall, at least one of the mass function fits we adopt throughout is 
adjusted specifically to match the SFR versus galaxy mass and redshift 
relation; as such, agreement with these observations is ensured by 
construction). 
And despite its simplicity, our approach effectively guarantees a match to 
various other halo occupation statistics including stellar 
mass functions and the fraction of active/passive galaxies 
as a function of mass \citep{yang:obs.clf,weinmann:obs.hod,
weinmann:group.cat.vs.sam,gerke:blue.frac.evol}.

\subsection{Step 6 (Optional): The Effects of Mergers on Morphology}

Finally, we can couple the model to what is seen in high-resolution 
simulations of galaxy-galaxy mergers, to say what effects mergers 
will have on the galaxy morphologies. 

In the model here, the galaxies are initially (at high redshift) disks; 
when a merger occurs, we use the model results 
from \citet{hopkins:disk.survival} to determine how much of the galaxy is 
converted from disk to bulge. The models used a suite of 
several hundred high-resolution 
hydrodynamic simulations, including star formation, black hole 
growth, and feedback from both to quantify the efficiency of 
bulge formation as a function of merger mass ratio,\footnote{
It is important to use the ``appropriate'' mass 
ratio, for which the scalings presented in \citet{hopkins:disk.survival} 
are calibrated. In detail, the authors find that the most dynamically 
relevant mass ratio is not strictly the baryonic galaxy-galaxy 
ratio $\mu_{\rm gal}$ nor the halo-halo ratio $\mu_{\rm halo}$. 
Rather, the important quantity is the tightly bound material that survives 
stripping to strongly perturb the primary. Generally speaking, this 
is the baryonic plus tightly bound dark matter mass \citep[the central 
dark matter mass, being tightly bound in the baryonic potential well, 
is robust to stripping in simulations;][]{quinn86:dynfric.on.sats,
benson:heating.model,kazantzidis:mw.merger.hist.sim,
purcell:minor.merger.thindisk.destruction}. 
This can be reasonably approximated as the baryonic mass 
plus dark matter mass within a small radius of one NFW scale 
length ($r_{s}=r_{\rm vir}/c$, where $c\sim10$ is the halo 
concentration; i.e.\ a few disk scale lengths). We find that around 
this range in radii, our results are not very sensitive to the precise 
definition, and in general, the baryonic mass ratio $\mu_{\rm gal}$ 
is a good proxy for the mass ratio calculated with this baryonic+tightly 
bound dark matter mass. However, in e.g.\ dark-matter dominated systems 
such as low-mass disks, the difference is important in particular for the 
absolute efficiency of bulge formation. Therefore, since this is what the 
simulation results are calibrated for, we adopt this mass ratio definition to 
calculate the dynamics in a given merger. However, this is not observable; 
we therefore present the predicted merger rates and their consequences 
in terms of an observable 
and easily-interpreted ratio $\mu_{\rm gal}$
(this also makes it possible to compare to other results in the literature). 
Again, the qualitative 
scalings are the same, but it is important to use the full information 
available in the model to calculate the merger dynamics. 
We include all mergers 
above a minimum mass ratio $\mu_{\rm gal}\sim0.01$, although 
our results are not sensitive to this limit so long as it is small.
} 
orbital parameters, 
gas fraction, and other properties. The large suite of simulations 
spans the parameter space of interest, so there are good 
simulation analogues to the cosmologically anticipated mergers here, 
for which we can simply adopt the fits therein to model their bulge formation. 

The details are presented in \citet{hopkins:disk.survival} -- specifically 
they provide fitting functions (their Equations 7-10 \&\ 27) that 
give the exact fraction of both the stellar and gaseous disks converted to 
bulge, as a function of merger mass ratio, gas fraction, and orbital 
parameters (all quantities that 
are determined in our model).\footnote{The efficiency of bulge formation 
does depend on merger orbital parameters -- namely the
relative inclination angles of the merging disks --- so we
simply draw them at random assuming an isotropic distribution of
inclinations (allowing some moderate inclination bias makes no difference). 
} These scalings are all tested against the suite of simulations therein, 
and shown to provide accurate fits with $<0.3\,$dex scatter over the 
entire dynamic range of merger types of interest for this paper 
(see their Figure~16). 
In detail, the authors there define ``bulge'' as the stellar population that is 
dynamically supported by dispersion (as compared to ``disk'' stars supported 
by rotation, and all cold gas, hence star formation, which is part of the disk). 
They show that this agrees fairly well with decomposition of the observable 
stellar mass profile into an exponential disk and $r^{1/4}$-law bulge. 
But the bulges thus defined are generally ``classical'' bulges, by various 
observational definitions; as such, this is the ``bulge mass'' that we 
predict in this paper (pseudobulges, in any case, are believed to form 
primarily via disk instabilities, which we are not modeling here). 
We refer to that paper for the full equations, but in an approximate sense, 
in a merger of mass ratio $\mu$, a fraction $\sim\mu$ of the primary stellar 
disk is violently relaxed and adds its low-density 
material to the bulge, and a fraction
$\sim(1-\fgas)\,\mu$ of the gas loses its angular momentum
and participates in a nuclear starburst, adding high-density starburst 
material to the bulge. The factor $(1-\fgas)$ represents the fact that 
the gas cannot be violently relaxed as stars are (because it is collisional); 
rather, it must lose angular momentum to the stars (which have a mass 
fraction in the disk $\propto 1-\fgas$) -- so in systems with increasing gas 
fractions, the efficiency of gas angular momentum loss decreases. 
This has very important consequences both for understanding what has 
now been seen in essentially all simulations of sufficiently high gas-fraction 
disks \citep[both high-resolution and cosmological; see e.g.][]{springel:spiral.in.merger,
robertson:disk.formation,
cox:massratio.starbursts,robertsonbullock:highz.disk.vs.model,
okamoto:feedback.vs.disk.morphology,scannapieco:fb.disk.sims,
governato:disk.formation,governato:disk.rebuilding}, 
and for the global census of bulge and disk populations and 
survival of e.g.\ thin disks \citep{stewart:disk.survival.vs.mergerrates,
hopkins:disk.survival.cosmo}.
The true scalings are more detailed \citep[see][]{hopkins:disk.survival}
but this represents the important physics. 

In \S~\ref{sec:robustness}, we discuss the consequences of 
using other, less accurate approximations to the behavior seen in 
full high-resolution simulations of galaxy mergers. We find that 
so long as the key scalings with mass ratio are at least qualitatively 
similar, our conclusions are unchanged.

\subsection{Summary}

Together, these simple assumptions are sufficient to define a
``background'' galaxy population. There are degeneracies in 
the model -- however, we are not claiming that this is unique nor that 
it contains any physics other than the number and effects of mergers. For our
purposes, the precise construction of the empirical model is not
important -- our results are unchanged so long as the same galaxy mass-halo 
mass relation and gas
fraction distributions are reproduced as a function of galaxy mass and
redshift. The importance of all the above is the following.
{\bf (1)} Observed galaxy stellar mass 
functions, the galaxy stellar mass and halo or subhalo mass relations, 
and the distributions of galaxy gas fractions as a function of 
stellar mass and redshift 
are all reproduced exactly as observed, at all redshifts $z=0-4$ where 
observational constraints exist, {\em by construction} in the model. 
{\bf (2)} Observed galaxy-galaxy clustering, both on large scales 
(the ``two-halo'' term, where it is primarily a function of halo mass) and 
small scales (the ``one-halo'' term, where it reflects subhalo/satellite 
halo occupation statistics), at those redshifts $z=0-4$, are 
also reproduced {\em by construction}. Since this is built into the model 
explicitly, we do not show such a comparison, but \citet{conroy:monotonic.hod}, 
\citet{wang:sdss.hod}, and \citet{zheng:hod.evolution} 
all show illustrations demonstrating that the simple methodology here yields 
(again by construction) excellent agreement with observed galaxy-galaxy 
correlation functions from scales of $50\,$kpc through $10\,$Mpc, 
at redshifts $z=0,\,1,\,2,\,3$ and $4$, as a function of galaxy mass or 
luminosity, and galaxy color. 
{\bf (3)} The only purpose of simulations ultimately, in this model, is (as in 
all halo occupation models) to provide a means of linking (statistically) 
galaxies at two different times. In other words, knowing the observed distribution of 
galaxy stellar masses and their separations, we use the dynamics which 
can be followed in the simulation to say which of these galaxies will 
merge in a given time interval (or again, some statistically approximation such 
as which fraction of the systems inside some distance). Likewise, we can 
ask where the systems that have such mergers ``end up'' in halo or stellar mass.

Together, this defines the suite of models adopted here. 
This is, by construction, a minimal model, and may leave out important 
details. However, in \S~\ref{sec:robustness}, we systematically 
vary the model assumptions, and find that our conclusions are robust.

\breaker
\section{The Consequences: How Halo Occupation Statistics Change 
Galaxy-Galaxy Merger Rates}
\label{sec:hod}

\begin{figure*}
    \centering
    \plotter{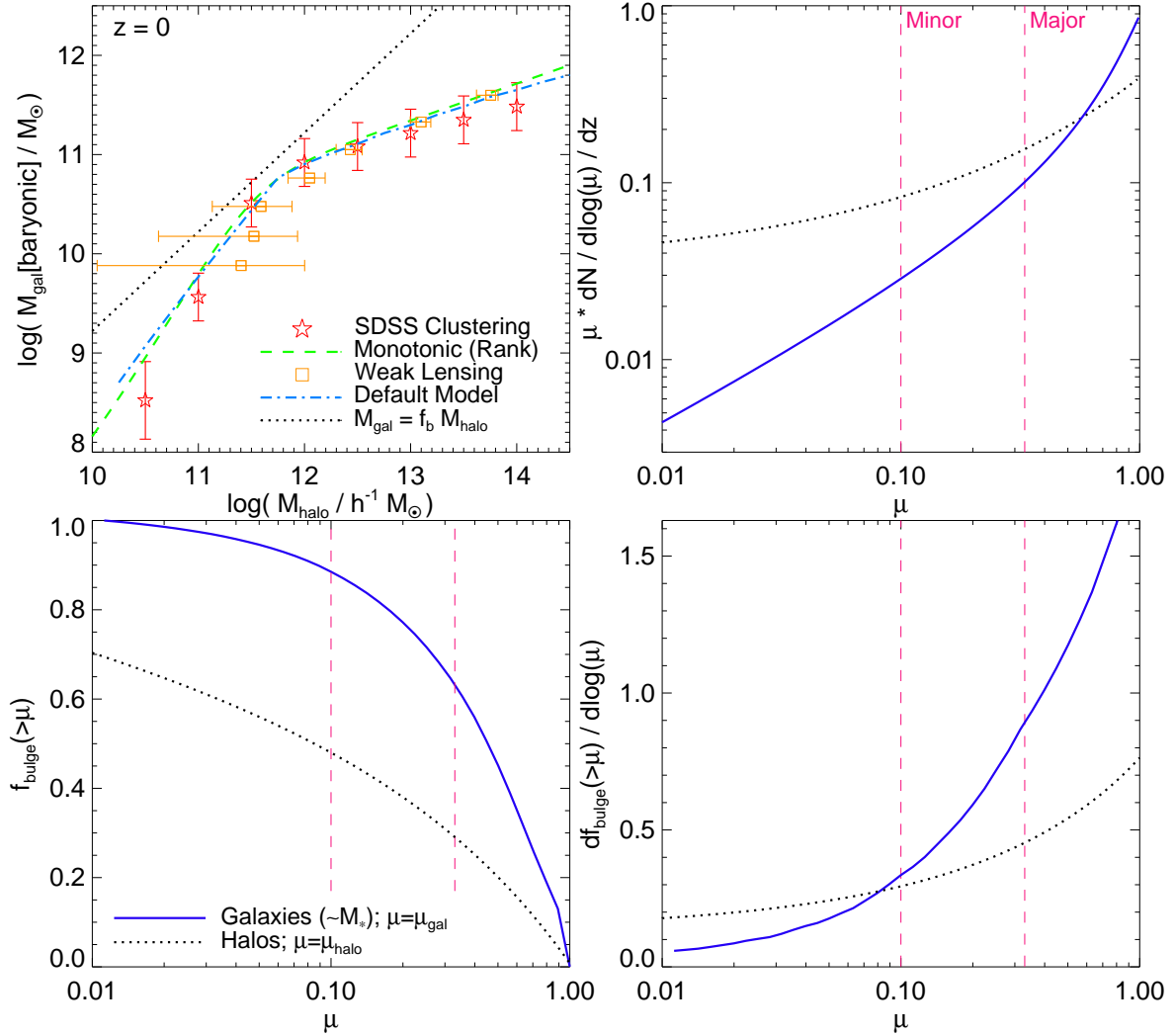}    
    \caption{{\em Top Left:} 
    Halo occupation: median baryonic 
    galaxy mass of the primary galaxy in a halo or subhalo, 
    as a function of that halo or subhalo's mass (at $z=0$). For {\em central} 
    galaxies, this refers to the total (primary) halo mass; for {\em satellite} galaxies, 
    to their subhalo ``infall'' mass (i.e.\ mass the now-subhalo had 
    at the last time it was a primary halo). 
    Dotted line represents maximally efficient star formation ($f_{b}$ is the Universal 
    baryon fraction). We compare empirical constraints from 
    clustering \citep{wang:sdss.hod}, weak lensing \citep{mandelbaum:mhalo}, 
    and abundance matching \citep[`monotonic';][]{conroy:monotonic.hod}. 
    Our default model is constructed to match these constraints. 
    {\em Top Right:} 
    Differential contribution to growth from different mass ratio mergers, i.e.\ 
    merger rate per logarithmic interval in mass ratio and unit redshift, 
    ${\rm d}N_{\rm mergers}\,{\rm d}\log{\mu}^{-1}\,{\rm d}\,z^{-1}$, weighted by $\mu$. Dotted line 
    is for halos or, equivalently, would be for galaxies if $M_{\rm gal}\propto M_{\rm halo}$ 
    were the actual HOD. Blue solid line is the result for $\sim L_{\ast}$ 
    ($M_{\ast}\approx10^{11}\,\msun$) galaxies, 
    given the observed HOD.
    {\em Bottom Left:} Cumulative contribution of different mass ratio mergers to the 
    $z=0$ spheroid mass density (or halo mass density), integrated over all 
    galaxy (halo) masses. 
    {\em Bottom Right:} Differential version of the same. 
    Because galaxy mass is not simply proportional to halo mass, 
    bulge growth is dominated by major mergers while halo growth is 
    contributed to by a wide range of mass ratios. 
    We focus on $M_{\rm gal}$ rather than just the stellar mass $M_{\ast}$ because the 
    former, not the latter, matters for the dynamics in mergers; 
    however the conclusions using $M_{\ast}$ are qualitatively identical. 
    \label{fig:fmu.gal.vs.halo}}
\end{figure*}

Figure~\ref{fig:fmu.gal.vs.halo} illustrates how the combination of halo-halo 
merger rates and the observed halo occupation distribution determines 
galaxy-galaxy merger rates. 

First, we show the halo occupation function itself 
(top left of Figure~\ref{fig:fmu.gal.vs.halo}): for our purposes, this function 
is summarized in the most important quantity, the average galaxy baryonic mass 
hosted by a halo of a given mass $M_{\rm gal}(M_{\rm halo})$. 
If galaxy formation were efficient 
this would simply be $M_{\rm gal}=f_{b}\,M_{\rm halo}$, 
where $f_{b}$ is the Universal baryon fraction, and galaxy-galaxy mergers would 
directly reflect halo-halo mergers. However, 
the relation between $M_{\rm gal}$ and 
$M_{\rm halo}$ is non-trivial. 

We show several observational constraints on this quantity at redshift $z=0$ 
(although we note that it is not redshift-independent, according to the observational 
constraints used for the HOD herein). 
First, the combination of the observed abundance and clustering of galaxies 
of a given mass have long been known to set tight constraints on 
$M_{\rm halo}(M_{\rm gal})$. We show a recent determination of these 
constraints, from clustering and abundance of local SDSS galaxies 
in \citet{wang:sdss.hod}.\footnote{These authors and others 
determine constraints in terms of galaxy stellar mass $M_{\ast}$; 
where necessary, we use our standard fit to observed gas 
fractions as a function of galaxy stellar mass and redshift 
(see \S~\ref{sec:model}) to convert freely between 
baryonic ($M_{\rm gal}$) and stellar ($M_{\ast}$) 
galaxy masses at all redshifts.}
Second, we show the empirical 
``monotonic'' or ``rank ordering'' results: it has been shown that good fits 
to halo occupation statistics (group counts, correlation functions as a function of 
galaxy mass and redshift, etc.) over a range of 
redshifts $z=0-4$ are obtained by simply rank-ordering all galaxies 
and halos+subhalos in a given volume and assigning one to another in a monotonic 
one-to-one manner \citep[see e.g.][]{conroy:monotonic.hod,valeostriker:monotonic.hod}. 
Here we plot the results of this exercise using the $z=0$ stellar mass 
functions from \citet{bell:mfs}. Third, we compare this to independent 
estimates of the average $M_{\rm halo}(M_{\rm gal})$ from weak lensing 
studies in \citet{mandelbaum:mhalo}. 
Other independent constraints give nearly identical results: these include 
halo mass estimates in low-mass systems from rotation curve 
fitting \citep[see e.g.][]{persic88,persic96,borriello01,avilareese:baryonic.tf}, 
or in high-mass systems from X-ray gas or group kinematics 
\citep{eke:groups,yang:obs.clf,brough:group.dynamics,
vandenbosch:concordance.hod}. 

Because it is the total baryonic mass, not just the stellar mass, that matters 
for e.g.\ defining the dynamics in a merger (a galaxy can be 
very massive and, in principle, contain little stellar mass), we will focus throughout this 
paper on that quantity ($M_{\rm gal}$ and $\mu_{\rm gal}$). 
However, Figure~\ref{fig:fmu.gal.vs.halo} is qualitatively identical 
if we use just the stellar mass $M_{\ast}$ instead. 
In future work (in preparation), we will compare the consequences for the 
HOD, merger rates, and observable quantities such as the 
merger fraction resulting from different choices $M_{\ast}$, 
$M_{\rm gal}$, $M_{\rm halo}$ etc.\ in the definitions of merger 
mass ratio. 

The second ingredient in predicting merger rates is the halo-halo 
merger rate determined from 
$N$-body simulations. Defined as 
the number of halo-halo mergers per primary halo, per logarithmic interval 
in mass ratio $\mu_{\rm halo}\equiv M_{\rm halo,\,2}/M_{\rm halo\,1}$, 
per unit redshift (or per Hubble time), 
the halo merger rate function can be 
approximated as \citep{fakhouri:halo.merger.rates} 
\begin{equation}
\frac{{\rm d}N_{\rm mergers}}{{\rm Halo}\ {{\rm d}\log{\mu_{\rm halo}}}\ {{\rm d}z}} \approx 
F(M_{\rm halo})\,G(z)\,\mu_{\rm halo}^{-1}\,
\exp{{\Bigl\{}{\Bigl(}\frac{\mu_{\rm halo}}{0.1}{\Bigr)}^{0.4}{\Bigr\}}}. 
\end{equation}
In these units, halo merger rates are nearly mass and redshift-independent: 
$F(M_{\rm halo})\approx0.03\,(M_{\rm halo}/1.2\times10^{12}\,\msun)^{0.08}$ and 
$G(z)\approx({\rm d}\delta_{c}/{\rm d}z)^{0.37}$ are weak functions of $M_{\rm halo}$ 
and $z$, respectively (in terms of mergers per unit {\em time}, 
this rate increases roughly as $(1+z)^{2}$). 
A number of other authors give alternative fits 
\citep[see e.g.][]{stewart:merger.rates,genel:merger.rates.perprogenitor}, 
but the salient features are similar: weak redshift and halo mass dependence 
(in these units), power law-like behavior at low mass ratios with a slope 
of roughly $\mu_{\rm halo}^{-1}$ and an excess above 
this power-law extrapolation at high $\mu_{\rm halo}$. 

The (fractional) contribution to halo growth from 
each interval is just $\mu_{\rm halo}$ times the merger rate; we plot this quantity in 
Figure~\ref{fig:fmu.gal.vs.halo} (top right) as we are ultimately interested in 
which mergers contribute to bulge growth. 
Because halo-halo merger rates go roughly as 
$dN_{\rm merger}/d\log{\mu_{\rm halo}}\propto\mu_{\rm halo}^{-1}$ (reflecting the shape of the 
halo mass function itself, and the nearly scale-free nature of 
CDM cosmologies), 
similar mass is contributed to the halo from each logarithmic interval in $\mu_{\rm halo}$. 

Convolving the halo-halo merger rates with the 
HOD (i.e.\ populating each halo with a galaxy of the appropriate mass), and the 
appropriate time lag between halo-halo and galaxy-galaxy merger, 
we obtain the galaxy-galaxy merger rate, now in terms of the {\em galaxy-galaxy} 
mass ratio $\mu_{\rm gal} = M_{\rm gal,\, 2}/M_{\rm gal,\, 1}$. 
In Figure~\ref{fig:fmu.gal.vs.halo}, we compare 
this function (evaluated at $\sim L_{\ast}$ 
or $M_{\rm gal}\approx10^{11}\,\msun$, where most of the stellar mass in the 
Universe is concentrated) to that obtained for halos. 

Integrating over the galaxy history in our models, Figure~\ref{fig:fmu.gal.vs.halo} (bottom left)
shows the fraction of the total $z=0$ bulge/halo mass contributed by mergers with a 
mass ratio above some $\mu_{\rm gal}$, i.e.\ 
\begin{equation}
f_{\rm bulge}(>\mu_{\rm gal}) \equiv \frac{1}{M_{\rm bulge}} 
\int \Theta{(\mu_{\rm gal}^{\prime}/\mu_{\rm gal})}\,{\rm d}m_{\rm bulge}\ , 
\label{eqn:fbul.defn}
\end{equation}
where $\mu_{\rm gal}^{\prime}$ refers to the mass ratio 
of the merger that formed each differential unit ${\rm d}m_{\rm bulge}$ 
of the final bulge mass $M_{\rm bulge}$, and 
$\Theta(x)=1$ for $x>1$ ($\mu_{\rm gal}^{\prime}>\mu_{\rm gal}$) 
and $\Theta(x)=0$ for $x<1$ ($\mu_{\rm gal}^{\prime}<\mu_{\rm gal}$). 
Note that this can be defined over the bulge mass of an individual galaxy, 
over all bulge mass in galaxies in a narrow interval in mass $M_{\rm gal}$, 
or over all galaxies (i.e.\ integrating over the bulge mass function). 
We show the latter.\footnote{In principle, some bulge mass could come 
from redshifts before our ``initial'' tracking of each halo, but in practice 
at any redshift, most of the mass has assembled relatively recently, so our 
results do not depend sensitively on the initial conditions.} 
We also show the differential version (bottom right): the fraction of 
$z=0$ bulge mass contributed per logarithmic interval in merger mass ratio 
${\rm d}f_{\rm bulge}/{\rm d}\log{\mu_{\rm gal}}$. 

For halos (or equivalently for galaxies if the trivial mapping $M_{\rm gal}\propto M_{\rm halo}$ 
were true), the distribution of bulge mass fraction from mergers of different mass ratio is quite broad, 
as expected: only $\sim50\%$ of halo mass comes from mergers with $\mu>0.1$. 
Because halo-halo merger rates are nearly self-similar, 
the differential version of this reflects the instantaneous rate also shown, 
with similar contributions per logarithmic interval in halo mass ratio. 
It is still the case, though, that ten 1:10 mergers are less common 
than a 1:1 merger, meaning that
major mergers dominate \citep{stewart:mw.minor.accretion}. 

In contrast, galaxy bulge assembly is biased much more towards 
high-mass ratio mergers, at least for 
$\sim L_{\ast}$ systems which dominate the mass density. 
This owes to the nature of halo 
occupation statistics: at low masses, galaxy mass grows rapidly with halo mass 
(galaxy formation is increasingly efficient as one moves from low masses 
closer to $\sim L_{\ast}$). Upon reaching $\sim L_{\ast}$, however, star formation 
shuts down relative to halo growth -- in terms of the HOD, the scaling of galaxy 
mass with halo mass transitions from steep $M_{\rm gal}\propto M_{\rm halo}^{1.5-2.0}$ 
at low masses to shallow $M_{\rm gal}\propto M_{\rm halo}^{0.2-0.5}$ at high 
masses. In short, as halos grow in mass past $\sim 10^{12}\,\msun$, 
galaxy masses ``pile up'' near $\sim L_{\ast}$ ($M_{\rm gal}\sim 10^{11}\,\msun$), 
as can be seen in Figure~\ref{fig:fmu.gal.vs.halo}. Since halo-halo merger 
rates are nearly self-similar in terms of halo-halo mass ratio, the ``pileup'' of 
galaxies near this mass 
means that a wide range of halo-halo mass ratios will 
be compressed into a narrow range of galaxy-galaxy 
mass ratios near $\mu_{\rm gal}\sim1$ 
\citep[see also][]{maller:sph.merger.rates,stewart:merger.rates,
stewart:massratio.defn.conf.proc}.

This is a general statement: in {\em any} scenario where 
$M_{\rm halo}/M_{\rm gal}$ has a minimum, the galaxy-galaxy merger 
rate will be weighted more towards 
major mergers than the halo-halo merger rate around 
(in particular at masses slightly above) that minimum. 
The minimum in $M_{\rm halo}/M_{\rm gal}$ is empirically well-established, 
and occurs near $\sim L_{\ast}$, 
where most of the mass density lies. It is therefore inevitable that 
the contribution to the integrated bulge mass will be 
more weighted towards major mergers than to the 
halo mass.  

This is also easily understandable in terms of the mass functions of 
galaxies and halos. The halo mass function does not feature a sharp 
break, so the mass density of halos is broadly distributed over several 
orders of magnitude in halo mass. In contrast, the galaxy mass function 
reflects inefficient star formation at low and high masses, with a sharp 
break, and so the galaxy mass density is concentrated in a narrow 
range (a factor $\sim3$) around the break $L_{\ast}$. The mass of 
subunits (which broadly reflects the global mass function) 
in halos therefore includes contributions from a wide range of 
mass ratios. In contrast, the bulge growth of 
a galaxy is dominated by systems near $\sim L_{\ast}$. 
At masses $\lesssim$ a few $L_{\ast}$, this means major mergers 
will be most important. It is not until an $\sim L_{\ast}$ 
galaxy represents a minor merger (i.e.\ galaxy masses 
$\gtrsim 3\,L_{\ast}$) that minor mergers (again, mergers of those 
$\sim L_{\ast}$ galaxies) begin to dominate the mass assembly.

\breaker
\section{Galaxy-Galaxy Merger Rates}
\label{sec:mgr.rates}

\subsection{Scaling with Mass and Redshift}
\label{sec:mgr.rates.scaling}

\begin{figure}
    \centering
    \scaleup
    \plotter{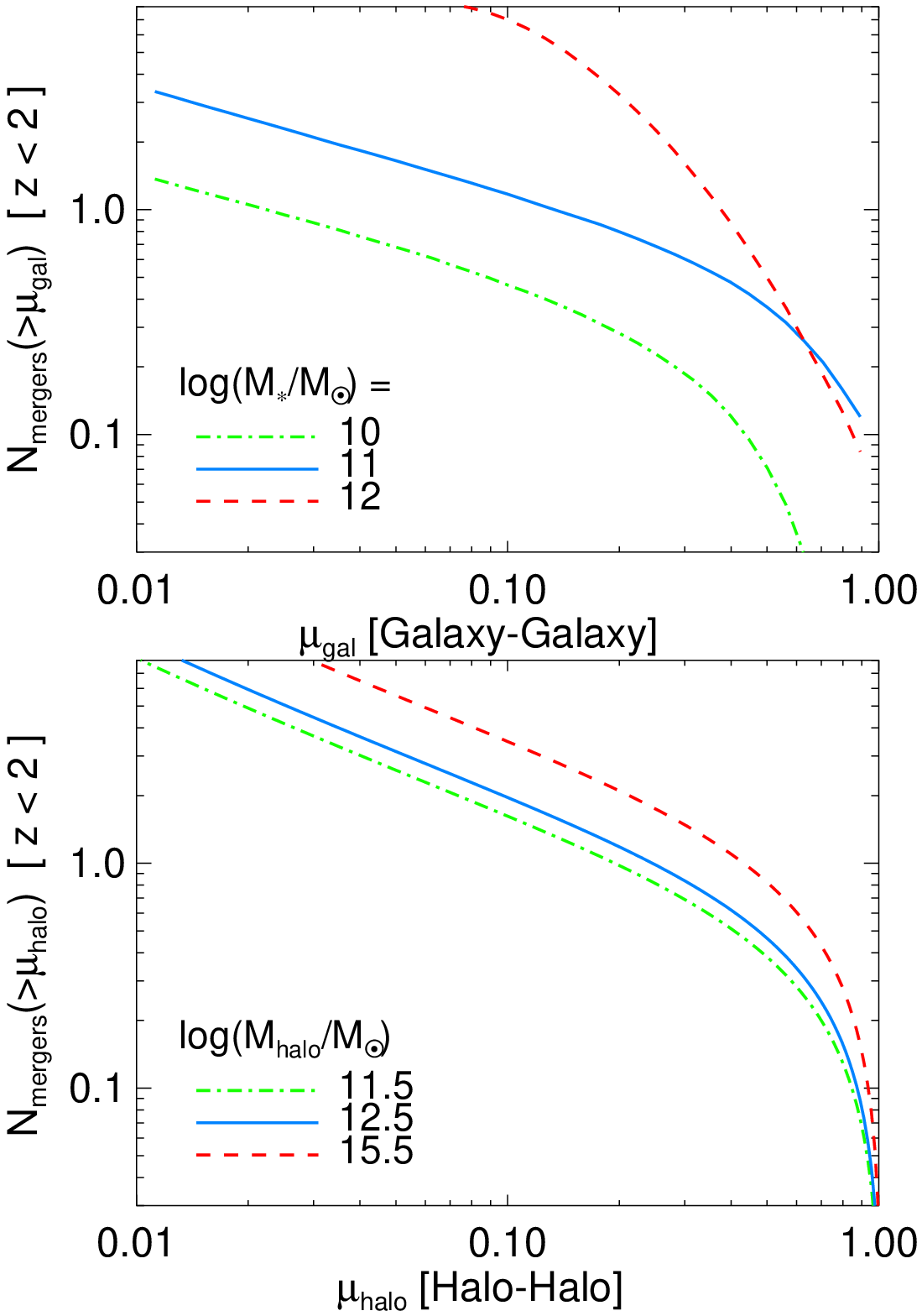}
    \caption{{\em Top:} Median number of galaxy-galaxy mergers since $z=2$ 
    above a given baryonic mass ratio $\mu_{\rm gal}$ for a $z=0$ 
    galaxy of the given stellar mass. The same mass selection at higher 
    redshift intervals will systematically increase the number of mergers. 
    The predictions here agree well with observational estimates 
    \citep{deravel:merger.fraction.to.z1,conselice:mgr.pairs.updated,lin:mergers.by.type}. 
    {\em Bottom:} Same, for halo-halo mergers. 
    Owing to the shape of the HOD, the merger rate is a steep function of 
    galaxy mass around $\sim L_{\ast}$, whereas the halo-halo merger 
    rate is only weakly mass-dependent. 
    Galaxies near $\sim L_{\ast}$ have had $\sim1$ 
    major merger since $z=2$. 
    \label{fig:nmu}}
\end{figure}

\begin{figure*}
    \centering
    \plotter{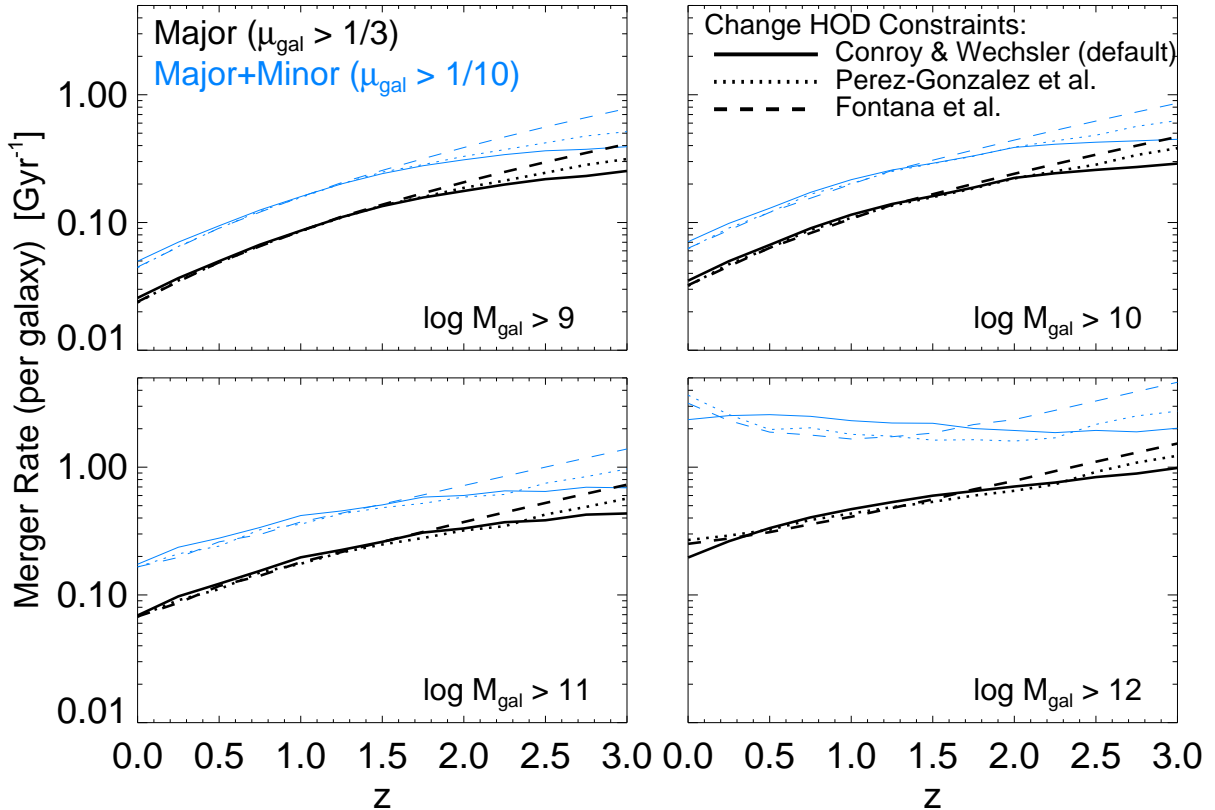}
    \caption{Merger rate (number of mergers per galaxy per Gyr) 
    as a function of redshift for different galaxy mass intervals (all $M_{\rm gal}$ 
    above some minimum baryonic mass in $\msun$ at each redshift). 
    Solid, dashed, and dot-dashed lines 
    correspond to three different halo occupation 
    model constraints: 
    uncertainties are small at $z<2$. 
    Colors correspond to the range of merger mass ratios. 
    \label{fig:rate.demo}}
\end{figure*}

We first examine the galaxy-galaxy merger rate, given these empirical 
constraints. Figure~\ref{fig:nmu} shows the number of mergers, as a 
function of mass ratio, that a typical galaxy of a given $z=0$ stellar 
mass has experienced since $z=2$. 
We emphasize that this is for a sample mass-selected based on their $z=0$ 
masses; a sample selected at the same mass at higher redshift will 
have a systematically larger number of mergers in a similar time or 
$\Delta z$ interval (and will be higher mass by $z=0$). 
Below, we discuss how these predictions compare with observations; the 
two generally agree well. 
We compare with the number of 
halo-halo mergers as a function of halo mass ratio, for 
typical corresponding halo masses. 
The two are quite different, for the reasons discussed in \S~\ref{sec:hod}: 
essentially, at low masses, $M_{\rm gal}\propto M_{\rm halo}^{2}$, 
so a 1:3 $\mu_{\rm halo}$ merger becomes a 1:9 $\mu_{\rm gal}$ 
merger, and merger rates at each $\mu_{\rm gal}$ are suppressed. At 
high masses, $M_{\rm gal}\propto M_{\rm halo}^{0.5}$, and 
rates are correspondingly enhanced. At 
$\sim10^{11}\,\msun$, where most of the spheroid mass density of the 
Universe resides, the typical galaxy has experienced $\sim0.5-0.7$ major 
($\mu_{\rm gal}>1/3$) mergers since $z=2$, a fraction that corresponds well 
to the observed fraction of bulge-dominated early-type systems at these 
masses \citep[see e.g.][]{bell:mfs}. At most masses (excepting the highest 
masses, where the shape of the HOD yields a strong preference towards minor 
mergers), the total number of mergers with $\mu_{\rm gal}\gtrsim1/10$ is a factor 
$\sim2-3$ larger than the number with $\mu_{\rm gal}>1/3$, and at low 
mass ratios $\mu_{\rm gal}\ll 1/10$, the merger rate asymptotes to a power-law 
with $N(>\mu_{\rm gal}) \propto \mu_{\rm gal}^{-(0.25-0.5)}$.

Figure~\ref{fig:rate.demo} shows how the median merger rates evolve with 
redshift, for four different intervals in mass. 
We compare the rate of major ($\mu_{\rm gal}>1/3$) and 
major+minor ($\mu_{\rm gal}>1/10$) 
mergers. We also compare different constraints on the HOD, from 
fitting different galaxy mass functions and clustering data. 
Specifically, we show the default model here, 
where the function $M_{\rm gal}(M_{\rm halo})$ is determined 
from fits to observed 
clustering, stellar mass functions, and star formation rate distributions 
in \citet{conroy:hod.vs.z}; we compare the results adopting a monotonic 
ranking between galaxy and halo mass and using the redshift-dependent 
stellar mass functions from \citet{fontana:highz.mfs} 
or \citet{perezgonzalez:mf.compilation}. Further variations are 
discussed in detail in \S~\ref{sec:robustness}. These illustrate the robustness 
of the model: empirical 
halo occupation constraints are sufficiently tight that they contribute little 
ambiguity in the resulting merger rate at $z<2$. Above $z=2$, the results 
begin to diverge, as the stellar mass function is less well-determined 
(and few clustering measurements are available); however  
these higher redshifts have relatively little impact on the predictions at low-$z$. 
The major merger rate increases with mass with a slope of roughly 
$\propto (1+z)^{1.5-2.0}$, but this is mass-dependent. 
The evolution in the galaxy-galaxy merger rate is somewhat shallower 
than the redshift evolution of the halo-halo merger rate (which 
scales with the Hubble time) as a consequence of the redshift 
evolution of the HOD.

\subsection{Comparison with Observations}
\label{sec:mgr.rates.obs}

{\textwidth 3.0in  
\tableclear
\begin{deluxetable}{lcl}
\tablecolumns{3}
\tabletypesize{\scriptsize}
\tablecaption{Observed Merger Rates\label{tbl:obs}}
\tablewidth{0pt}
\tablehead{
\colhead{Reference} &
\colhead{Selection\tablenotemark{1}} &
\colhead{Symbol\tablenotemark{2}} 
}
\startdata
 & Pairs & \\ 
 \hline
\citet{kartaltepe:pair.fractions} & $20\,h^{-1}\,{\rm kpc}$ & blue triangles \\ 
\citet{lin:merger.fraction,lin:mergers.by.type} & $30\,h^{-1}\,{\rm kpc}$ & pink circles \\
\citet{xu:merger.mf} & $20\,h^{-1}\,{\rm kpc}$ & blue circle \\
\citet{depropris:merger.fraction} & $20\,h^{-1}\,{\rm kpc}$ & black asterisk \\
\citet{bluck:highz.merger.fraction} & $20\,h^{-1}\,{\rm kpc}$ & orange squares \\
\citet{bundy:merger.fraction.new} & $20\,h^{-1}\,{\rm kpc}$ & green stars \\
\citet{bell:dry.mergers,bell:merger.fraction} & $20\,h^{-1}\,{\rm kpc}$ & red pentagons \\
 &  & \\
 & Morphology & \\
 \hline
\citet{conselice:mgr.pairs.updated} & CAS & pink $\times$'s \\
\citet{conselice:highz.mgr.rate.vs.mass} & CAS & blue $\times$'s \\
\citet{cassata:merger.fraction} & CAS & cyan inverted triangles \\
\citet{jogee:merger.density.08.conf.proc} & visual & pink triangles \\
\citet{bundy:mfs} & visual & light green stars \\
\citet{wolf:merger.mf} & visual & orange diamonds \\
\citet{bridge:merger.fractions,bridge:merger.fraction.new} & visual & purple squares \\
\citet{lotz:morphology.evol,lotz:merger.fraction} & Gini-M20 & dark green $+$'s \\
\enddata
\tablenotetext{1}{Selection criterion used to identify merger 
candidates. For pair samples, this refers to the pair separation. 
For morphological samples, to the method used.} 
\tablenotetext{2}{Symbol used for each sample 
in Figures~\ref{fig:mgr.rate.vs.z} \&\ \ref{fig:mgr.rate.vs.obs}.}
\end{deluxetable}
\tableclear
}

\begin{figure}
    \centering
    \scaleup
    \plotter{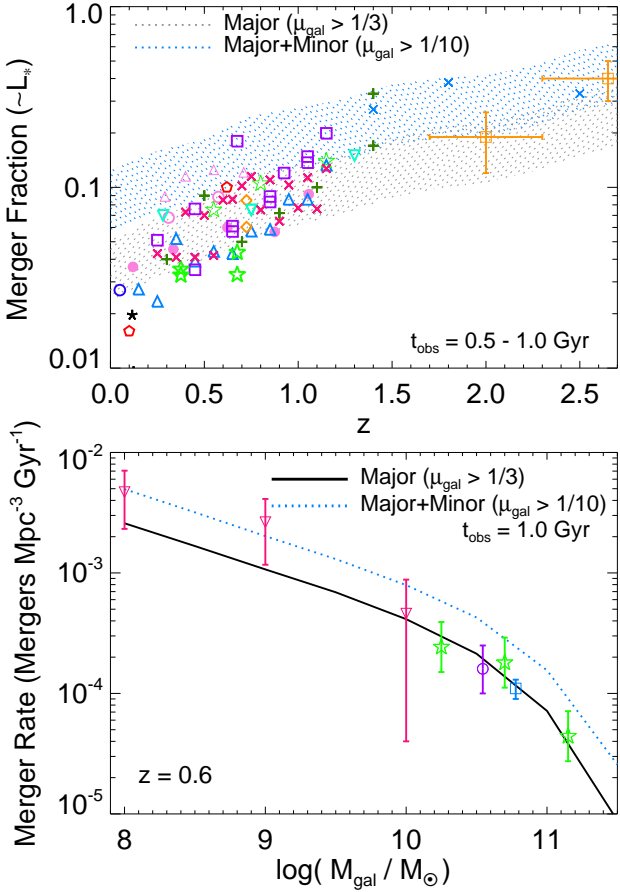}
    \caption{{\em Top:} Major merger fraction of $\sim L_{\ast}$ galaxies. 
    Observations (with the symbol type for each) 
    are listed in Table~\ref{tbl:obs}; for now, we treat them 
    all the same. Shaded range corresponds to the predicted merger 
    rate, convolved with an observable lifetime $t_{\rm obs}=0.5-1.0\,$Gyr 
    (color denoted the mass ratio range, as labeled). 
    {\em Bottom:} Integrated merger rate as a function of 
    galaxy mass at $z=0.6$ (in absolute units, mergers per Mpc$^{3}$ per Gyr above some 
    minimum mass), where observations span the greatest dynamic range. 
    Observations are from \citet[][blue square]{bell:merger.fraction}, 
    \citet[][pink triangles]{conselice:mgr.pairs.updated},
    \citet[][green stars]{bundy:merger.fraction.new}, and 
    \citet[][purple circle]{lopezsanjuan:mgr.rate.pairs}. 
    Here, morphological samples are converted to rates with $t_{\rm obs}=1\,$Gyr; pair samples 
    the appropriate timescales for their separation (see text). 
    Predicted rates agree with those observed, but there is considerable scatter between 
    various observational selection methods. 
    \label{fig:mgr.rate.vs.z}}
\end{figure}

In Figure~\ref{fig:mgr.rate.vs.z}, we compare these predictions to observed 
major merger fractions. As most measurements of the merger fraction 
do not have a well-defined mass selection, we first simply 
consider a large compilation of observational results 
compared to the predicted merger rate of $\sim L_{\ast}$ galaxies. 
For now, because we are considering a range of mass and 
observational methodologies, we simply convert the predicted merger rate to 
an observed merger fraction assuming a constant observable lifetime $t_{\rm obs}$, 
here showing $t_{\rm obs}=0.5$\,Gyr and $t_{\rm obs}=1\,$Gyr, typical 
values in the literature. We also compare the number density of mergers versus 
mass, at a given redshift. The agreement appears reasonable, but there is a large 
scatter in the observations, mostly owing to different selection  
and merger identification criteria. 

We can also compare the predicted integrated number of mergers in 
Figure~\ref{fig:nmu} to various observational estimates. 
For example, \citet{deravel:merger.fraction.to.z1} use pair-selected samples to estimate that 
$\sim20-25\%$ of the $M_{\ast}>10^{10}\,\msun$ population has experienced a 
merger with mass ratio $\mu>0.25$ since $z=1$; \citet{conselice:mgr.pairs.updated} 
estimate a similar number of mergers for 
$M_{\ast}>10^{10}\,\msun$ galaxies since $z=1$ and about again as many 
since $z=2$, they also find that the number of mergers increases significantly 
with galaxy mass; \citet{lin:mergers.by.type}
estimate that $\sim54\%$ of $L_{\ast}$ ($M_{\ast}\sim10^{11}\,\msun$) 
galaxies have experienced a $\mu>0.25$ merger since $z=1.2$. 
All of these predictions can be compared with our predictions in Figure~\ref{fig:nmu}, 
and they agree well (especially given different methodologies, masses, 
and redshift ranges involved).

\begin{figure*}
    \centering
    \plotter{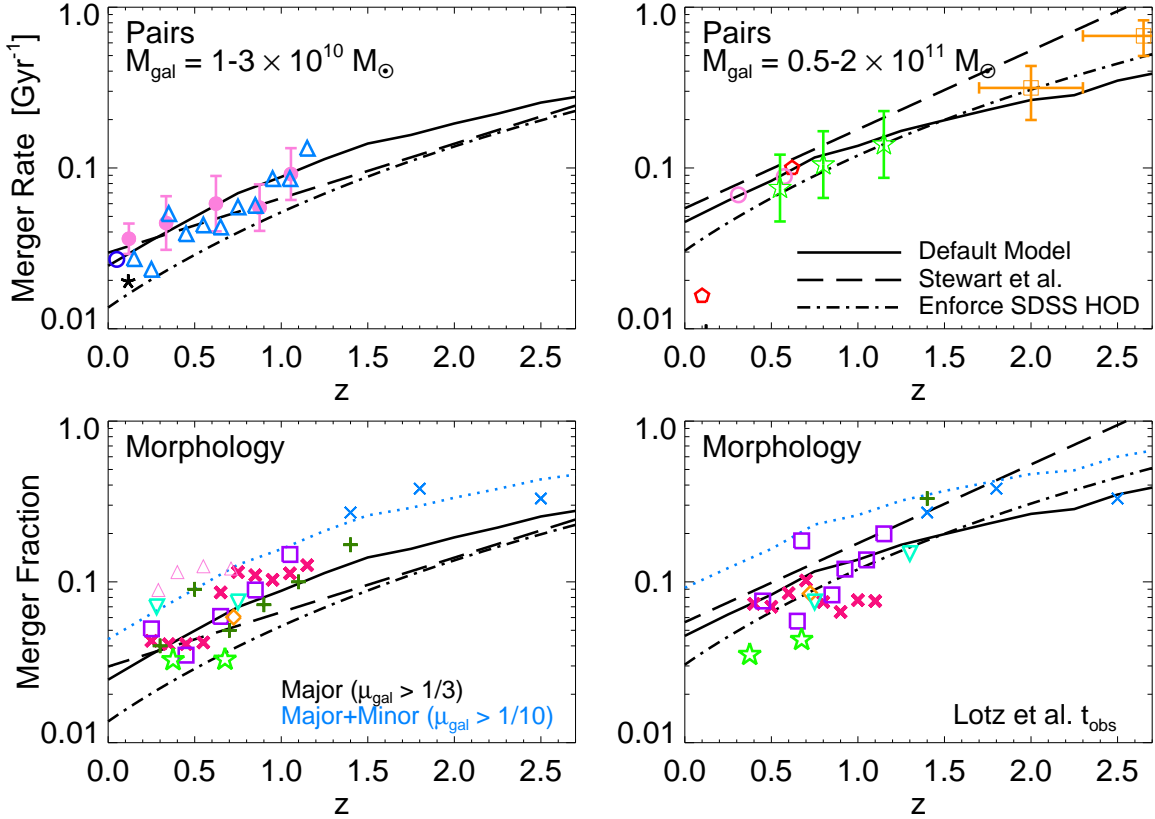}
    \caption{{\em Top:} Predicted merger rate in different stellar mass intervals, 
    compared to that 
    inferred from close pair studies (projected $r<20\,h^{-1}\,{\rm kpc}$). 
    Observations (with point types) are listed in Table~\ref{tbl:obs}; they are 
    converted to merger rates given the calibration of merger times as a function of 
    pair separation from high-resolution N-body simulations in 
    \citet{lotz:merger.selection,
    lotz:mgr.timescale.vs.massratio.morphology,lotz:gasfraction.vs.mergertime}. 
    We show our default model, as well 
    as the ``instantaneous'' calculation obtained by assuming all galaxies 
    lie on the $z=0$ (non-evolving) SDSS HOD shown in Figure~\ref{fig:fmu.gal.vs.halo}.
    We also show the results from \citet{stewart:merger.rates}, using different 
    simulation merger rates, merger trees, and HOD constraints. 
    {\em Bottom:} Morphologically identified merger fraction in the same mass 
    intervals, compared with predictions. The assumed timescale for identification as 
    morphologically disturbed is  $\sim1\,$Gyr, again calibrated from 
    simulations in \citet{lotz:merger.selection}. Observations in this case could be 
    contaminated by minor mergers; we therefore show both major and major+minor 
    merger fractions.     
    The scatter is larger in morphological samples (with some 
    probably contamination), but predictions agree within a factor $\sim2$. 
    \label{fig:mgr.rate.vs.obs}}
\end{figure*}

In order to test more strictly, and to take advantage of 
where observable merger timescales 
have been rigorously calibrated, in 
Figure~\ref{fig:mgr.rate.vs.obs} we restrict our comparison to 
galaxy-galaxy mergers identified observationally using a consistent 
methodology and covering a well-defined mass range. 

First, we consider 
pair fractions: specifically the fraction of {\em major} ($\mu_{\rm gal}>1/3$) 
pairs with small projected separations $r_{p}<20\,h^{-1}\,{\rm kpc}$ 
(often with the additional requirement of a
line-of-sight velocity separation $<500\,{\rm km\,s^{-1}}$), 
and stellar masses $M_{\ast}\sim1-3\times10^{10}\,\msun$ or 
$M_{\ast}\sim0.5-2\times10^{11}\,\msun$. For each mass bin, the pair fractions 
as a function of redshift can be empirically converted to a merger rate 
using the merger timescales at each radius. \citet{lotz:merger.selection,
lotz:mgr.timescale.vs.massratio.morphology,lotz:gasfraction.vs.mergertime}
specifically calibrate these timescales for the same projected separation and 
velocity selection from a detailed study of a large suite of hydrodynamic 
merger simulations (including a 
range of galaxy masses, orbital parameters, gas fractions and star formation 
rates) using mock images obtained by applying realistic radiative transfer models, 
with the identical observational 
criteria to classify mock observations of the galaxies at all times 
and sightlines during their evolution. 
For this specific pair selection 
criterion (if we average over the typical distribution of mass ratios 
for mergers selected in this interval), they find a median 
merger timescale of $t_{\rm merger}\approx 0.35\,$Gyr, with relatively 
small scatter and very little dependence on simulation parameters ($\pm0.15\,$Gyr).\footnote{
The merger timescale from simulations at this radius is shorter than 
the time obtained assuming dynamical friction and circular orbits in e.g.\ an 
isothermal sphere, as has commonly been done 
\citep[this is assumed in e.g.\ both][]{patton:merger.fraction,
kitzbichler:mgr.rate.pair.calibration}. This 
owes to two effects: first, angular momentum loss at these radii is {\em not} 
dominated by dynamical friction, but rather by exchange in strong 
resonances that act much more efficiently
(even allowing for e.g.\ mass loss by the secondary inside the primary halo, 
which is of course included in the high-resolution simulations, 
this has the net effect of significantly accelerating most mergers). 
Second, by these radii, even initially 
circular orbits have become highly radial, leading to shorter merger times. 
Because of these effects, the remaining merger time at this scale depends 
only weakly on initial conditions or orbital parameters --
essentially, these processes have erased most of the ``memory'' of 
the original orbital configuration. 
This emphasizes the importance of using full simulations with 
baryonic effects in calibrating these timescales. 
}
We use their median $t_{\rm merger}$ to convert the observations to a 
merger rate. Because of the weighting over merger mass ratio and orbital 
parameters, for which the explicit dependence is presented in these papers, 
we obtain the same result (within the observational error bars) if we convolve 
our predicted merger rates with the explicit observable merger 
timescale as a function of merger gas fraction, galaxy mass, 
redshift, mass ratio, and orbital parameters, as presented in 
\citet{lotz:mgr.timescale.vs.massratio.morphology,lotz:gasfraction.vs.mergertime}. 
Completeness corrections are discussed in the various papers; 
we also adopt the standard correction from \citet{patton:mgr.rate.vs.rmag}, 
calibrated to high-resolution simulations, for the fraction of systems on early 
or non-merging passages (to prevent double-counting systems on multiple 
passages); but this is relatively small \citep[$20-40\%$; see also][]{lotz:merger.selection}. 

The advantages of these pair fractions are that:
(1) the mass ratio can be determined, 
leading to little contamination from minor mergers,\footnote{Note that many older 
studies adopt the galaxy-galaxy luminosity ratio as a proxy for mass ratio. 
This is not a bad approximation in e.g.\ numerical simulations, but could be 
subject to bias from e.g.\ differential enhancement in star formation. Obviously 
it is preferable to use an actual stellar mass ratio where possible. Restricting our 
samples to just studies with stellar masses, however, we obtain similar conclusions.} 
(2) at such small separations, most such pairs 
will eventually merge, and (3) there is little ambiguity in the merger 
timescale, with only a factor $\sim10-20\%$ systematic 
uncertainty in the median/average merger timescale 
in high-resolution calibrations (with a 
$\sim25-50\%$ dispersion or variation always present 
about that median, owing 
to cosmological variation in e.g.\ the exact orbital parameters).  

Second, we consider morphologically-selected mergers, identified on the 
basis of by-eye classification or automated morphological criteria such as 
the concentration-asymmetry (CAS) or Gini-M20 planes 
\citep[see e.g.][]{conselice:merger.fraction,lotz:gini-m20}. \citet{lotz:merger.selection} 
also attempt to calibrate the observable timescale for classification of 
major mergers via the Gini-M20 criterion, at rest-frame wavelengths and masses 
of the observations. They find an observable timescale $t_{\rm obs}({\rm Gini-M20})\sim1\,$Gyr, 
and we adopt that here, but note that the predicted timescale in this case 
depends much more sensitively on the depth of the observations, 
the waveband adopted, and properties such as the gas-richness 
of the merging systems.\footnote{For the explicit 
dependence on these parameters, see 
\citet{lotz:mgr.timescale.vs.massratio.morphology,lotz:gasfraction.vs.mergertime}. 
Note that the timescale of $1\,$Gyr here is slightly longer than the 
$0.6\,$Gyr estimated directly 
from observations in \citet{conselice:merger.timescale.obs.est} 
(although within their quoted $1\sigma$ error bars) and in the original 
\citet{lotz:merger.selection} for radial 1:1 mergers. The difference 
in the latter owes to the dependence of observable timescale on gas fraction 
as calibrated in \citet{lotz:gasfraction.vs.mergertime} (here we 
adopt a median appropriate for the median gas fractions of the 
model galaxies), and from the dependence on merger mass ratio 
as calibrated in \citet{lotz:mgr.timescale.vs.massratio.morphology} 
(where our value here represents a weighting over the mass ratio distribution 
down to mass ratio $\mu=1/3$ as appropriate for the observations here). 
In any case, the difference is generally 
smaller than the scatter between different observational estimates.
} Moreover, although, by definition, this methodology is 
complete to events that have violently disturbed the galaxy, the level of 
disturbance at a given merger mass ratio depends on orbital parameters 
and galaxy gas fractions, so a fixed level of disturbance does not correspond 
to a fixed merger mass ratio. Some contamination from minor mergers is likely. 
\citet{jogee:merger.density.08} 
estimate empirically that $\sim30-40\%$ of 
their (by-eye) morphologically-identified sample represent 
contamination from $1/10 < \mu_{\rm gal} < 1/3$ minor mergers. 
We therefore compare the predicted merger fraction for $\mu_{\rm gal}>1/3$ 
and $\mu_{\rm gal}>1/10$. Allowing for this range, the observations 
agree well with the predictions. In particular, the observations using a 
calibration of Gini-M20 or CAS specifically matched to high-resolution 
hydrodynamic major merger simulations \citep[e.g.][]{lotz:merger.fraction,
conselice:mgr.pairs.updated} agree reasonably well with 
the predicted $\mu_{\rm gal}>1/3$ fractions. And external, purely 
empirical indicators favor similar merger timescales 
\citep[see e.g.][]{conselice:merger.timescale.obs.est}.

A quantity closely related to the pair fraction on small scales is the 
galaxy-galaxy autocorrelation function (specifically that on small scales, 
inside the ``one halo term'' where it reflects galaxies inside the same 
parent halo). Effectively this generalizes the predicted pair 
fraction from $<20\,h^{-1}\,$kpc to all scales. But recall, the adopted 
halo occupation-based methodology is designed, by construction, to 
match the observed correlation functions as a function of mass. 
It is therefore guaranteed that the clustering at scales $\sim100\,$kpc 
through $\gtrsim10\,$Mpc is reproduced as a function of 
galaxy mass and redshift 
\citep[for explicit illustrations, see e.g.][]{conroy:monotonic.hod,
zheng:hod.evolution,wang:sdss.hod}.

\subsection{Analytic Fits}
\label{sec:mgr.rates.analytic}

It is useful to quantify the predicted 
merger rates with simple analytic fitting functions.

\begin{figure}
    \centering
    \scaleupp
    \plotter{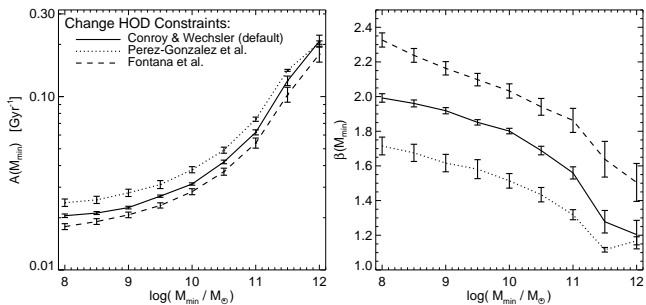}
    \caption{Results of fitting the major merger rate 
    as a function of mass and redshift to a function of 
    the form of Equation~\ref{eqn:mgrrate.form}.  
    Linestyles correspond to 
    different HOD choices, as in Figures~\ref{fig:rate.demo}-\ref{fig:mgr.rate.vs.obs}. 
    {\em Left:} Normalization, i.e.\ number of mergers 
    per galaxy per Gyr at $z=0$, for galaxies above a 
    minimum mass $M_{\rm min}$. 
    {\em Right:} Redshift dependence, i.e.\ slope 
    $\beta$ given a merger rate per unit time $\propto (1+z)^{\beta}$, 
    as a function of minimum mass. 
    The mass-dependent $A(M_{\rm min})$ and 
    $\beta$ can be approximated with 
    Equations~\ref{eqn:amin.major.fit}-\ref{eqn:beta.major.fit}, 
    with systematic uncertainties of $0.3$\,dex and $0.2$, respectively. 
    \label{fig:mgr.rate.fits}}
\end{figure}

First, consider major mergers. We find that the major merger rate 
(number of $\mu_{\rm gal}>1/3$ mergers per galaxy per unit time), for 
galaxies above a given minimum stellar 
mass threshold ($M_{\ast}>M_{\rm min}$), 
can be well-fitted by the following simple function: 
\begin{equation}
\frac{{\rm d}N_{\rm major}}{{\rm d}t} = A(M_{\rm min})\,
(1+z)^{\beta(M_{\rm min})}\ \  [{\rm per\ galaxy}],
\label{eqn:mgrrate.form} 
\end{equation}
i.e.\ a $z=0$ normalization $A(M_{\rm min})$ and 
simple power-law scaling with redshift with slope $\beta(M_{\rm min})$. 
Figure~\ref{fig:mgr.rate.fits} shows these quantities, fitted to 
the predictions shown in 
Figures~\ref{fig:rate.demo}-\ref{fig:mgr.rate.vs.obs}, as a function of the 
mass $M_{\rm min}$. The trends discussed above are evident: the normalization of 
merger rates increases with mass, and (albeit more weakly), the dependence on 
redshift decreases with mass. 
This normalization and redshift variation can be approximated with 
the scalings: 
\begin{equation}
A(M_{\rm min})_{\rm major} 
\approx 0.02\,{\Bigl [}1 + {\Bigl(}\frac{M_{\rm min}}{M_{0}}{\Bigr)}^{0.5} {\Bigr ]}\,
{\rm Gyr^{-1}}
\label{eqn:amin.major.fit}
\end{equation}
and 
\begin{equation}
\beta(M_{\rm min})_{\rm major} 
\approx 1.65 - 0.15\,\log{{\Bigl(}\frac{M_{\rm min}}{M_{0}}{\Bigr)}}. 
\label{eqn:beta.major.fit}
\end{equation}
where $M_{0}\equiv 2\times10^{10}\,\msun$ is fixed. 
There is a systematic factor $\sim2$ uncertainty in the merger rate 
normalization $A(M_{\rm min})$ at all $M_{\rm min}$, considering the range of 
models discussed in detail in \S~\ref{sec:robustness} below. The uncertainty in 
$\beta(M_{\rm min})$ is illustrated in Figure~\ref{fig:mgr.rate.fits}, 
approximately a systematic $\Delta\beta\sim0.15 - 0.20$. 

These fits are for major ($\mu_{\rm gal}>1/3$) mergers. To rough approximation, 
the number of mergers as a function of mass ratio 
scales with the approximate form 
\begin{equation}
\frac{{\rm d}N(>\mu_{\rm gal})}{{\rm d}t} \propto \mu_{\rm gal}^{-0.3}\,(1 - \mu_{\rm gal})\ \ [{\rm per\ galaxy}]
\end{equation}
\citep[derived and discussed in more detail in][]{stewart:merger.rates}. 
This is a good approximation as long as the galaxy is within an order of 
magnitude of $\sim L_{\ast}$. For the range where this 
function is a good approximation, it implies an approximately constant 
ratio of a factor $\approx2$ ($0.3-0.4$\,dex) of major+minor ($\mu_{\rm gal}>1/10$) 
to major ($\mu_{\rm gal}>1/3$) mergers. 

More specifically, we can fit a function of the form 
of Equation~\ref{eqn:mgrrate.form} to the 
major+minor ($\mu_{\rm gal}>1/10$) merger rate of galaxies 
above some $M_{\rm min}$, 
and obtain the best-fit scalings 
\begin{equation}
A(M_{\rm min})_{\rm minor} 
\approx 0.04\,{\Bigl [}1 + {\Bigl(}\frac{M_{\rm min}}{M_{0}}{\Bigr)}^{0.8} {\Bigr ]}\,
{\rm Gyr^{-1}}
\end{equation}
and 
\begin{equation}
\beta(M_{\rm min})_{\rm minor} 
\approx 1.50 - 0.25\,\log{{\Bigl(}\frac{M_{\rm min}}{M_{0}}{\Bigr)}}, 
\label{eqn:mgrrate.minor.beta}
\end{equation}
with similar systematic uncertainties in both $A(M_{\rm min})$ 
and $\beta(M_{\rm min})$ to the major merger rate. 
Note that these equations should be treated with caution for the most 
massive systems -- the simple fitting functions 
do not extrapolate to arbitrarily high mass and the direct numerical 
results (e.g.\ Figure~\ref{fig:rate.demo}) should be used.

\breaker
\section{The Relative Contributions To Bulge Growth from Different Mergers}
\label{sec:compare}

\subsection{Overview}
\label{sec:compare.overview}

\begin{figure*}
    \centering
    \scaleupp
    \plotter{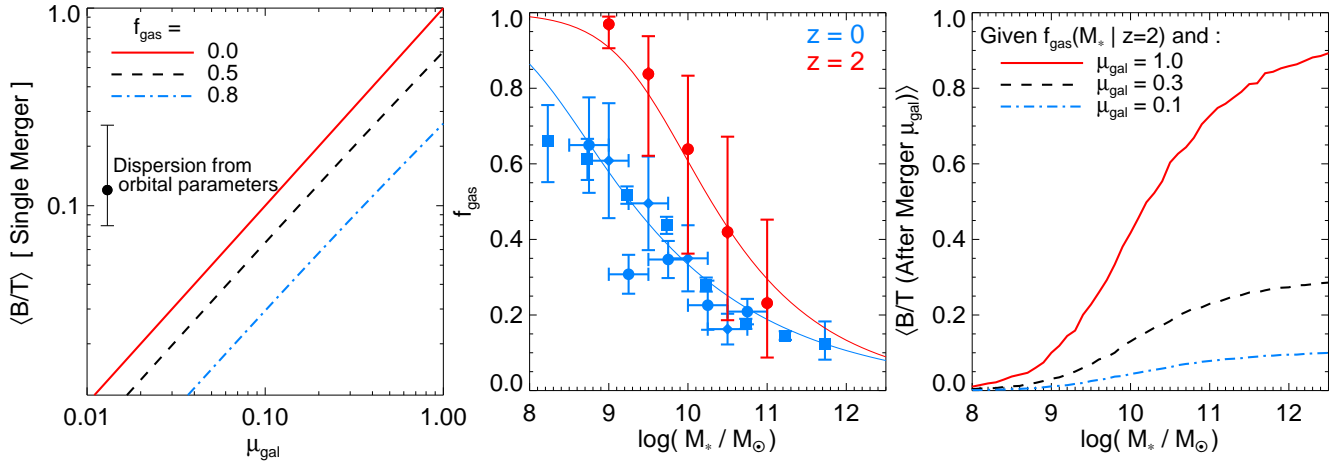}
    \caption{{\em Left:} Average bulge-to-total ratio $B/T$ 
    (of the {\em classical} bulge)
    resulting from a single merger of 
    mass ratio $\sim\mu_{\rm gal}$ between galaxies with gas fraction $f_{\rm gas}$ 
    \citep[from simulations in][]{hopkins:disk.survival}. To lowest 
    order $B/T$ scales as $\propto \mu_{\rm gal}\,(1-\fgas)$. 
    Error bar shows scatter owing to the cosmological range of orbital parameters. 
    {\em Center:} Median observed gas fraction and 
    scatter (error bars) 
    for disks of a given stellar mass at $z=0$ (blue) from 
    \citet[][diamonds]{belldejong:tf}, \citet[][squares]{kannappan:gfs}, and 
    \citet[][circles]{mcgaugh:tf}, and at $z=2$ (red) from \citet{erb:lbg.gasmasses}. 
    Solid lines show fits to the median at each redshift. 
    {\em Right:} Corresponding median $B/T$ expected from mergers 
    at $z=2$ with primary of a given stellar mass and mass ratio $\mu_{\rm gal}$ (given 
    the observed $\langle f_{\rm gas}[M_{\ast}\,|\,z=2] \rangle$). 
    Suppression of bulge formation by gas-richness is important for the absolute bulge mass 
    formed (especially at low masses), but because it is $\mu_{\rm gal}$-independent, 
    does not affect the relative contribution of major/minor mergers. 
    \label{fig:BT.fgas.mstar}}
\end{figure*}

Figure~\ref{fig:BT.fgas.mstar} illustrates how the efficiency of bulge formation 
scales in simulations. We show how 
the average $B/T$ resulting from disk-disk mergers 
scales with mass ratio (approximately $\mu_{\rm gal}$, but see \S~\ref{sec:model}), 
gas fraction $f_{\rm gas}$, and merger orbital parameters, 
according to the fits to the 
hydrodynamic simulations in \citet{hopkins:disk.survival}. 
To lowest order, as discussed in \S~\ref{sec:hod}, the amount of 
bulge formed (the amount of stellar disk of the primary 
galaxy that is violently relaxed, and amount of gas disk that is drained of 
angular momentum and participates in the nuclear starburst) 
scales linearly\footnote{Note that 
this refers to the mass fraction which is violently relaxed, thus adding to 
the bulge. Disk heating and resonant processes that contribute to 
the thick disk or disk substructure are different. 
For example, \citet{hopkins:disk.heating} show that disk heating in minor mergers 
is second-order in mass ratio; \citet{purcell:minor.merger.thindisk.destruction} 
and \citet{kazantzidis:thin.disk.thickening} 
reach similar conclusions (albeit with slightly different 
absolute normalization/efficiency).}
with the mass ratio of 
the encounter, $\propto \mu_{\rm gal}$. This conclusion -- ultimately 
the important statement for our analysis -- has been reached by 
numerous independent simulation studies, adopting different 
methodologies and numerical techniques, and 
naturally follows from the simple gravitational dynamics involved 
in violent relaxation \citep[see e.g.][]{hernquist.89,barneshernquist92,
mihos:starbursts.94,mihos:starbursts.96,
naab:minor.mergers,
bournaud:minor.mergers,younger:minor.mergers,
dimatteo:merger.induced.sb.sims,
hopkins:disk.survival,cox:massratio.starbursts}.
In addition, observational constraints on the efficiency of merger-induced 
star formation support these estimates \citep{woods:tidal.triggering,barton:triggered.sf,
woods:minor.mergers}. 

At fixed mass ratio, the bulge formed (remnant $B/T$) can vary 
considerably depending on orbital parameters of the merger, 
in particular the relative inclinations of the disks (prograde or retrograde). This 
variation is shown in Figure~\ref{fig:BT.fgas.mstar}. However, in a 
cosmological ensemble, this will average out. Here, we assume 
random inclinations, but allowing for some preferred inclinations amounts to a systematic 
offset in the $B/T$ predicted and will not change our conclusions regarding the 
relative importance of different mass ratios for bulge formation. 

Another important parameter determining $B/T$ is the merger 
gas fraction. To lowest order, angular momentum loss in gas is suppressed by a 
factor $\sim(1-\fgas)$; as a result, the efficiency of bulge formation 
($B/T$ expected for a merger of a given mass ratio) is suppressed by 
the same factor. This can have dramatic cosmological implications, because 
$\fgas$ is a strong function of galaxy mass; these are discussed in 
detail in \citet{hopkins:disk.survival.cosmo}. 
Figure~\ref{fig:BT.fgas.mstar} shows the observed dependence of 
disk/star-forming galaxy gas fractions on stellar mass at $z=0$ and $z=2$; 
if we assume mock galaxies on the $z=2$ relation each undergo a merger of a 
given mass ratio, then the resulting $B/T$ at each mass is shown. 
Bulge formation will be significantly suppressed by high gas fractions 
in low-mass galaxies, giving rise to e.g.\ a strong 
mass-morphology relation similar to that observed 
\citep{stewart:disk.survival.vs.mergerrates,hopkins:disk.survival.cosmo}. However, it is clear 
in Figure~\ref{fig:BT.fgas.mstar} that the effect of $\fgas$ is a systematic 
offset in $B/T$, independent of mass ratio. Because 
in what follows we will generally 
examine the {\em relative} importance of mergers of different mass 
ratios (independent of gas fraction), the inclusion or exclusion 
of the effects of gas on merger dynamics makes little difference 
to our conclusions.

\begin{figure}
    \centering
    \plotter{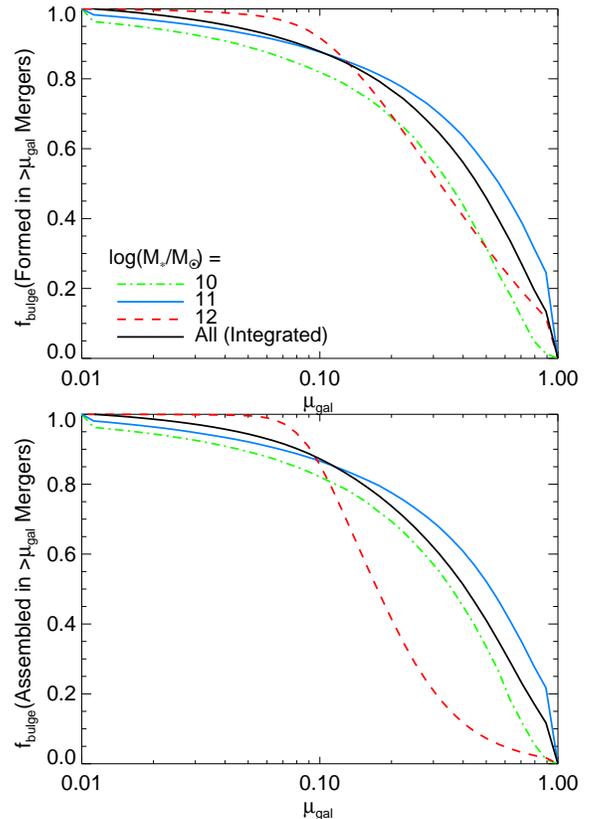}
    \caption{{\em Top:} Cumulative contribution to original 
    {\em formation} of spheroid mass 
    density from mergers of different mass ratios, for galaxies of given stellar 
    mass at $z=0$ (as in Figure~\ref{fig:fmu.gal.vs.halo}). 
    ``Integrated'' curve refers to the integral over all bulge masses (net 
    contribution to global spheroid mass density). 
    {\em Bottom:} Same, but showing the contribution to 
    the integrated spheroid {\em assembly}. 
    Major mergers dominate near $\sim L_{\ast}$; 
    minor mergers become more important at lower/higher masses. 
    Assembly by minor dry mergers (of bulges first formed in more 
    major mergers) occurs at the highest masses. 
    \label{fig:fmu}}
\end{figure}

Given these constraints from simulations on the amount of bulge formed 
in a given merger, and the merger rates predicted in \S~\ref{sec:mgr.rates}, 
Figure~\ref{fig:fmu} shows the contribution of mergers of different mass 
ratios to the $z=0$ stellar mass in bulges (as Figure~\ref{fig:fmu.gal.vs.halo}). 
We show this for galaxies in a narrow range of 
{\em total} stellar mass around three different values, and for the 
entire galaxy population (integrated over bulges of all masses). 
We specifically define this as the fraction of bulge mass formed or assembled 
in mergers that were above a given galaxy-galaxy mass ratio $\mu_{\rm gal}$. 
Considering different 
variations to the empirical and simulation constraints 
(see \S~\ref{sec:robustness}), about $60-70\%$ of 
the globally integrated bulge mass 
is assembled by major mergers $\mu_{\rm gal}>0.3$, with another $\sim30\%$ 
assembled by 
minor mergers $0.1<\mu_{\rm gal}<0.3$, and the remaining $\sim0-10\%$ from a wide 
range of mass ratios $\mu_{\rm gal}\sim0.01-0.1$. 

Note that, for massive galaxies ($\gtrsim$a few $L_{\ast}$), 
where ``dry'' mergers become an important 
channel (assembling mass already in massive bulges), there is an 
ambiguity in the fraction $f_{\rm bulge}(>\mu_{\rm gal})$ 
``contributed'' by mergers above a given $\mu_{\rm gal}$. 
Figure~\ref{fig:fmu} illustrates this. 
We therefore introduce the distinction between bulge {\em formation} 
and bulge {\em assembly}, terms we will use throughout. 

First, formation: we can define 
$f_{\rm bulge}(>\mu_{\rm gal})$ as the fraction of bulge mass 
originally {\em formed}, i.e.\ initially converted into bulge mass 
from disk mass (gas or stars) by mergers with some mass ratio 
$\mu_{\rm gal}$. In other words, taking 
Equation~\ref{eqn:fbul.defn}, where for each parcel of mass 
in the final bulge (${\rm d}m_{\rm bulge}$), the 
``contributing'' mass ratio $\mu_{\rm gal}^{\prime}$ is defined by 
$\mu_{\rm gal}$ of the merger that first made the mass into bulge 
(regardless of whatever merger ultimately brings it into the final galaxy). 
This quantity answers the question ``what kind of merger destroyed the 
progenitor disks of these galaxies?'' or ``what kind of merger created 
most bulges in the first place?''. 

Second, assembly: we can define $f_{\rm bulge}(>\mu_{\rm gal})$ 
as the fraction of bulge mass {\em assembled} into the main 
branch of the galaxy by mergers with mass ratios $>\mu_{\rm gal}$. 
Here, we take Equation~\ref{eqn:fbul.defn} where, for each 
parcel of mass, $\mu_{\rm gal}^{\prime}$ is defined 
by $\mu_{\rm gal}$ of the merger that brought it into the main progenitor 
branch of the final galaxy. 
This answers the question, ``what kind of merger brought most of the 
present-day bulge together?'' or ``what kind of merger has affected most 
of the mass in the bulge?''.

Clearly, the two are equal if all bulge mass is formed ``in situ'' in the 
main progenitor -- i.e.\ if the secondary galaxies have no pre-existing 
bulges. 
Indeed, at low masses, where most galaxies have little bulge, 
there is little difference. 

At high masses, however, there is a dramatic 
difference. This makes sense: high-mass galaxies grow 
primarily by dry mergers. Galaxies first become bulge-dominated 
around $\sim L_{\ast}$, then assemble hierarchically. Since most bulge mass is first 
formed around $\sim L_{\ast}$, where we see bulge formation is dominated 
by major mergers, we obtain the result that bulges in high-mass 
galaxies are primarily {\em formed} in major mergers. 
However, as they grow in mass via dry mergers, minor mergers 
become increasingly important to the {\em assembly} of the most massive 
systems (minor mergers bring together bulges that have already been formed). 
Both definitions are clearly important and have their applications; 
however, because of e.g.\ the importance for remnant kinematics 
and growth histories, we will generally adopt the latter 
(assembly-based) definition in what follows.

\begin{figure}
    \centering
    \scaleupp
    \plotter{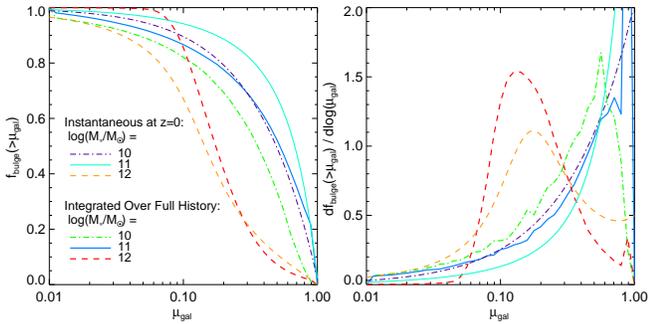}
    \caption{As Figure~\ref{fig:fmu} (assembly), comparing the full 
    result to an instantaneous calculation from convolving the $z=0$ 
    HOD of \citet{wang:sdss.hod} with instantaneous $z=0$ merger rates. 
    Because the {\em shapes} of the merger rate versus $\mu_{\rm halo}$ 
    and $M_{\rm gal}(M_{\rm halo})$ functions do not evolve strongly 
    with redshift, both yield similar results. 
    \label{fig:fmu.int.vs.inst}}
\end{figure}

The most important determinants of Figure~\ref{fig:fmu} are the 
{\em shapes} (logarithmic slopes), not normalizations, of the halo merger 
rate versus $\mu_{\rm halo}$ and 
function $M_{\rm gal}(M_{\rm halo})$. These shapes evolve 
weakly with redshift, as such our results can be reasonably understood 
with a simple instantaneous calculation. 
Taking the $z=0$ known $M_{\rm gal}(M_{\rm halo})$ 
alone, we can calculate the relative 
contribution to the differential growth rate of bulges at $z=0$ as a function of 
mass (essentially this amounts to populating a $z=0$ simulation with the observed 
HOD, and evolving it forward for some arbitrarily small amount of time, then 
calculating the relative importance of mergers as a function of their mass 
ratio). Figure~\ref{fig:fmu.int.vs.inst} compares the results from this 
simple procedure with an integration over merger history
(of course, without the full merger history, we can only define this in 
terms of the contribution to bulge assembly, not formation). 
There is little difference between the two. Because at any given 
stellar mass the most important mergers are those that happened while 
the system was relatively near that mass (not mergers that happened 
when the system was much lower mass, since those by definition will contribute 
little to the present total mass of the system), the ``memory'' of early formation 
or growth at low masses is effectively erased. 
This makes our results robust to details of 
the model at low masses and/or high redshifts, where 
empirical constraints are more uncertain.

\subsection{Dependence on Galaxy Mass}
\label{sec:compare.mass}

\begin{figure*}
    \centering
    \scaledown
    \plotter{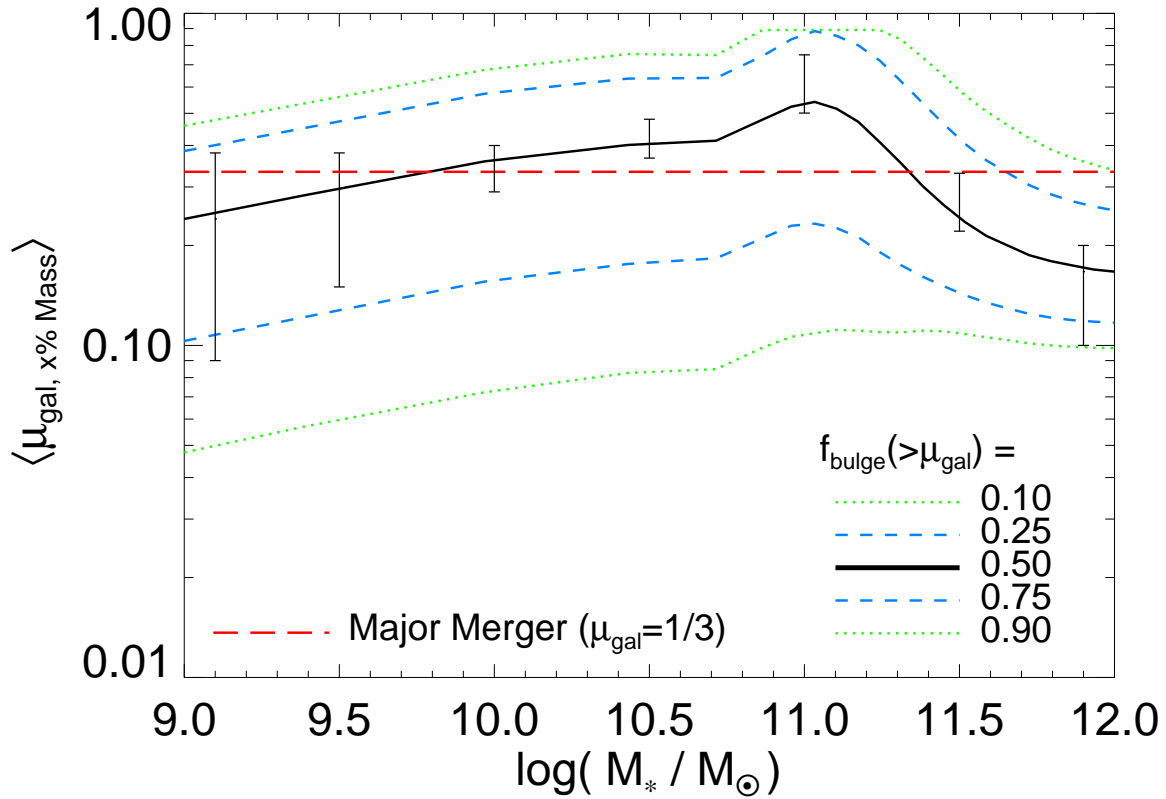}
    \caption{The relative importance of different mass ratios to bulge assembly  
    as a function of $z=0$ stellar mass. 
    We plot the (mass-weighted) median $\mu_{\rm gal}$ contributing to the assembly of  
    bulge mass density (defined as 
    $\mu_{\rm gal}$ where $f_{\rm bulge}[>\mu_{\rm gal}]=0.5$) 
    as a function of 
    stellar mass, along with the interquartile range and 
    $10-90\%$ range (lines bracketing different 
    ranges as labeled; horizontal dashed line 
    denotes the standard ``major merger'' definition $\mu>1/3$). 
    Error bars show the range resulting from 
    variations to the model (see \S~\ref{sec:robustness}). 
    Recall, the number of major mergers increases monotonically with 
    mass; the decrease here at high mass is simply because the number of 
    minor mergers increases yet more rapidly. 
    \label{fig:fmu.vs.m}}
\end{figure*}

We have shown how the relative importance of different mass ratio mergers 
depends on mass in Figures~\ref{fig:fmu} \&\ \ref{fig:fmu.int.vs.inst}. 
Figure~\ref{fig:fmu.vs.m} summarizes these results. We plot the 
median $\langle\mu_{\rm gal}\rangle$ (specifically the 
mass-weighted median $\mu_{\rm gal}$, corresponding to the 
merger mass ratio above which $50\%$ of the mass in bulges was 
assembled; i.e.\ where 
$f_{\rm bulge}(>\mu_{\rm gal})=0.5$ in Figure~\ref{fig:fmu.int.vs.inst}) as a 
function of stellar mass, at $z=0$. 
We also plot the corresponding $\pm1\,\sigma$ and $10-90\%$ ranges. 
As demonstrated in Figure~\ref{fig:fmu.int.vs.inst}, similar results 
are obtained with an ``instantaneous'' calculation; in 
\S~\ref{sec:robustness}, we show similar results varying a number of choices 
in the model. 

As seen before, major mergers dominate near $\sim10^{10}-3\times10^{11}\,\msun$, 
with minor mergers increasingly important at lower and higher masses. 
(In terms of the initial bulge formation, rather than assembly, the prediction would 
be the same but without the ``turnover'' at high masses -- i.e.\ asymptoting 
to a constant $\langle\mu_{\rm gal}\rangle\sim0.5$ at high masses.) 
Most of the variance comes from differences in merger histories at 
fixed mass. At all masses, the range of contributing mergers is quite large -- 
there is always a non-negligible contribution from minor mergers 
with mass ratios $\sim 0.1-0.3$. 

We stress that the turnover at high masses does not come because 
of fewer major mergers. In fact, we have shown explicitly that the number of major 
mergers in the primary history increases monotonically with galaxy mass. 
Rather, at high masses, the number of minor mergers increases even faster. 
Thus the {\em relative} importance of minor mergers is enhanced at the highest 
masses (picture for example the growth of a BCG via accretion of many satellites 
in a cluster). This is why the turnover would not appear if we made Figure~\ref{fig:fmu.vs.m} 
in terms of the mass ratios important for bulge formation, instead of bulge assembly.
So we expect these systems to be more bulge dominated 
as compared to $\sim L_{\ast}$ galaxies, but with interesting second-order effects 
in e.g.\ their kinematics and light profile shapes that indicate the role of 
many minor mergers in their recent history.

\subsection{Dependence on Bulge-to-Disk Ratios}
\label{sec:compare.BT}


\begin{figure}
    \centering
    \scaleupp
    \plotterr{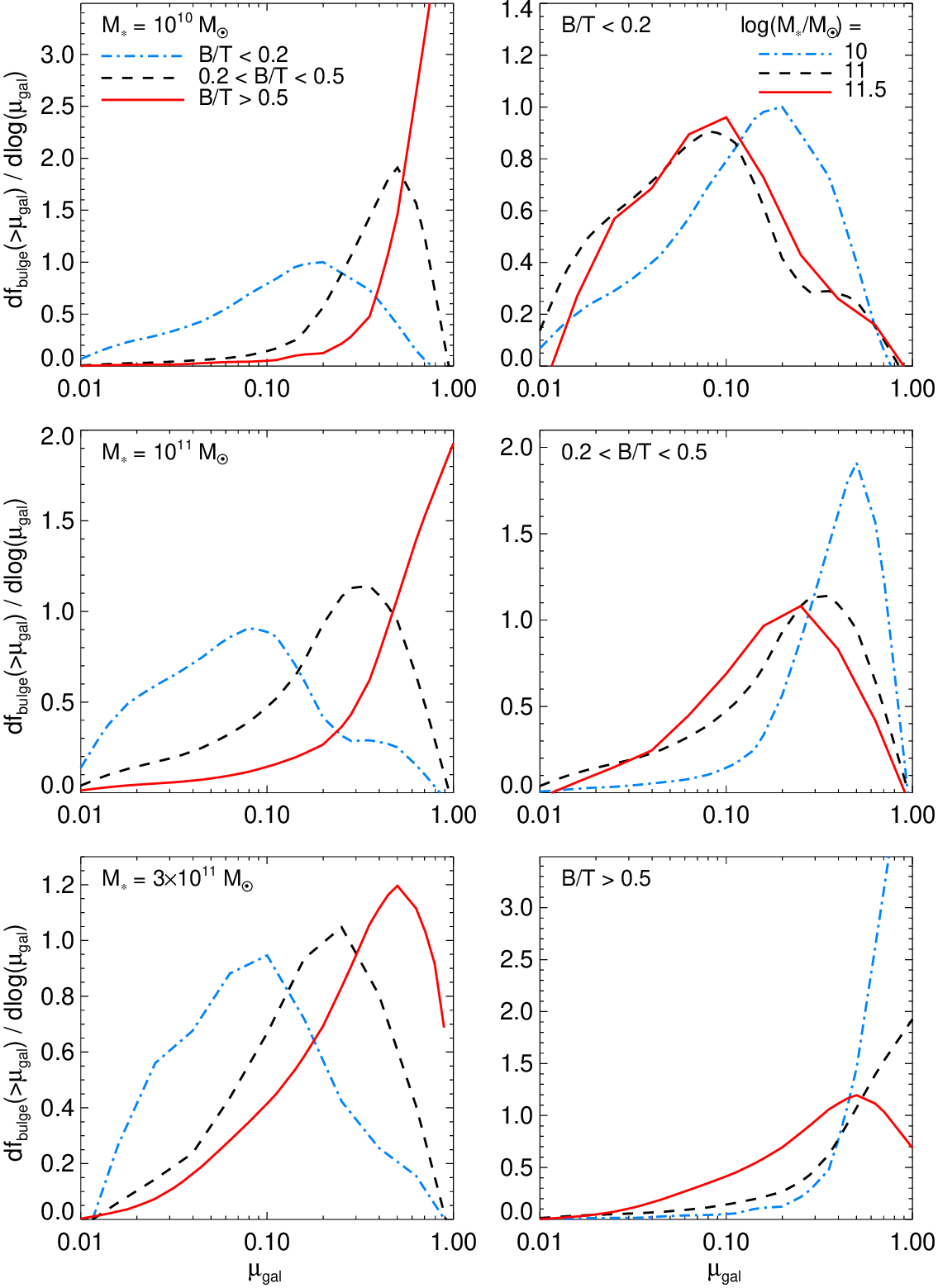}
    \caption{{\em Left:} Contribution of mergers of different $\mu_{\rm gal}$ 
    to bulge assembly (as Figure~\ref{fig:fmu.int.vs.inst}; differential 
    version), as a function of 
    bulge-to-total ratio $B/T$ at fixed galaxy 
    mass. Line type denotes the final $B/T$ value (as labeled), 
    and each panel shows galaxies of a different mass. 
    At all masses, more bulge-dominated systems 
    are formed by more major mergers. Ellipticals 
    and S0's are dominated by major merger remnants; 
    late-type disk bulges are preferentially formed in situ 
    in minor mergers. {\em Right:} Same, as a function of galaxy 
    stellar mass at fixed $B/T$ (lines denote different 
    galaxy masses, panels show results for systems with different 
    final $B/T$). 
    At fixed $B/T$, the residual dependence 
    on mass is weak; low-mass galaxies are more gas-rich, so require 
    more major mergers to reach the same $B/T$. 
    \label{fig:fmu.vs.BT.mbin}}
\end{figure}

Figure~\ref{fig:fmu.vs.BT.mbin} examines how the distribution of contributing 
$\mu_{\rm gal}$ (in terms 
of bulge assembly) scales with the bulge-to-total stellar mass ratio $B/T$ of 
galaxies. At fixed mass, galaxies with higher $B/T$ are formed in preferentially 
more major mergers, and the trend is similar at all masses. This is 
the natural expectation: a more major merger yields a system with higher 
$B/T$. Because $\mu_{\rm gal}\,{\rm d}N_{\rm merger}/{\rm d}\log{\mu_{\rm gal}}$ is not 
quite flat in $\mu_{\rm gal}$ (rising to larger $\mu$), and $\sim1$ significant mergers are expected 
since $z\sim2$ (i.e.\ at times when the galaxy is near is present mass), the 
local $B/T$ will be dominated by the largest merger the system has experienced in 
recent times. Many objects all have some 
amount of bulge built by $\mu_{\rm gal}\sim 0.1$ mergers -- the question is which will have larger 
mergers that convert more mass to bulge.

\begin{figure}
    \centering
    \scaleup
    \plotter{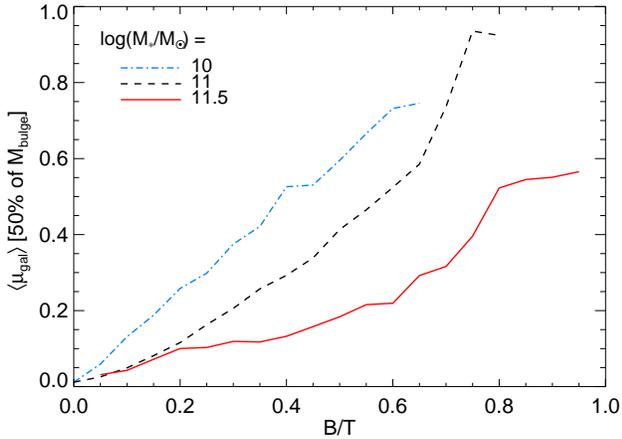}
    \caption{Summary of Figure~\ref{fig:fmu.vs.BT.mbin}: median mass ratio 
    of mergers contributing to bulge growth (defined as in 
    Figure~\ref{fig:fmu.vs.m}) as a function of the final $B/T$ 
    and galaxy stellar mass (different lines). Roughly speaking, the 
    correlation reflects the instantaneous scaling for a single merger, 
    $B/T\propto\mu_{\rm gal}$ (as expected if galaxies grow in a manner such that the 
    most recent mergers, at e.g.\ $z\lesssim2$, are most important). 
    \label{fig:mumean.vs.BT}}
\end{figure}

Figure~\ref{fig:mumean.vs.BT} summarizes these results, showing the (mass-weighted) median 
merger mass ratio $\langle\mu_{\rm gal}\rangle$ 
contributing to bulge formation as a function of $B/T$ 
at different stellar masses. 
At all masses, even masses where the {\em global} bulge population is predominantly 
formed in minor mergers, galaxies that are bulge-dominated (the E/S0 population) 
are predominantly 
assembled (and formed) in major mergers. 
In principle ten 1:10 mergers in a short time will form as much bulge 
as a single 1:1 merger. However, 1:10 mergers are {\em not} ten times more common, 
and as such are not an important or efficient channel for 
the formation of bulge-dominated galaxies. 
Recall that the average galaxy still experiences only $\sim1$ minor 
1:10 mergers since 
$z\sim2$ (see Figure~\ref{fig:rate.demo} \&\ \citet{stewart:mw.minor.accretion,
stewart:merger.rates}); the 
case of ten 1:10 mergers is then a $\sim 5-10\sigma$ outlier. 
Moreover, even if a system has several such mergers, they will 
be spaced widely in time (they essentially never occur 
simultaneously), so 
the galaxy disk will re-grow, reducing $B/T$ after each and offsetting 
the bulge growth from mergers. In contrast, 
$\sim 1/2$ of all galaxies undergo a single 1:3 
merger since $z\sim2$; these will 
immediately form a large $B/T$ system. In short, minor 
mergers are not {\em so much more common} than major mergers
as to dominate the formation 
of high $B/T$ systems. 

This is also important for reproducing the existence of 
disks (especially ``bulge-less'' disks) -- 
if minor mergers were so common as to dominate the formation of 
high $B/T$ systems (if e.g.\ half of $\sim L_{\ast}$ galaxies had formed through the 
channel of $\sim10$ rapid 1:10 mergers), it would be 
correspondingly much more rare for a system to 
have undergone very {\em few} 1:10 mergers, necessary to explain the existence of 
at least some significant number of low-mass systems with $B/T<0.1$. Moreover, 
in practice any system with such an extreme merger history is likely to have 
also experienced an enhanced major merger rate -- so the major mergers 
will still dominate bulge formation (it is unlikely to contrive an 
environment with so many $\sim1:10$ mergers in a short time and 
no major mergers). 

On the other hand, this implies that minor mergers do dominate the formation of 
bulges in {\em low} $B/T$ galaxies (Sb/Sc/Sd galaxies). 
This is for the same reason -- most galaxies 
have experienced $\sim 1$ 1:10 merger in recent times ($z<2$), whereas only some 
fraction have undergone more major mergers. The ``traditional'' scenario 
for bulge formation -- early formation in a major merger, followed by subsequent 
re-growth of a disk by new cooling -- is only responsible for a 
small fraction of the mass density in disk-dominated $B/T\lesssim0.2$ systems. 
It is very rare that a system would have such an early major merger 
but then {\em not} have a later $\sim1:10$ merger in the Hubble time required 
to grow the galaxy by a factor $\sim10$ in mass. 

These results are independent of all our model variations (\S~\ref{sec:robustness}), so long as 
we ensure that we reproduce a reasonable match to the observed HOD and 
halo merger rates. In fact, these conclusions 
appear to be quite general, similar to those found from other models 
that adopt different models for bulge formation in mergers 
\citep[see e.g.][]{khochfarsilk:new.sam.dry.mergers,weinzirl:b.t.dist}. 

Figures~\ref{fig:fmu.vs.BT.mbin} \&\ \ref{fig:mumean.vs.BT} also show how, 
at fixed $B/T$, the contributing mass ratio distribution depends on stellar mass. 
At fixed $B/T$, the trend with mass 
is much weaker than seen comparing all galaxies as a function of mass 
(Figure~\ref{fig:fmu.vs.m}). Moreover, the trend at 
fixed $B/T$ appears to 
have an {\em opposite} sense: low-mass galaxies require {\em higher} 
mass ratio $\mu_{\rm gal}$ mergers to reach the same $B/T$. 
This is priarily a consequence of the dependence of gas fraction 
on stellar mass and the effects of gas on bulge formation 
(Figure~\ref{fig:BT.fgas.mstar}). A low-mass galaxy, being 
very gas-rich, might require a major merger to even get to 
$B/T\sim0.2$ (if, say, $f_{\rm gas}\sim0.8$) -- so low-mass systems 
will require more major mergers. On the other hand, mergers in a 
massive $\sim10^{12}\,\msun$ system, being gas-poor ($f_{\rm gas}\lesssim0.05$), 
will yield $B/T>0.2$ for any mergers with $\mu_{\rm gal}>0.2$; 
the $\sim10^{12}\,\msun$ galaxies with low $B/T$ (what few there are) 
must be those that had only minor mergers in the last few Gyr. 

Overall, however, we wish to stress that 
systems with large $B/T$ at low masses and 
low $B/T$ at high masses are rare -- {\em most} low mass 
systems, having low $B/T$, have had relatively more contribution to their bulge 
growth from minor mergers, and most higher mass systems, having high $B/T$, 
have had increasing contributions from major mergers.

\subsection{Dependence on Galaxy Gas Fractions}
\label{sec:compare.fgas}

\begin{figure}
    \centering
    \scaleup
    \plotter{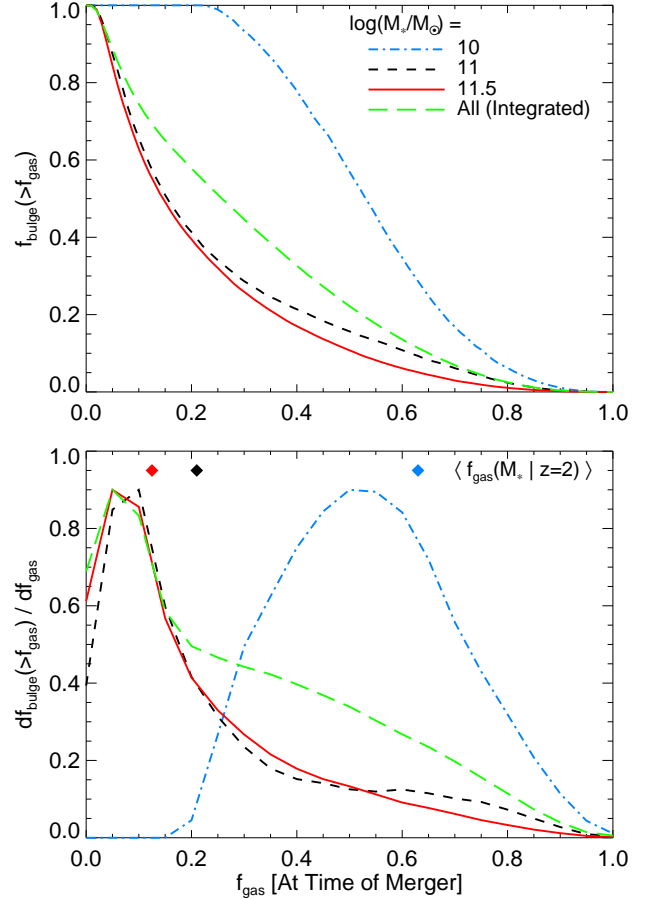}
    \caption{As Figure~\ref{fig:fmu.vs.BT.mbin}, but showing the contribution to 
    bulge growth from mergers with different gas fractions $\fgas$. {\em Top:} 
    Integrated distribution (fraction of bulge mass formed in mergers with 
    $\fgas$ above the given value; lines show results at different stellar masses, 
    with the ``all'' line integrated over the entire bulge mass function). 
    {\em Bottom:} Same, in differential form (contribution per unit $\fgas$). 
    We compare the median $z=2$ gas fractions of disks of the 
    same mass (diamonds, color corresponds to the 
    mass as labeled). To lowest order, the gas fractions of bulge-forming mergers 
    simply reflect the gas fractions of disks at the time of merger ($z=2$ is just representative; 
    the distribution of merger times is broad). 
    The bulge mass density is 
    dominated by mergers with $f_{\rm gas}\sim0.1-0.2$, with a tail towards 
    more dissipational mergers in lower-mass systems. 
    \label{fig:fmu.fgas}}
\end{figure}

Figure~\ref{fig:fmu.fgas} compares the contribution to bulge formation from mergers 
not as a function of mass ratio, but as a function of the gas-richness of the merger, 
where $f_{\rm gas}$ is here defined as the sum gas-richness just before 
the merger ($=(M_{\rm gas,\, 1}+M_{\rm gas,\, 2})/(M_{\rm gal,\, 1}+M_{\rm gal,\, 2})$).
In an integrated sense, the most important gas fractions for bulge formation are 
$\fgas\sim0.1-0.2$.\footnote{Recall, this is the gas fraction {\em at the time of the 
merger}, and can be different from the ``initial'' gas fraction at the beginning 
of an interaction, depending on e.g.\ the efficiency of star formation 
and stellar feedback. For example, in hydrodynamic simulations of idealized mergers 
without ongoing continuous accretion, a gas fraction at the time of merger of 
$\sim0.1-0.2$ corresponds to an ``initial'' gas fraction $\sim 2$\,Gyr before 
merger of $\sim0.3-0.4$.}. This agrees well with estimates from numerical simulations of the 
gas fractions required to form realistic $\sim L_{\ast}$ ellipticals 
\citep[in terms of their profile shapes, effective radii and fundamental plane 
correlations, rotation and higher-order kinematics, isophotal shapes, triaxiality, 
and other properties; see e.g.][]{naab:gas,
cox:kinematics,jesseit:kinematics, jesseit:merger.rem.spin.vs.gas,
hopkins:cusps.mergers,hopkins:cusps.fp,
hopkins:cusps.evol}\footnote{In fact, only mergers with these properties 
have been shown to yield a good match to these quantities: mergers with significantly 
less or more gas, as well as secular instabilities and 
dissipational collapse have been shown to yield remnants with properties 
unlike observed ellipticals.}. 
In a cosmological sense, it simply reflects 
the gas fractions of $\sim L_{\ast}$ disks, the progenitors of $\sim L_{\ast}$ 
ellipticals. 
We stress that this does not mean the bulges are made purely from this 
(relatively small) gas mass -- 
rather, this represents the typical mass fraction formed in a central starburst 
in the bulge-forming merger; the majority of the bulge mass is 
formed via violent relaxation of the 
pre-merger disk stars. 

As a function of mass, the typical merger contributing to bulge formation 
is more gas-rich at low masses. But as shown in Figure~\ref{fig:fmu.fgas}, this 
largely reflects the trend of 
gas fractions in late type or star-forming galaxies as a function of mass. 
At a given mass, in particular at the lowest masses where gas fractions can 
be sufficiently high as to significantly suppress bulge formation, there is a 
weak tendency for the dominant mergers contributing to bulge formation to 
be less gas-rich (since such mergers will, for the same mass ratio, form 
more bulge). However, the effect is not large.

\begin{figure}
    \centering
    \scaleup
    \plotter{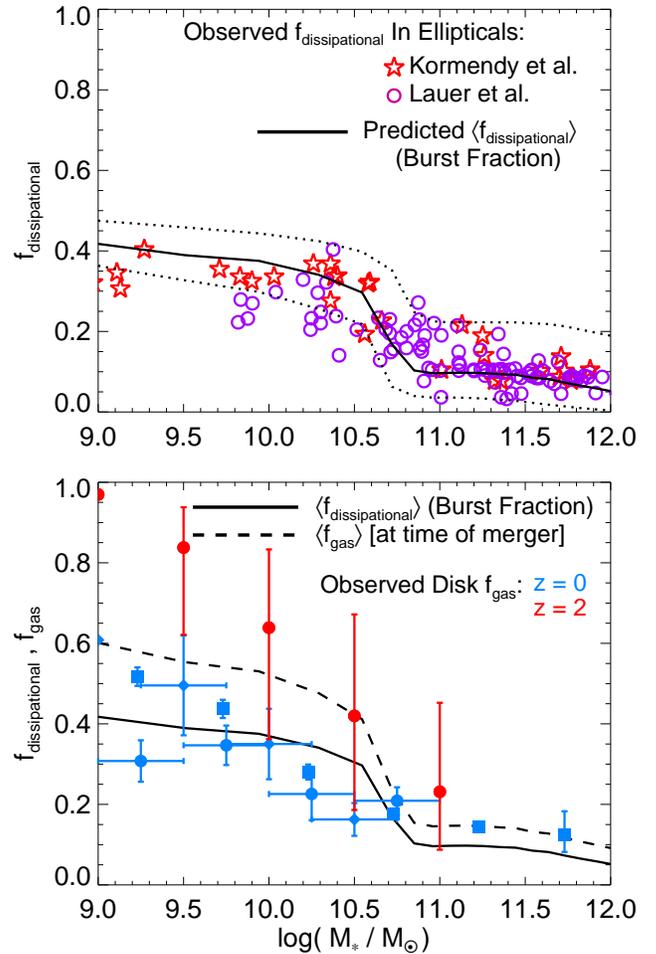}
    \caption{{\em Top:} Dissipational fraction (mass fraction of bulges formed in starbursts 
    from gas that has lost its angular momentum in mergers, relative to the total -- 
    starburst plus violently relaxed former stellar disk -- bulge mass) as a function of 
    stellar mass (solid line is the predicted median; dotted the $\pm1\,\sigma$ scatter). 
    We compare to empirically inferred $f_{\rm dissipational}$ from 
    decomposition of high-resolution 
    surface brightness profiles and kinematics of observed ellipticals, 
    presented in \citet{hopkins:cusps.ell,hopkins:cores} 
    with samples from \citet{jk:profiles} and \citet{lauer:bimodal.profiles}. 
    {\em Bottom:} Predicted dissipational fraction from above (solid line), 
    and median gas fraction $f_{\rm gas}$ of mergers contributing to bulge 
    growth (from Figure~\ref{fig:fmu.fgas}; dashed line), compared to observed disk gas 
    fractions (from Figure~\ref{fig:BT.fgas.mstar}; points in the same 
    style, with values at $z=0$ and $z=2$ in blue and red, respectively). Dissipational fractions 
    reflect the gas fractions of progenitor disks, 
    but with an asymptotic upper limit of $f_{\rm dissipational}\sim0.4$ 
    that reflects the suppression of angular momentum loss
    in very gas-rich mergers. 
    \label{fig:fdiss}}
\end{figure}

An important check of this is that it reproduce the 
``dissipational'' mass fraction in observed ellipticals, as a function of mass. 
This is the mass fraction of the spheroid formed in a dissipational 
starburst, rather than violently relaxed from the progenitor stellar disks. 
Being compact, this component is the primary element that determines the 
effective radii, profile shape, and ellipticity of a merger remnant. 
\citet{hopkins:cusps.ell,hopkins:cores,hopkins:cusps.fp} develop and test an empirical method 
to estimate the dissipational mass fraction in observed local ellipticals, 
and apply this to a wide range of observed ellipticals with a combination of HST and 
ground-based data from \citet{jk:profiles} 
and \citet{lauer:bimodal.profiles}. As shown therein, resolved stellar 
population studies yield supporting conclusions. 
Figure~\ref{fig:fdiss} shows these empirically inferred dissipational fractions as a 
function of mass, and compares the predictions from the models here. 
The agreement is reasonable. Similar conclusions are reached even by 
models with significantly different bulge formation prescriptions 
\citep{khochfar:size.evolution.model}. Note that the 
observed systems here are all classical bulges, appropriate for 
comparison to our predictions. 

We also compare observed disk gas fractions. To lowest order, the 
dissipational fractions simply trace these gas fractions, but at low masses, 
the predicted and observed dissipational fractions asymptote to a 
maximum $\sim0.3-0.4$. 
This is because angular momentum loss in the gas becomes less efficient at 
these high gas fractions; 
if the fraction of gas losing angular momentum scales as 
adopted here, then the dissipational fraction of the bulge formed 
from disks with gas fraction $f_{\rm gas}$ is not $\sim f_{\rm gas}$, 
but $\sim f_{\rm gas}/(1+f_{\rm gas})$, i.e.\ asymptoting 
to the values observed for all $f_{\rm gas}\sim0.5-0.9$.

\subsection{Redshift Evolution: Can Mergers Account for the Mass Density in Bulges?}
\label{sec:compare.evol}

\begin{figure}
    \centering
    \scaleupp
    \plotterr{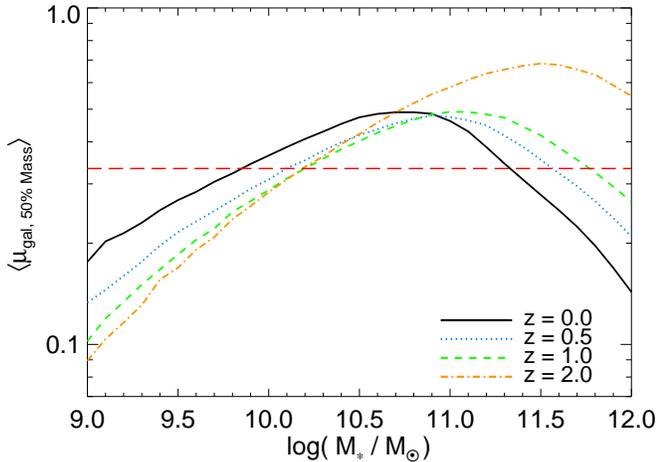}
    \caption{As Figure~\ref{fig:fmu.vs.m}, showing the 
    (mass-weighted) median mass ratio $\mu_{\rm gal}$ 
    contributing to bulge assembly as a function of mass for samples at different 
    observed redshifts, $z=0-2$ (different lines, 
    as labeled; again the red horizontal line denotes the 
    traditional major merger definition $\mu>1/3$). The qualitative trends are similar; 
    at high redshifts the break/turnover moves to higher masses. 
    \label{fig:fmu.vs.m.z}}
\end{figure}

At all redshifts, the distribution of $\mu_{\rm gal}$ contributing to bulges is similar 
to that at $z=0$, shown in 
Figure~\ref{fig:fmu.vs.m.z}. The only significant evolution is that the ``turnover'' 
in $\langle\mu_{\rm gal}\rangle$ at high masses for the mass ratios contributing 
to bulge assembly (Figure~\ref{fig:fmu.vs.m}) becomes less pronounced 
and moves to higher 
masses. Technically, this relates to how $M_{\rm gal}(M_{\rm halo})$ (empirically 
constrained) evolves, but physically it is simply understood: at higher 
redshifts, ``dry'' assembly is less important, so the 
assembly $\langle\mu_{\rm gal}\rangle$ increasingly 
resembles the formation $\langle \mu_{\rm gal} \rangle$ 
(Figure~\ref{fig:fmu}). By $z\sim2$, there is no difference -- dry assembly 
is negligible, and all high-mass, high-$B/T$ 
systems are preferentially formed in major mergers.

\begin{figure}
    \centering
    \scaleup
    \plotter{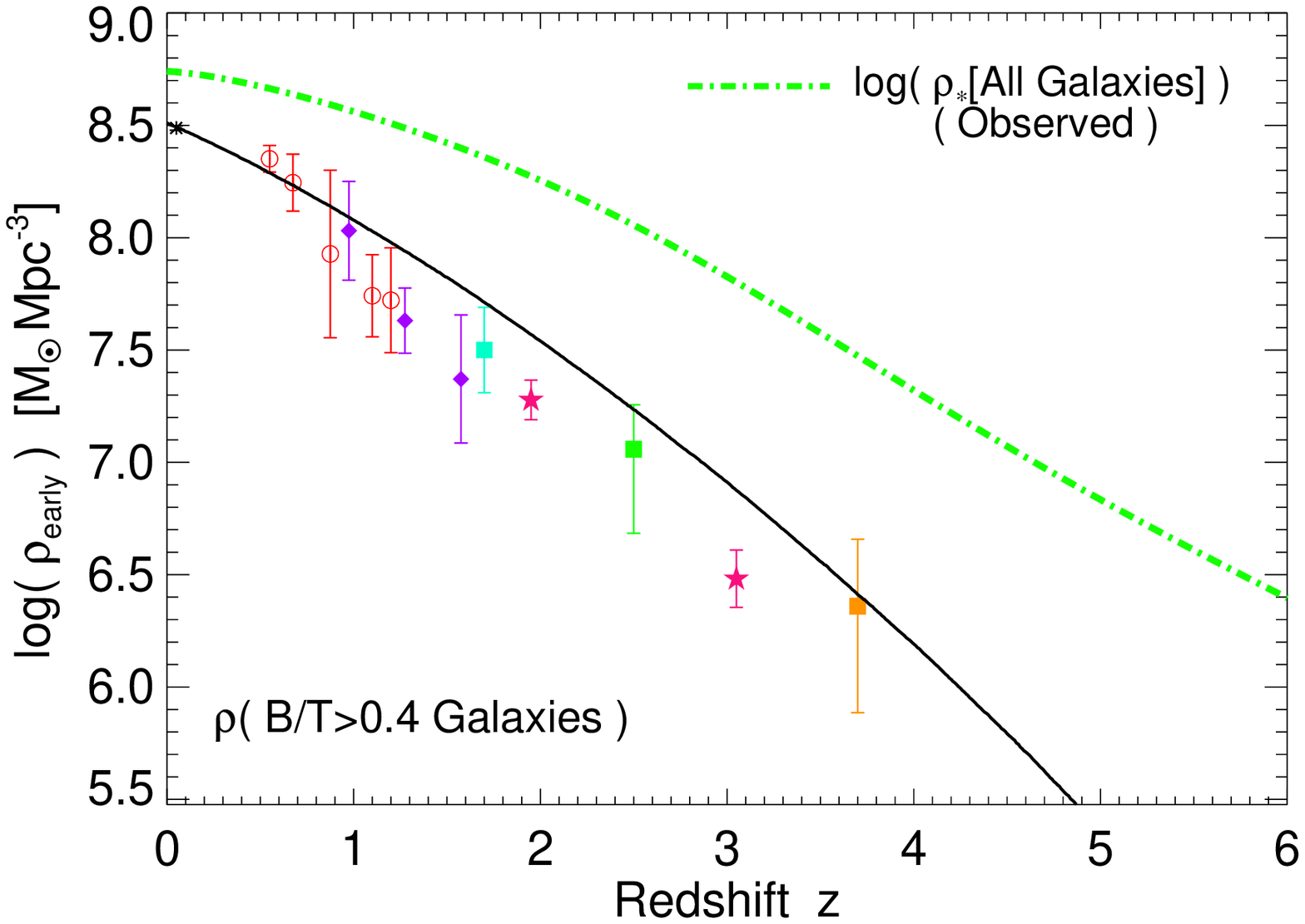}
    \caption{Predicted 
    integrated mass density in bulge-dominated galaxies ($B/T>0.4$; black solid line) as a 
    function of redshift, compared to observations (points). 
    Observations are from the morphologically-selected 
    samples of \citet[][black $\times$]{bell:mfs}, 
    \citet[][red circles]{bundy:mfs,bundy:mtrans}, 
    \citet[][violet diamonds]{abraham:red.mass.density}, 
    and \citet[][cyan square]{daddi05:drgs}, and color-selected 
    samples of 
    \citet[][green square]{labbe05:drgs},
    \citet[][orange square]{vandokkum06:drgs}, and 
    \citet[][magenta stars]{grazian:drg.comparisons}.
    We compare the total stellar density observed
    \citep[green dot-dashed line;][]{hopkinsbeacom:sfh}. 
    Increasing gas fractions and cooling rates at high redshift 
    suppress the bulge mass density relative to the total stellar 
    mass density. 
    The predicted number of mergers is sufficient to account for the 
    $z=0$ and high-redshift evolution in the global bulge mass budget, 
    but with factor $\sim2$ uncertainties. 
    \label{fig:rhobul.evol}}
\end{figure}

Given these predicted merger rates and $B/T$ distributions, we also obtain a 
prediction for the mass density in bulge-dominated galaxies as a function of 
redshift. 
Figure~\ref{fig:rhobul.evol} compares the redshift 
evolution of the bulge mass density to that observed.
\footnote{
Specifically,
  we plot the mass density in bulge-dominated galaxies,
  which is not the same as the absolute mass density in all bulges,
  but is closer to the observed quantity. At high redshifts 
  $z>1.5$ observed morphologies are ambiguous; we show 
  the mass density in passively
  evolving red galaxies as a proxy. This may not be appropriate, 
  but at $z<1$ the two correspond well, and 
  the compactness, size, and kinematics of the ``passive'' objects 
  do appear distinct from star-forming objects 
  \citep{kriek:drg.seds,toft:z2.sizes.vs.sfr,trujillo:ell.size.evol.update,
  franx:size.evol,genzel:highz.rapid.secular}. Also, 
  the observations in some cases do not distinguish ``classical'' and 
  ``pseudo'' bulges; but as we discuss in \S~\ref{sec:discuss}, the 
  latter where measured appear to contribute only $\sim10\%$ of the 
  mass density.
}
The agreement is good: not only are 
there a sufficient number of major mergers to account for the observed 
merger fractions, but also to account for the observed buildup of the bulge 
population with redshift. This should not be surprising, given the agreement 
with observed merger fractions demonstrated in \S~\ref{sec:mgr.rates.obs} 
above; \citet{hopkins:transition.mass,hopkins:groups.ell}, \citet{bundy:merger.fraction.new}, 
and \citet{bell:merger.fraction} have 
demonstrated that observed major merger fractions are sufficient, 
within a factor $\sim2$, to account for 
the observed growth of the bulge population over the same redshift interval
(given the observable lifetime calibrations that we adopt in \S~\ref{sec:mgr.rates.obs}; 
note that some of these works use different merger timescale estimates and reach 
different conclusions, 
but they are consistent using a {\em uniform}, simulation-calibrated timescale). 
Similar results as a function of galaxy morphology are suggested by local 
observations \citep[e.g.][]{darg:galzoo.merger.frac.by.morph}. 
Likewise the agreement between the predicted integrated number of mergers 
and various observational estimates suggests this is plausible 
\citep{deravel:merger.fraction.to.z1,conselice:mgr.pairs.updated,lin:mergers.by.type}. 

For a more detailed comparison, as a function of 
e.g.\ galaxy stellar mass, we refer to \citet{hopkins:disk.survival.cosmo}, 
who use the same merger rates as modeled here to predict e.g.\ the 
morphology ($B/T$)-mass relation and bulge/disk mass functions as a 
function of redshift. Provided proper account of galaxy gas fractions is 
taken, good agreement is obtained. \citet{stewart:disk.survival.vs.mergerrates} 
perform a similar calculation (with a basic criteria for bulge formation), 
with merger rates in close agreement as a function of galaxy 
mass to those measured observationally in \citet{bundy:merger.fraction.new}, 
and obtain similar good agreement with the bulge 
mass function as a function of redshift. They actually find 
that bulge mass is somewhat over-produced, without accounting 
for the role of gas-rich mergers. 
\citet{hopkins:groups.qso} and \citet{hopkins:groups.ell} 
consider a range of model parameter space 
(with several of the specific model variations discussed in \S~\ref{sec:robustness}) 
and perform similar calculations; they explicitly show the predicted 
spheroid mass function and mass fraction as a function of 
stellar mass, halo mass, environment, and redshift, for the different models 
considered. They likewise conclude that, for all the model variations 
considered (with scatter in merger rates as a function of mass similar 
to that discussed below), good agreement with the 
mass function and mass density 
of classical bulges, at masses $>10^{10}\,\msun$, is obtained. At lower 
masses, however, uncertainties grow rapidly. 

Relative to the {\em total} stellar mass density, the mass density in 
bulge-dominated galaxies decreases with redshift. This is discussed in 
detail in \citet{hopkins:disk.survival.cosmo}, but the reason is simply that 
at higher redshifts, higher gas fractions 
suppress bulges (and the suppression moves to 
higher mass, relative to the galaxy mass function break. 
This trend agrees with that observed and is not trivial 
(models neglecting the importance of gas-richness in affecting 
bulge formation efficiency in mergers, for example, may predict 
the opposite).

\subsection{Analytic Fits}
\label{sec:compare.analytic}

It is convenient to fit the distribution of merger mass ratios contributing to bulge 
formation at a given mass.
The average $\mu_{\rm gal}$ contributing to bulge 
{\em assembly}, i.e.\ $\langle\mu_{\rm gal}\rangle$ where 
$f_{\rm bulge}(>\mu_{\rm gal})=0.5$, as a function of 
$z=0$ galaxy stellar mass $M_{\ast}$ (Figure~\ref{fig:fmu.vs.m}) 
can be approximated as 
\begin{equation}
\langle\mu_{\rm gal}\rangle = \frac{1}{(M_{\ast}/10^{11}\,\msun)^{-0.5}+
(M_{\ast}/10^{11}\,\msun)^{0.8}} \ .
\end{equation}
If instead the $\langle\mu_{\rm gal}\rangle$ contribution to bulge 
{\em formation} is desired, a similar formula applies, but with a 
weaker turnover at high $M_{\ast}$, i.e.\ 
\begin{equation}
\langle\mu_{\rm gal}\rangle = \frac{1}{(M_{\ast}/10^{11}\,\msun)^{-0.5}+
(M_{\ast}/10^{11}\,\msun)^{0.2}} \ .
\end{equation}

In greater detail, Figures~\ref{fig:fmu.vs.BT.mbin}-\ref{fig:mumean.vs.BT} 
demonstrate that the typical $\langle\mu_{\rm gal}\rangle$ contributing to 
bulge assembly depends on the bulge 
mass fraction $B/T$ at a given mass. 
As a bivariate 
function of $B/T$ and mass, $\langle\mu_{\rm gal}\rangle$ 
can be approximated by: 
\begin{equation}
\langle\mu_{\rm gal}\rangle = {\Bigl[}\frac{B}{T}{\Bigr]} \times \frac{1}{1 +
(M_{\ast}/10^{11}\,\msun)^{0.5}} \ .
\label{eqn:mugal.vs.bt.mass}
\end{equation}

Finally, knowing $\langle\mu_{\rm gal}\rangle$, we find that the complete 
distribution $f_{\rm bulge}(>\mu_{\rm gal})$ (e.g.\ Figure~\ref{fig:fmu}) 
can be simply approximated by 
\begin{equation}
f_{\rm bulge}(>\mu_{\rm gal}) = (1 - \mu_{\rm gal})^{\gamma}
\end{equation}
where (since $f_{\rm bulge}(>\langle\mu_{\rm gal}\rangle)=0.5$), 
the slope $\gamma$ is trivially related to $\langle\mu_{\rm gal}\rangle$ as: 
\begin{equation}
\gamma = \frac{-\ln{2}}{\ln{(1-\langle\mu_{\rm gal}\rangle)}} \ .
\end{equation}

\breaker
\section{How Robust Are These Conclusions? A Comparison of Models}
\label{sec:robustness}

The relative importance of e.g.\ minor and major mergers in bulge assembly 
owes to the combination of 
reasonably well-determined halo merger rates and halo occupation 
statistics. Nevertheless, there are still uncertainties in this 
approach. We therefore examine how robust the conclusions here are 
to a variety of possible model differences. 
A much more detailed investigation of e.g.\ differences in predicted 
merger rates between semi-empirical models, semi-analytic models, 
and simulations will be the subject of a companion paper 
\citep{hopkins:merger.rates.methods}. Here, we wish to examine 
differences arising within the semi-empirical framework.

\begin{figure*}
    \centering
    \scaleup
    \plotter{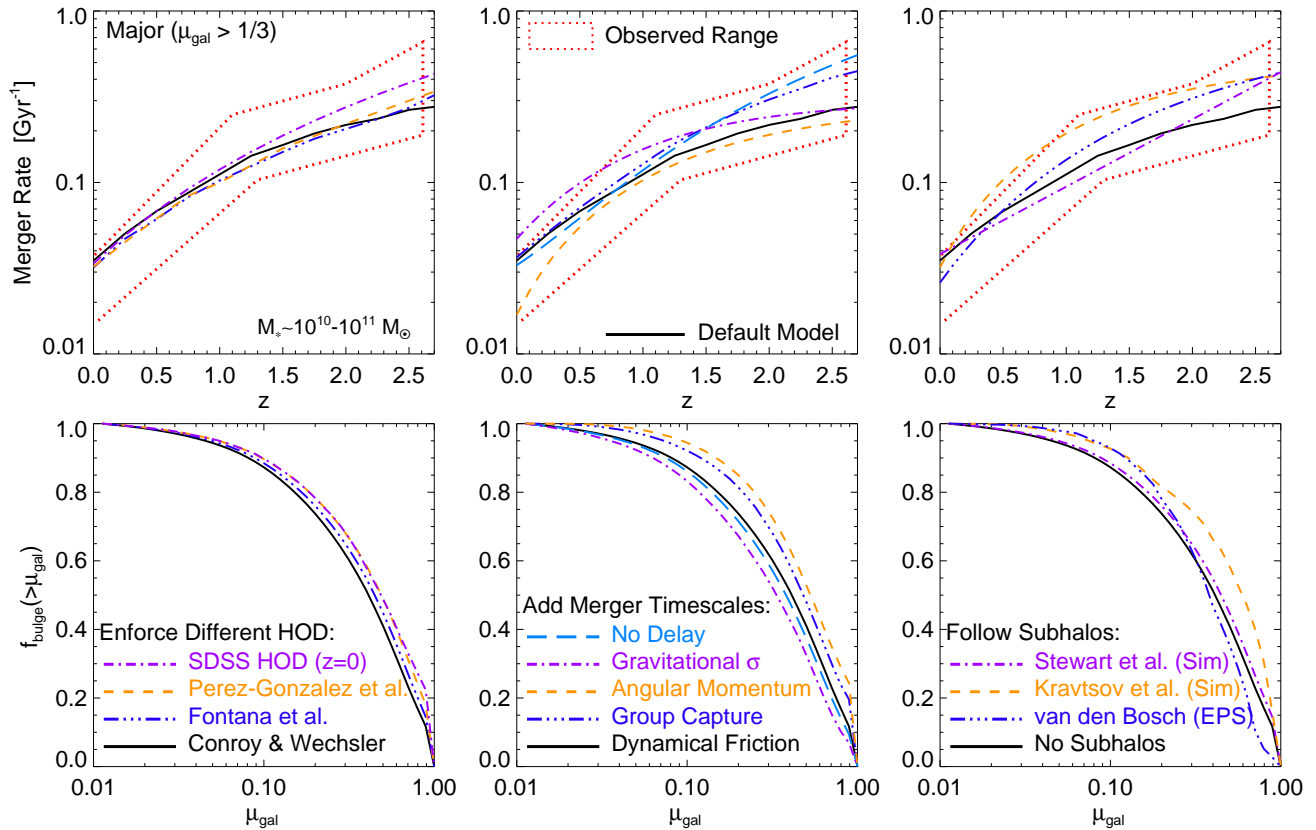}
    \caption{{\em Top:} Predicted major merger rate
    ($\mu_{\rm gal}>1/3$), varying the assumptions 
    in the models (different lines). Black line is the ``default'' model assumption. Red dotted 
    range is the approximate range allowed by observations 
    (the compiled points in Table~\ref{tbl:obs} \&\ Figure~\ref{fig:mgr.rate.vs.z}; 
    for clarity, we show the range of these points rather than each individual measurement). 
    {\em Bottom:} Corresponding cumulative contribution of different mass 
    ratio mergers to the assembly of the bulge mass density (as 
    Figure~\ref{fig:fmu}, integrated over all bulge masses). 
    {\em Left:} Changing the halo occupation constraints: 
    the three cases from Figure~\ref{fig:rate.demo} are shown, together 
    with adopting the $z=0$ SDSS fits from \citet{wang:sdss.hod} at 
    all redshifts. 
    {\em Center:} Changing the merger timescale between halo-halo 
    and galaxy-galaxy merger: using dynamical 
    friction times \citep[calibrated in][]{boylankolchin:merger.time.calibration}; 
    angular momentum-space capture cross sections 
    from \citet{binneytremaine}; 
    collisional group capture cross sections from 
    \citet{mamon:groups.review}; gravitational 
    capture cross-sections for field/small group crossings from 
    \citet{krivitsky.kontorovich}; or no delay. 
    {\em Right:} Tracking subhalos to assign merger times 
    or using subhalo 
    mass functions instead of halo merger 
    trees as a starting point: 
    using the subhalo merger trees from cosmological 
    simulations in \citet{stewart:merger.rates}; 
    adopting the subhalo mass functions from simulations 
    in \citet{kravtsov:subhalo.mfs}; or the 
    same from extended Press-Schechter trees constructed following 
    \citet{vandenbosch:subhalo.mf}. 
    \label{fig:mgr.rate.vs.model.1}}
\end{figure*}

\begin{figure*}
    \centering
    \scaleup
    \plotter{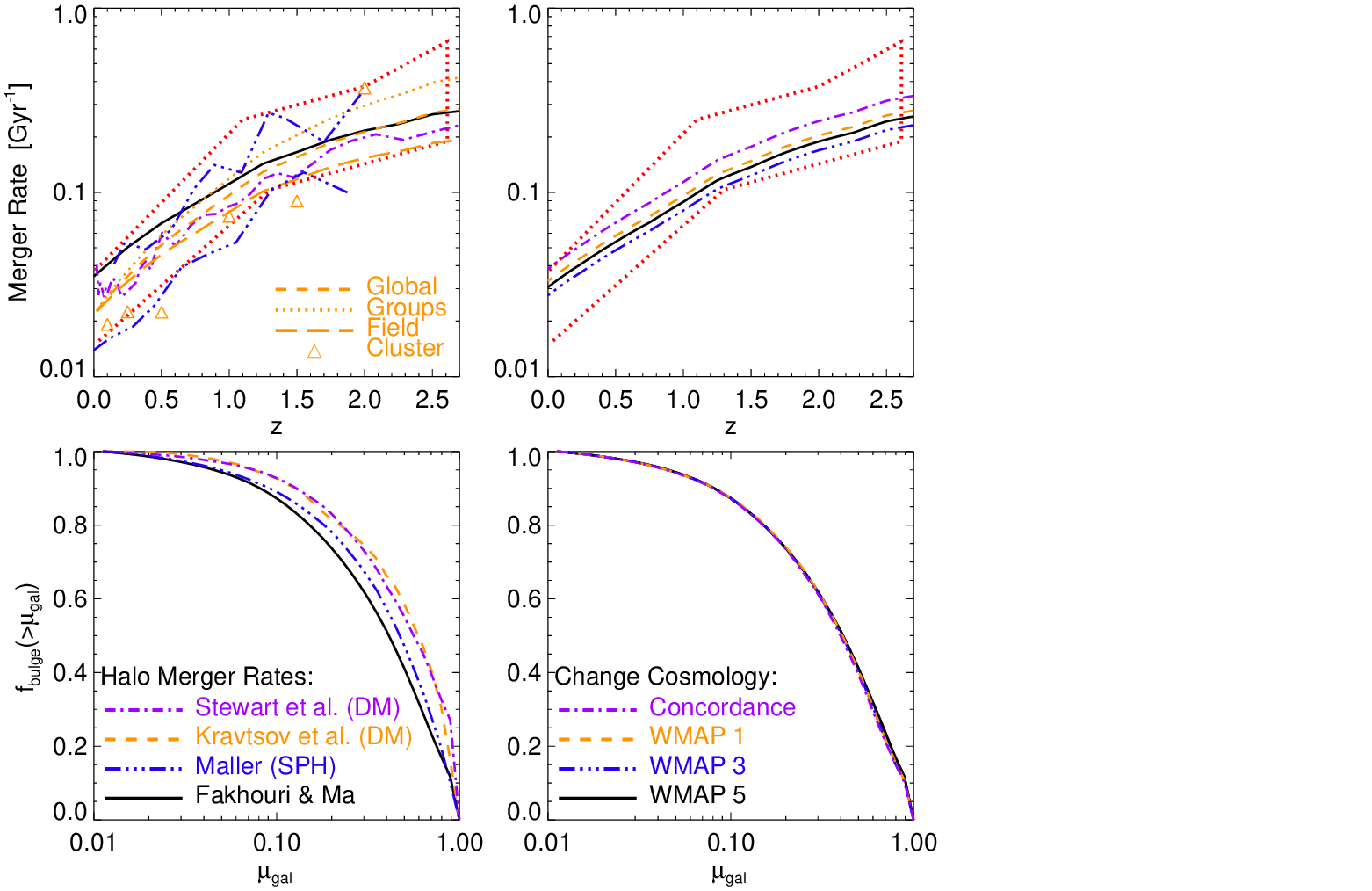}
    \caption{Figure~\ref{fig:mgr.rate.vs.model.1}, continued. 
    {\em Left:} Changing the halo-halo merger trees: our default choice from 
    the Millenium simulation analyzed in 
    \citet{fakhouri:halo.merger.rates}; 
    numerical trees from an alternative high-resolution cosmological simulation 
    \citep{stewart:mw.minor.accretion,stewart:merger.rates};  
    cosmological DM only simulations of field, group, and cluster 
    environments \citep[as labeled;][]{gottlober:merger.rate.vs.env}; or 
    merger rates from 
    cosmological SPH simulations \citep[tracking galaxy-galaxy mergers but still 
    re-populating them appropriately for the observed HOD;][upper and lower 
    correspond to their high and medium-mass primary sample, 
    respectively]{maller:sph.merger.rates}. 
    {\em Center:} Changing the cosmology: our default WMAP5 
    ($\Omega_{m}$,\,$\sigma_{8}$)=($0.27$,\,$0.81$) cosmology, versus 
    a WMAP1 ($0.27$,\,$0.84$), 
    WMAP3 ($0.27$,\,$0.77$), and 
    ``concordance'' ($0.3$,\,$0.9$) cosmology. 
    \label{fig:mgr.rate.vs.model.2}}
\end{figure*}

Figures~\ref{fig:mgr.rate.vs.model.1}-\ref{fig:mgr.rate.vs.model.2} 
compare our ``default'' model with a number of 
alternatives. For each alternative, we show the 
the merger rate as a function of 
redshift, and the 
integrated contribution of different mass ratios
$f_{\rm bulge}(>\mu_{\rm gal})$ to bulge 
assembly (integrated over all bulge masses). 
For the merger rates, we 
also compare with the observational constraints: the dotted region in the 
Figure shows the approximate 
$\pm1\,\sigma$ allowed range from the observational compilation 
in Figures~\ref{fig:mgr.rate.vs.z}-\ref{fig:mgr.rate.vs.obs} 
(fitting a piecewise broken power-law to the 
observations), given the merger lifetime calibrations discussed 
in \S~\ref{sec:mgr.rates.obs}. 

The model variations we consider include:

\subsection{Halo Occupation Models}
\label{sec:robustness:hod}

Here, we consider an otherwise identical 
model, but adopt a different set of halo occupation constraints 
to determine $M_{\ast}(M_{\rm halo})$ 
(for now, we keep $M_{\rm gas}(M_{\ast})$ fixed, but varying 
that is very similar to varying $M_{\ast}(M_{\rm halo})$). 
First, our default model, using the fits from 
\citet{conroy:hod.vs.z}. Second, the first to $M_{\ast}(M_{\rm halo})$ 
and its scatter for central and satellite galaxies from the observed SDSS 
clustering at $z=0$ \citep{wang:sdss.hod}; here, we simply adopt 
the $z=0$ fit at all redshifts -- we do not allow for evolution. 
Third, assigning galaxies to halos and subhalos based 
on a monotonic rank-ordering method 
\citep[see][]{conroy:monotonic.hod}, fitting to the 
redshift-dependent stellar mass function from \citet{fontana:highz.mfs}. 
Fourth, the same, with the mass functions from 
\citet{perezgonzalez:mf.compilation}. 

We have also considered various fits directly taken from other sources, including 
\citet{yan:clf.evolution,cooray:highz,cooray:hod.clf,
conroy:monotonic.hod,conroy:mhalo-mgal.evol,zheng:hod.evolution} 
and \citet{perezgonzalez:hod.ell.evol}; 
these lie within the range shown in Figure~\ref{fig:mgr.rate.vs.model.1}. 
Using the HOD predicted by semi-analytic models, at least for 
central galaxies, also appears to give similar results 
\citep[we have compared the results in][]{croton:sam,bower:sam,
delucia:sam}; given that these models are constrained to match the 
observed stellar mass function, this appears to be sufficient for the 
level of convergence shown.

\subsection{Merger Timescales}
\label{sec:robustness:timescales}

In our ``default'' model, we assume a delay between halo-halo 
and galaxy-galaxy mergers, given by the dynamical friction time 
calibrated to simulations in \citet{boylankolchin:merger.time.calibration}. 
We now allow this to vary according to five different scalings, 
described in detail in \citet{hopkins:groups.qso}. 
{\bf (a)} Dynamical Friction: the traditional dynamical friction time, 
using the calibration from numerical simulations in 
\citet{boylankolchin:merger.time.calibration} 
\citep[see also][]{jiang:dynfric.calibration}. 
{\bf (b)} Group Capture: the characteristic timescale for pair-pair gravitational 
capture in group environments, calibrated to simulations following 
\citet{mamon:groups.review} \citep[see also][]{white:cross.section,
makino:merger.cross.sections}
{\bf (c)} Angular Momentum Capture: as group capture, but considering 
capture in angular momentum space rather than gravitationally, 
following \citet{binneytremaine}. 
{\bf (d)} Gravitational Cross-Sections: similar to the group capture timescale, 
this is the timescale for gravitational capture between passages in e.g.\ 
loose group or field environments, calibrated from simulations 
in \citet{krivitsky.kontorovich}. {\bf (e)} No Delay: simply 
assuming galaxy-galaxy mergers occur when their parent halo-halo 
mergers do. 

Although the dynamical friction time is most commonly adopted in 
e.g.\ semi-analytic models, each of these 
timescales depends on certain assumptions and is relevant in different 
regimes. A dynamical friction time is appropriate for a 
small, dense satellite at large radii; it becomes less so at small radii. 
A group capture or 
gravitational capture cross section, 
on the other hand, is appropriate for collisions in small 
groups or field environments where ``inspiral'' is not well-defined. The 
angular momentum calibration from 
\citet{binneytremaine} is more appropriate for 
satellite-satellite mergers. 
In any case, we see the choice makes little difference to our conclusions. 
And the range between these choices is generally 
much larger than other, more subtle details (e.g.\ allowing for continuous 
satellite mass loss in inspiral, or 
resonant baryonic effects that speed up the coalescence). The reason is simply that we are largely 
focused on fairly major mergers, for which the merger time is short 
relative to the Hubble time. The rate-limiting step is the accretion of such a 
companion, not the inspiral time.

\subsection{SubHalo Mass Functions/Substructure-Based Methodologies}
\label{sec:robustness:subhalos}

Instead of using a halo-halo merger rate with some ``delay'' 
applied, we can attempt to follow subhalos directly after the halo-halo 
merger, and define the galaxy-galaxy merger when the subhalos 
are fully merged/destroyed. 
This will self-consistently allow for 
some distribution of merger times owing to e.g.\ a range of orbital parameters, 
and will include satellite-satellite mergers 
(neglected in our default model), which \citet{wetzel:mgr.rate.subhalos} show 
can be important at the $\sim10-20\%$ level independent of halo mass. 

Here, we compare our default model to those obtained 
tracking the halo+subhalo populations in cosmological simulations from
\citet{stewart:merger.rates} (populating subhalos according to our default HOD). 
\citet{wetzel:mgr.rate.subhalos,wetzel:merger.bias} 
also analyze subhalo merger rates, with a different 
methodology. They reach similar conclusions, but with a systematic 
factor $\sim1.5-2$ lower merger rate. As they discuss, this is 
quite sensitive to how one defines e.g.\ subhalo versus friends-of-friends group 
masses; some of those choices of definition will be ``normalized out'' 
by the appropriate HOD (renormalized for whatever subhalo populations 
are identified in a simulation so as to reproduce the observed clustering 
and mass functions), but it also 
reflects inherent physical uncertainties in the instantaneous mass and 
time of subhalo merger. 

We also compare 
with the results using the different subhalo-based methodology 
described in \citet{hopkins:groups.qso} (essentially, beginning from the subhalo 
mass function constructed from cosmological simulations and evolving this 
forward in short time intervals after populating it, at each time, according to 
the HOD constraints). We compare two different 
constructions of the subhalo mass functions: that from 
cosmological dark-matter only simulations in 
\citet{kravtsov:subhalo.mfs} \citep[see also][]{zentner:substructure.sam.hod} 
and that from the extended Press-Schechter formalism coupled to 
basic prescriptions for subhalo dynamical evolution, 
described in \citet{vandenbosch:subhalo.mf}. 
Alternative subhalo mass functions from e.g.\ 
\citet{delucia:subhalos,gao:subhalo.mf,nurmi:subhalo.mf} 
are consistent.

\subsection{Halo Merger Rates}
\label{sec:robustness:halomergers}

We can next vary the halo-halo merger rates adopted. 
Our default model uses the merger rates in 
\citet{fakhouri:halo.merger.rates}, calibrated from the Millenium 
dark-matter only cosmological simulation \citep{springel:millenium,
springel:millenium.review}. Since we use the full history, 
this is equivalent to the ``per progenitor'' merger rates defined 
from the same simulation in \citet{genel:merger.rates.perprogenitor}. 
An alternative dark-matter simulation, of comparable resolution, with 
halo merger rates determined using a different methodology, 
is described in \citet{stewart:merger.rates}. Another is found 
in \citet{gottlober:merger.rate.vs.env} 
\citep[see also][]{kravtsov:subhalo.mfs,zentner:substructure.sam.hod}; they 
quantify the fit separately to field, group, and cluster environments. 

We can also compare with the merger rates from 
\citet{maller:sph.merger.rates}, determined 
from cosmological hydrodynamic simulations.  
Although it is well known that, without proper implementations of 
feedback from various sources, cosmological hydrodynamic simulations yield 
galaxies that suffer from overcooling (and do not reproduce the observed 
halo occupation statistics), the galaxies in these simulations can 
still serve as ``tracers'' of halos and subhalos.
This provides a means to avoid the 
considerable ambiguities in defining a halo merger 
(moreover in considering the delay between halo-halo 
and galaxy-galaxy mergers). Although the galaxy {\em masses} may not be correct, 
they are still tracers of where in the halo real galaxies should be, 
and therefore can be used to measure the halo merger rate. 
We do so by recalculating their merger rates after re-populating the 
galaxies appropriately (essentially renormalizing their predicted 
mass function to match that observed).

\subsection{Cosmological Parameters}
\label{sec:robustness:cosmology}

We can also vary the 
cosmological parameters and see if this makes a significant difference 
to our conclusions. We consider four sets of cosmological 
parameters: a ``concordance'' model with 
$(\Omega_{\rm M},\,\Omega_{\Lambda},\,h,\,\sigma_{8},\,n_{s})$=$(0.3,\,0.7,\,0.7,\,0.9,\,1.0)$, 
the WMAP1 $(0.27,\,0.73,\,0.71,\,0.84,\,0.96)$ results of \citet{spergel:wmap1}, 
WMAP3 $(0.268,\,0.732,\,0.704,\,0.776,\,0.947)$ \citep{spergel:wmap3}, 
and WMAP5 $(0.274,\,0.726,\,0.705,\,0.812,\,0.96)$ \citep{komatsu:wmap5}. 
It is prohibitively expensive to re-run the simulations for each case, and moreover 
the qualitative behavior is not expected to change (seen in e.g.\ lower-resolution 
dark-matter simulations). We simply renormalize the halo masses at all times 
to match the halo mass function and accretion history appropriate for the 
revised cosmological parameters \citep[see e.g.][]{neistein:natural.downsizing} -- 
the dominant effect is the predicted halo mass function shifting to higher masses 
with larger $\sigma_{8}$. However, because we use a halo occupation-based 
approach, the model is re-normalized to yield the same observed galaxy mass 
function and clustering, so these differences are largely normalized out. 
\citet{elahi:substructure.mf.vs.powerspectrum.ns} show that the quantity of greatest 
importance for our conclusions, the normalized substructure mass function or 
(equivalently) dimensionless merger rate (mergers per halo per Hubble time 
per unit mass ratio) is almost completely independent of 
cosmological parameters including e.g.\ the power spectrum shape and 
amplitude over the range of variations here (not until 
one goes to much larger effective $n_{s}\sim3$ does one see 
this function change shape).

\subsection{Bulge Formation Prescriptions}
\label{sec:robustness:bulform}

We can also consider variations in the physical prescription by which bulge mass 
is formed in mergers. Obviously this will not change the merger rates, but it could 
change the relative importance of mergers of different mass ratios. However, 
we are tightly constrained by the results of $N$-body simulations; since the physics 
determining gas angular momentum loss and violent relaxation are predominantly 
gravitational, there is little uncertainty in how much bulge should be formed 
in a given merger (given the appropriately normalized initial conditions of interest). 
Still, there are some differences in fitted prescriptions: 
we have re-calculated the results from our default model according to the 
approximate results of 
simulations from \citet{naab:minor.mergers} 
and \citet{naab:etg.formation}, as well as
\citet{bournaud:minor.mergers} and \citet{dimatteo:merger.induced.sb.sims}. 
We have also used the fits to the same suite of simulations in \citet{hopkins:disk.survival} 
as presented in \citet{cox:feedback} and \citet{cox:massratio.starbursts}. Note that 
the results in several of these works do not necessarily include a complete survey 
of parameters such as e.g.\ mass ratio, orbital parameters, and gas fraction; where 
not given we interpolate between the results presented based on the model outlined 
in \citet{hopkins:disk.survival}. In any case, the differences in quantities such as 
the absolute bulge mass (especially in gas-rich, low mass systems) and 
dissipational fractions (fraction of mass formed in starbursts, rather than 
violently relaxed from stellar disks) 
of ellipticals can be non-negligible, but the relative contribution 
of major and minor mergers is almost identical. 

This will be true, it turns out, in any model where the 
amount of bulge formed in a given merger scales roughly in linear fashion with 
the mass ratio. As such, even highly simplified models which ignore the role 
of gas fraction and orbital parameters, and/or only violently relax the primary 
in major mergers (but do destroy the secondary in minor mergers), 
will still obtain the same qualitative features in $f_{\rm bulge}(>\mu_{\rm gal})$; 
see e.g.\ \citet{khochfar:size.evolution.model}.

\subsection{Combinations of the Above: Typical ``Scatter''}
\label{sec:robustness:combinations}

We have considered various permutations of the above 
models, amounting to $\sim700$ total models; our 
conclusions are robust to these combinations. The interquartile range 
between this sampling of models lies within the observationally allowed 
range in terms of the merger rate, and yields very little scatter in 
$f_{\rm bulge}(>\mu_{\rm gal})$. 

To the extent that comparison of these models can be considered ``scatter'' 
or reflective of uncertainties in the theoretical predictions, the corresponding 
typical ``uncertainties'' are as follows: 
around $\sim L_{\ast}$ and at slightly higher masses, uncertainties are 
small -- a factor of $\sim1.5$ in merger rates (at $z<2$; uncertainties 
grow at higher redshifts as in Figure~\ref{fig:rate.demo}) and 
smaller in $f_{\rm bulge}(>\mu_{\rm gal})$, 
with $f_{\rm bulge}(\mu_{\rm gal}>1/3)\sim60-80\%$. At factors 
less than a few higher and lower masses, 
these uncertainties 
increase to a factor $\sim2$ in merger rate, and factor $\sim1.5$ in the importance of 
major versus minor mergers. At much lower masses ($\sim10^{9}\,\msun$), uncertainties 
in both grow rapidly -- here, the halo occupation statistics are not strongly constrained. 
Moreover, $\mu_{\rm gal}$ is the 
{\em baryonic} (not just stellar) mass ratio, and systems at these masses 
are increasingly gas-dominated, so uncertainties in $M_{\rm gas}(M_{\ast})$ 
can strongly affect our predictions.

\breaker
\section{Discussion}
\label{sec:discuss}

\subsection{Approach}
\label{sec:discuss:approach}

We have used an extensive set of models to examine galaxy-galaxy mergers 
and to identify robust predictions for the relative importance of mergers 
of different mass ratios for bulge formation. 
Although halo-halo merger rates have been relatively well-understood, 
mapping halo-halo mergers to galaxy-galaxy mergers is not trivial. 
There can be significantly more or fewer major or 
minor galaxy-galaxy mergers, relative to halo-halo mergers; likewise, 
bulge growth can be dominated by preferentially more major or minor 
mergers than the growth of the host halo. 

However, there is hope. Numerical simulations are converging
in predicting how the efficiency of bulge formation 
scales with merger mass ratio (and what the ``appropriate'' mass ratio 
to use in these calculations should be), giving a 
straightforward set of predictions for how much bulge 
should be formed in a given galaxy-galaxy encounter. 
To lowest order, the amount of bulge formed scales linearly in the 
merger mass ratio, close to the maximal efficiency 
possible for minor mergers \citep{hopkins:disk.survival}. 

Meanwhile, observations are converging on relatively tight 
constraints on halo occupation models: namely, the stellar and gas 
mass of the average galaxy hosted by a halo/subhalo of a given 
mass. The correlation between galaxy stellar mass and halo 
mass is monotonic and, to lowest order, amounts to a simple 
matched rank-ordering of the two, with small 
scatter \citep[e.g.][]{conroy:monotonic.hod}.

\subsection{Conclusions}
\label{sec:discuss:conclusions}

This convergence makes the time ripe to examine the consequences of 
galaxy-galaxy mergers on bulge formation. To good approximation, 
the salient features of the merger rate distribution can be captured by 
convolving the theoretically determined halo-halo merger 
rate with the empirically determined halo occupation statistics. 
Given this simple, well-constrained 
approach, there are 
some robust predictions that are insensitive to most if not all model details: 
\\

{\bf (1)} {\em Major-merger ($\mu_{\rm gal}>1/3$) remnants 
dominate the integrated 
mass density of merger/interaction-induced bulges at all redshifts} 
(Figures~\ref{fig:fmu}-\ref{fig:fmu.vs.BT.mbin}).
Minor mergers ($1/10<\mu_{\rm gal}<1/3$) do contribute a significant, albeit 
not dominant, fraction ($\sim 30\%$) to 
the assembly of the total mass density. 
More minor mergers $\mu_{\rm gal}<1/10$ are not 
important (contributing $<5-10\%$). 
\\

{\bf (2)} This statement is significantly mass-dependent 
(Figures~\ref{fig:fmu.int.vs.inst}-\ref{fig:fmu.vs.m}). Although the 
relative major/minor contribution to halos is nearly mass-independent, the 
mass-dependent HOD shape leads to a galaxy mass dependence: 
major mergers strongly dominate bulge production around $\sim L_{\ast}$ (where 
most of the bulge and stellar mass density of the Universe lies). 
At masses $\ll L_{\ast}$, merger rates at all mass ratios are suppressed, 
and minor mergers are relatively more important 
(since $M_{\rm gal}\propto M_{\rm halo}^{2}$ at these masses, approximately, 
a 1:3 halo-halo merger becomes a 1:9 galaxy-galaxy merger; so all mergers 
are shifted to lower mass ratio, suppressing the number at some fixed $\mu_{\rm gal}$ 
and relatively suppressing major mergers). 
At higher masses $\gtrsim L_{\ast}$, merger rates are higher, and 
major mergers are relatively more important (here $M_{\rm gal}\propto M_{\rm halo}^{1/2}$, 
so a 1:9 halo-halo merger becomes a 1:3 galaxy-galaxy merger; making 
mergers at all significant mass ratios more abundant and relatively increasing the 
importance of major mergers). This is true until $\gtrsim\,$a few $L_{\ast}$, where 
minor mergers again become relatively more important owing to the rapid 
dropoff in the number of ``major'' companions (equivalently, since most of the 
galaxy mass is concentrated near $\sim L_{\ast}$, most of the incoming mass 
density is weighted near this region as well, which is a major merger 
at $\lesssim$ a few $L_{\ast}$ and minor merger above). These trends are 
quite general, and the relative ``increased weight'' of major mergers will 
occur wherever the ``mass-to-light ratio'' or formation 
efficiency $M_{\rm halo}/M_{\rm gal}$ has a minimum ($\sim L_{\ast}$). 
\\

{\bf (3)} For massive galaxies, there is a difference between the mass 
ratios important for bulge {\em formation} 
(the mergers which initially converted some disk mass into bulge mass) 
and those important 
for bulge {\em assembly} (the mergers that brought together the present-day 
bulge from any combination of pre-existing bulges and/or disks). 
At low masses, the two are equivalent (they are only different where 
``dry mergers'' are significant). At high masses, 
the description above applies to assembly. 
Most mergers onto $\gg L_{\ast}$ systems are of already 
bulge-dominated galaxies (i.e.\ dry), systems which first turned their 
disk mass into bulge (``formed'' the bulges) 
at progenitor masses near $\sim L_{\ast}$, 
where major mergers are most efficient. As a consequence, 
most bulge mass at all $>L_{\ast}$ masses is {\em formed} 
in major mergers (albeit again with 
non-negligible contributions from minor mergers); 
however, bulges are assembled in increasingly 
minor (dry) mergers at larger masses. 
\\

{\bf (4)} The relative importance of major and minor mergers is 
also significantly morphology-dependent 
(Figures~\ref{fig:fmu.vs.BT.mbin}-\ref{fig:mumean.vs.BT}). Bulge-dominated 
(E/S0 or $B/T\gtrsim0.4$) 
galaxies are preferentially formed in major mergers; 
later-type (Sb/c/d or $B/T\lesssim0.2$) galaxies 
are preferentially formed in minor mergers. 
Despite the fact that simulations show that e.g.\ ten 1:10 mergers 
can yield just as much bulge mass as one 1:1 merger, 
cosmological models show that they are {\em not} ten times 
more common. Moreover, this many minor mergers 
would have to happen in a time much less than a Hubble time in 
order to successfully 
build a bulge-dominated galaxy, and this scenario is very unlikely 
(even at high redshift; minor merger rates may increase, but so 
do major merger rates). 
However, since just one or two 1:10 
mergers are sufficient to account for a $B/T<0.2$ bulge, 
this is a common formation channel for small bulges, in particular more common 
than a major merger at high redshifts that destroys the entire disk followed 
by a factor of $\sim10$ subsequent disk re-growth (even if this occurred, it 
would take $\sim$ a Hubble time, in which time a 1:10 merger would be 
very likely, and that merger would then dominate the final bulge mass). 
To lowest order, bulges of systems of bulge-to-total ratio $B/T$ 
are characteristically formed in mergers of mass ratio $\mu_{\rm gal} \sim B/T$ 
(Equation~\ref{eqn:mugal.vs.bt.mass}).  
\\

{\bf (5)} Gas-richness, with high gas fractions $f_{\rm gas}\gtrsim 0.5$, 
can dramatically suppress the {\em global} efficiency of bulge formation 
(from mergers at all mass ratios), and 
the important implications 
of this for establishing the morphology-mass relation and allowing 
for a significant population of low $B/T$ systems is discussed 
in \citet{hopkins:disk.survival.cosmo} and \citet{stewart:disk.survival.vs.mergerrates}. 
However, it does not affect the merger rate, and because the effects are 
not mass ratio-dependent, it does not significantly affect 
the relative importance of major/minor mergers. 
Because low-mass galaxies are typically more gas-rich, they 
require somewhat more violent merger histories to reach the 
same $B/T$ as a comparable high-mass galaxy 
(Figures~\ref{fig:fmu.vs.BT.mbin}-\ref{fig:mumean.vs.BT}). 
To lowest order the gas fractions of 
progenitor galaxies that contribute to the observed bulge population, 
and the fraction of bulge mass formed dissipationally (by gas losing 
angular momentum in mergers and forming stars in concentrated 
nuclear starbursts) simply reflect the cosmological average gas fractions of 
progenitor disks corresponding to the same stellar mass and assembly  
times (Figures~\ref{fig:fmu.fgas}-\ref{fig:fdiss}). 
We have included the effects of gas on merger dynamics because it is 
known to be very important; however, given the above, 
our key prediction in this paper specifically would be not be 
dramatically changed if we ignored these effects throughout. 
\\

{\bf (6)} The predicted major merger rate (mergers per galaxy per 
Gyr) agrees well with observed merger fractions from $z\sim0-2$ 
(Figures~\ref{fig:rate.demo}-\ref{fig:mgr.rate.vs.obs})
when one accounts for the observable merger timescale determined 
by applying the {\em same} observational methods directly to 
high-resolution galaxy-galaxy merger simulations 
\citep[see e.g.][]{lotz:merger.selection}. 
The corresponding rate is 
$\sim0.5$ major galaxy-galaxy mergers per central galaxy per 
unit redshift (in these units, nearly redshift-independent), 
around $\sim L_{\ast}$, and is mass-dependent as per conclusion {\bf (2)}: 
half to two-thirds of the $\sim L_{\ast}$ population has had a major merger 
since $z\sim2$, but the fraction is a factor $\sim 3-5$ lower at an order-of-magnitude 
lower stellar mass, and becomes one (with many galaxies 
having a couple such mergers) at a factor of a few higher stellar mass
(Figures~\ref{fig:nmu} \&\ \ref{fig:mgr.rate.fits}). 
The merger rate as a function of 
galaxy-galaxy baryonic mass ratio $\mu_{\rm gal}$, redshift $z$, and 
primary stellar mass $M_{\ast}$ can be reasonably well fit by the simple functions 
in Equations~\ref{eqn:mgrrate.form}-\ref{eqn:mgrrate.minor.beta}. 
\\

{\bf (7)} Integrating over all 
mergers, the predicted 
merger rates yield good agreement with the growth of the 
mass density in bulge-dominated 
galaxies, from redshifts $z=0-1.5$ and (to the extent that 
color and morphology are correlated) the passive/red sequence 
population from redshifts $z=0-4$ 
(Figure~\ref{fig:rhobul.evol}). 
The typical uncertainties in both theory and observations are at 
the factor $\sim2$ level; this is an interesting range discussed below. 
\\

\subsection{Robustness}
\label{sec:discuss:robust}

We have examined how these conclusions depend on a variety of 
choices, including the empirical HOD constraints, the global cosmology, 
halo merger rates, substructure tracking, and merger timescales, 
and find that they are robust (\S~\ref{sec:robustness}; 
Figures~\ref{fig:mgr.rate.vs.model.1}-\ref{fig:mgr.rate.vs.model.2}). 

Varying the halo occupation model within the range 
allowed by observations including weak lensing, clustering, group 
dynamics, and abundance matching methods all yield similar 
conclusions. A very different halo occupation model, for example 
simply assuming $M_{\rm gal}\propto M_{\rm halo}$, would yield 
very different conclusions, but observational constraints are sufficiently 
tight that within the range allowed, resulting variations are small. 
Varying the cosmological parameters primarily affects the absolute abundance 
of halos of a given mass, not e.g.\ the shape of the merger rate function, 
and since the observed galaxy mass function is fixed, this difference is simply 
folded into the halo occupation model and does not change our conclusions. 

Halo-halo merger rates, likewise, are sufficiently converged between different 
simulations such that they yield no large differences. 
However, a halo-halo merger is not a galaxy-galaxy merger. 
Typically, one attempts to better approximate the latter by adopting either 
some merger timescale, representing a delay corresponding to subhalo 
orbital decay before the galaxy-galaxy merger, or by following 
subhalos directly in high-resolution cosmological simulations. One can 
also use galaxy-galaxy mergers identified in cosmological 
hydrodynamic simulations, after re-normalizing their masses to agree 
with empirical constraints. Considering variations in each of these choices, 
we find that they have little effect on the {\em shape} of the merger 
rate function, hence little effect on the relative importance of 
major/minor mergers. Further, some apparent differences in the resulting 
merger rate owe purely to definitions, and are implicitly normalized out 
in the HOD. Nevertheless, these different approaches do yield  
important systematic differences in the {\em absolute} merger rate, 
at the factor $\sim2$ level. 

Independent models adopting the halo occupation methodology described 
here also obtain results in good agreement \citep[see e.g.][]{zheng:hod.evolution,
brown:hod.evol,perezgonzalez:hod.ell.evol,stewart:merger.rates}. 
However, models that attempt to predict galaxy formation and 
merger rates in an a priori manner, such as e.g.\ cosmological 
hydrodynamic simulations and semi-analytic models, have reached 
various mixed conclusions -- some in agreement with those here, 
some not, with significantly larger variation in the predicted 
galaxy-galaxy merger rates than the factor $\sim2$ above 
\citep[compare e.g.][]{weinzirl:b.t.dist,parry:sam.merger.vs.morph,
maller:sph.merger.rates,naab:etg.formation,governato:disk.formation,
guo:gal.growth.channels,somerville:new.sam}. 
The origins and implications of these differences is examined in detail in a 
companion paper \citep{hopkins:merger.rates.methods}. 

However, the 
important point for our conclusions is that these methods are fundamentally 
different; they are not strictly tied to observed halo occupation 
constraints as are the models here. As such, they 
can yield very different predictions. For example, it is well-known 
that cosmological simulations without feedback yield efficient star formation 
at all masses, such that the predicted halo occupation 
has a form more like $M_{\rm gal}\propto M_{\rm halo}$ , and so galaxy-galaxy 
mergers will, in such a model (without re-normalizing masses) 
trivially reflect halo-halo mergers. Some semi-analytic models, meanwhile, 
have well-known discrepancies between predicted and observed 
populations of satellite galaxies, which propagate to the predicted 
merger rates. It is increasingly clear that these semi-analytic models 
have considerable difficulty reproducing the observations of the 
merger history \citep[generally with the sense that the semi-analytic 
merger rates/fractions are lower than those observed; see e.g.][]{jogee:merger.density.08.conf.proc,
bertone:merger.rate.vs.stellar.wind.model,lopezsanjuan:merger.fraction.to.z1}. 
In \citet{hopkins:merger.rates.methods}, we show that 
this indeed primarily owes to well-known mis-matches between the 
predicted satellite galaxy properties and galaxy mass functions in these 
models, and those observed. As such, the semi-empirical approach 
can perform much better in explaining the observations (of course, 
the model here cannot predict satellite properties as can a semi-analytic 
model, but that is not its purpose). By adopting the halo occupation 
constraints directly from observations, the semi-empirical model 
simply bypasses a major, well-known theoretical uncertainty 
in attempts to predict merger rates directly from semi-analytic 
models or cosmological simulations. This does not mean the 
answer is ``built in'' implicitly somehow in our models here -- what it does mean, 
however, is that the apparent discrepancy between observations and 
other predictions of the merger rate owes not to some 
fundamental problem of $\Lambda$CDM, but rather to well-known 
difficulties in properly modeling the accretion and star-formation histories 
of galaxies. 

Carefully accounting for these distinctions, the 
different results in various models can be understood. 
And in fact, 
despite differences in some quantitative predictions, 
many of the qualitative conclusions are the 
same; \citet{parry:sam.merger.vs.morph} and \citet{weinzirl:b.t.dist} 
demonstrate that 
different SAMs reach similar conclusions regarding how merger 
rates and the relative importance of major versus minor mergers 
scale as a function of e.g.\ galaxy stellar mass and redshift.

\subsection{Outlook and Future Work}
\label{sec:discuss:outlook}

Convergence in predicted merger rates among different theoretical approaches, 
at the factor $\sim2$ level or better, is a remarkable achievement. Unfortunately, 
obtaining greater convergence in theoretical predictions will be difficult. 
Applying constraints from empirical halo occupation approaches to e.g.\ cosmological 
simulations and semi-analytic models is important. Tighter observational 
constraints on the halo occupation distribution, in particular at low masses and at 
high redshifts, will allow semi-empirical models such as those in this paper to 
greatly extend the dynamic range of predictions (as well as putting 
strong constraints on a priori models for galaxy formation at these masses and 
redshifts). 

However, we have shown that these differences only account for 
a fraction of the scatter in theoretical predictions -- subtle details of how e.g.\ halos 
are defined and followed become important at this level. Moreover 
resolution limits and the absence of baryons in simulations \citep[which does, at the 
level of uncertainty here, have potentially important effects on the longevity and 
merger timescales of subhalos; see e.g.][]{weinberg:baryons.and.substructure} 
limit all theoretical models. Ideally, high-resolution cosmological hydrodynamic 
simulations could form the basis for halo occupation models: avoiding ambiguity in 
identifying a galaxy-galaxy versus halo-halo merger by simply tracking 
the galaxies (even if their absolute masses are incorrect, and they need to be 
``repopulated'' in post-processing). Although some steps have been made in 
this direction, it remains prohibitively expensive to simulate large volumes 
at the desired high resolution with gas physics. 

It is also unclear whether a merger rate alone is meaningful at an accuracy 
much better than a factor $\sim2$. At this level, the question of e.g.\ the 
``proper'' mass ratio becomes important \citep[see e.g.][]{stewart:massratio.defn.conf.proc}. 
What matters, in detail, for galaxy 
dynamics and the effect of a given merger is a combination of 
several quantities in 
the merger ``mass ratio'' -- including stars, gas, and the tightly bound portion of the 
halo that has been robust to stripping; as such, the halo structure and history, 
as well as effects such as adiabatic contraction, become important. Moreover, 
at this level, the orbital parameters, galaxy gas fractions, and progenitor structure 
(relative bulge-to-disk ratios and disk scale lengths) become 
non-trivial corrections to the estimate of the effects ``per merger.'' 
Without models for all of these details, a merger rate constrained to 
arbitrarily high accuracy does not necessarily translate to a bulge formation 
model with accuracy better than a similar factor $\sim2$. 

In the meantime, however, 
there is considerable room for improvement in the 
comparison of model predictions and observations of the merger rate. 
The formal statistical 
errors in observed merger and close pair fractions are rapidly decreasing; 
even including cosmic variance, such observations at $z\sim0-1.2$ 
are converging to better than a factor of $\sim2$. However, 
as shown in Figure~\ref{fig:mgr.rate.vs.z}, simply putting all such observations 
on equal footing yields order-of-magnitude scatter; similar uncertainties 
plague the conversion of these quantities to merger rates (and it is further unclear 
what the sensitivity is to different mass ratio mergers). 
This is an area where considerable improvements can and should be made: 
most of the differences in observational estimates are attributable to 
different methodologies, observational depth, and selection effects. 
The conversion of some specific pair or morphologically identified 
sample to a merger rate should be calibrated to  
suites of high-resolution $N$-body simulations, {\em specifically 
with mock observations matched to the exact selection and methodology adopted}. 

Moreover, the merger rate is predicted to be a non-trivial function of 
galaxy mass: many samples identifying merger fractions 
have ambiguous luminosity selection; what is ultimately necessary are 
samples with well-defined stellar or baryonic mass selection. 
At the level of present data quality and theoretical 
convergence, order-of-magnitude estimates of merger lifetimes 
and lack of such calibration represent the dominant uncertainty in comparisons. 

Improvements are being made in this area: \citet{lotz:merger.selection} 
have calibrated the merger timescale for major pair samples of different separations 
and certain specific automated morphological selection criteria to mock observations 
of high-resolution hydrodynamic merger simulations. \citet{conselice:mgr.pairs.updated} 
adopt these and similar detailed calibrations to attempt to address consistency between 
merger populations identified with different methodologies. 
\citet{jogee:merger.density.08.conf.proc,jogee:merger.density.08} attempt to 
calibrate their morphological selection criteria as a function of merger 
mass ratio. Various works have attempted to quantify merger rates as a function of 
stellar mass, rather than in a pure magnitude-limited sample 
\citep[see e.g.][]{bell:merger.fraction,
conselice:highz.mgr.rate.vs.mass,bundy:merger.fraction.new,
lopezsanjuan:mgr.rate.pairs,darg:galzoo.merger.properties}. 

Together, these approaches will allow rigorous comparison of predicted and observed 
galaxy-galaxy merger rates as a function of galaxy stellar mass, redshift, 
and (ideally) mass ratio. Obviously, extension of observational constraints in 
any of this parameter space represents a valuable constraint on the models here. 
Using the calibrations above, we attempt such a comparison specific 
to different observational methods (at least in terms of 
pair versus morphological fractions), and find good agreement between predicted 
and observed merger rates, and the integrated buildup of the bulge population. 
Considering the most well-constrained 
observations and well-calibrated conversions, we find agreement within a similar 
factor $\sim2$ as that characteristic of the theoretical uncertainties. 

Far from implying that the problem is ``solved,'' such a factor of $\sim2$ is 
of great interest. There is a large parameter space where predicted merger rates are 
consistent with observed merger/pair fractions as a function of mass and redshift 
and can be tuned to precisely account for the entire bulge mass budget of the 
Universe. However, allowing for the factor $\sim2$ uncertainty in one direction 
would lead to ``too many'' mergers, implying 
that mergers must be less efficient than cosmologically predicted: this might 
mean that real gas fractions are in fact higher than what we have modeled here, 
or that tidal destruction of satellites is efficient, even in the major merger regime, or 
that there is some problem in our understanding of halo occupation statistics or 
cosmological dark matter merger rates. 

On the other hand, allowing for the same factor of $\sim2$ variation in the opposite 
sense would imply that $\sim$half the bulge mass density of the Universe could not 
be attributed to mergers as we understand them. This means that, within the present 
uncertainties, secular processes such as bar or disk instabilities 
might account for up to half of the bulge mass of the Universe. Since the uncertainties 
grow at low mass, the fraction could be even higher at lower masses. 

Independent observational tests can put complementary constraints on these 
possibilities. It must be emphasized, for example, that essentially all numerical studies 
of spheroid kinematics find that {\em only} mergers 
can reproduce the observed kinematic properties of observed elliptical 
galaxies and ``classical'' bulges \citep{hernquist.89,
hernquist:bulgeless.mergers,hernquist:bulge.mergers,
barnes:disk.halo.mergers,barnes:disk.disk.mergers,
schweizer92,bournaud:minor.mergers,
naab:gas,naab:profiles,cox:kinematics,
jesseit:kinematics}. These are, in general, the bulges whose formation history 
we predict here. 
Disk instabilities and
secular evolution (e.g.\ bar instabilities, harassment, and other 
isolated modes) can indeed produce bulges, but these are  
``pseudobulges'' \citep{pfenniger:bar.dynamics,
combes:pseudobulges,raha:bar.instabilities,
kuijken:pseudobulges.obs,oniell:bar.obs,athanassoula:peanuts}, 
with clearly distinct shapes, kinematics, structural properties, and colors 
from classical bulges 
\citep[for a review, see][]{kormendy.kennicutt:pseudobulge.review}. 

Observations at present indicate that 
pseudobulges constitute only a small fraction of the total mass density 
in spheroids \citep[$\lesssim10\%$; see][]{allen:bulge-disk,ball:bivariate.lfs,
driver:bulge.mfs}; they do, however, become a large fraction of the bulge 
population in small bulges in late-type hosts 
\citep[e.g.\ Sb/c, corresponding to typical $M_{\rm gal}\lesssim10^{10}\,\msun$; see][and 
references therein]{carollo98,kormendy.kennicutt:pseudobulge.review}. 
However, this is not to say that secular processes cannot, in principle, 
build some massive bulges \citep[see e.g.][]{debattista:pseudobulges.a,debattista:pseudobulges.b}. 
And it is not clear that mergers -- specifically minor mergers with 
mass ratios $\mu_{\rm gal}\lesssim1/10$ -- cannot build 
pseudobulges, depending on e.g.\ the structural properties of the secondary 
and orbital parameters of the merger \citep[see e.g.][]{gauthier:triggered.bar.from.substructure,
younger:minor.mergers,elichemoral:pseudo.from.minor}. 

Improvements in theoretical constraints (from high-resolution simulations) 
on how bulges with different structural 
properties are formed, combined with improved observational constraints on the 
distribution of these structural properties, can constrain the role of secular processes 
at better than a factor $\sim2$ level (at least at low redshifts) -- a level at which theoretical 
models cannot yet uniquely predict the importance of mergers. 
On the other hand, observational constraints on the mass budget in 
extended galaxy halos, intra-group and intra-cluster light 
can constrain satellite disruption \citep[see e.g.][]{lin:intracluster.light.measurements,
cypriano:icl.cluster.gal.stripping,brown:hod.evol,lagana:intracluster.light}, 
and observations of high redshift disk+bulge systems that may represent recent 
re-forming or relaxing merger remnants can constrain the efficiency of bulge 
formation in mergers \citep{hammer:obs.disks.w.mergers,zheng:morphological.sfr.evol,
trujillo:truncation.scale.evol,
flores:tf.evolution,puech:minor.merger.at.z06,
puech:highz.vsigma.disks,puech:tf.evol,
atkinson:scatter.in.tf.owes.to.increased.mergers}. 
Together, these improvements in observational constraints and theoretical 
models have the potential to enable precision tests of models for bulge formation 
in mergers, and allow a robust determination of the relative roles of 
secular processes, minor mergers, and major mergers in galaxy formation, 
as a function of cosmic time and galaxy properties. 

In order to facilitate comparison with future observations, we have provided 
fitting functions to both the predicted merger rates as a function of galaxy 
mass and mass ratio, and to the relative contributions of these mergers 
to bulge formation. However, for various applications, additional information 
is desired. We therefore make public a simple ``merger rate calculator'' 
code\footnote{\mergercalcurl} which can be used to obtain the predicted 
merger rates from the models as a function of e.g.\ galaxy mass, 
mass ratio, redshift, and galaxy gas fractions. 
The script can be used to determine merger rates as a function of 
halo, stellar, or total galaxy baryonic masses, 
and can be used to restrict to e.g.\ gas-rich (``wet'') or gas-poor 
(``dry'') mergers. It also allows for different choices with respect to e.g.\ the 
stellar mass functions used to normalize the HOD and 
$M_{\rm gal}(M_{\rm halo})$ distribution used in the models here, 
and different cosmological parameters. As desired, it can output 
the merger rate per galaxy, the volumetric total merger 
rate (mergers per unit volume per unit time), or merger fractions 
with the appropriate observable timescales used here as 
calibrated for pair or morphologically-selected samples. 
We note that, in the interest of running time and memory use, the 
script uses some of fitting functions and 
approximations to the full models discussed here -- however, 
we have tested extensively that the approximations and fitting functions used 
yield much smaller differences in the ultimate merger rates 
than the inherent uncertainties discussed here.

\acknowledgments We thank 
Andrew Benson, Owen Parry, Simon White, 
Volker Springel, Gabriella de Lucia, 
and Carlos Frenk for helpful discussions. 
Support for PFH was provided by the Miller Institute for Basic Research 
in Science, University of California Berkeley.
JDY acknowledges support from NASA through Hubble Fellowship grant HST-HF-01243.01
awarded by the Space Telescope Science Institute, which is operated
by the Association of Universities for Research in Astronomy, Inc.,
for NASA, under contract NAS 5-26555.

\bibliography{/Users/phopkins/Documents/lars_galaxies/papers/ms}

\begin{thebibliography}{290}
\expandafter\ifx\csname natexlab\endcsname\relax\def\natexlab#1{#1}\fi

\bibitem[{{Abraham} {et~al.}(2007)}]{abraham:red.mass.density}
{Abraham}, R.~G., {et~al.} 2007, \apj, 669, 184

\bibitem[{{Alexander} {et~al.}(2005){Alexander}, {Smail}, {Bauer}, {Chapman},
  {Blain}, {Brandt}, \& {Ivison}}]{alexander:bh.growth}
{Alexander}, D.~M., {Smail}, I., {Bauer}, F.~E., {Chapman}, S.~C., {Blain},
  A.~W., {Brandt}, W.~N., \& {Ivison}, R.~J. 2005, \nat, 434, 738

\bibitem[{{Allen} {et~al.}(2006){Allen}, {Driver}, {Graham}, {Cameron},
  {Liske}, \& {de Propris}}]{allen:bulge-disk}
{Allen}, P.~D., {Driver}, S.~P., {Graham}, A.~W., {Cameron}, E., {Liske}, J.,
  \& {de Propris}, R. 2006, \mnras, 371, 2

\bibitem[{{Aller} \& {Richstone}(2007)}]{aller:mbh.esph}
{Aller}, M.~C., \& {Richstone}, D.~O. 2007, \apj, 665, 120

\bibitem[{{Athanassoula}(2005)}]{athanassoula:peanuts}
{Athanassoula}, E. 2005, \mnras, 358, 1477

\bibitem[{{Atkinson} {et~al.}(2007){Atkinson}, {Conselice}, \&
  {Fox}}]{atkinson:scatter.in.tf.owes.to.increased.mergers}
{Atkinson}, N., {Conselice}, C.~J., \& {Fox}, N. 2007, \mnras, in press,
  arXiv:0712.1316 [astro-ph]

\bibitem[{{Avila-Reese} {et~al.}(2008){Avila-Reese}, {Zavala}, {Firmani}, \&
  {Hern{\'a}ndez-Toledo}}]{avilareese:baryonic.tf}
{Avila-Reese}, V., {Zavala}, J., {Firmani}, C., \& {Hern{\'a}ndez-Toledo},
  H.~M. 2008, \aj, 136, 1340

\bibitem[{{Ball} {et~al.}(2006){Ball}, {Loveday}, {Brunner}, {Baldry}, \&
  {Brinkmann}}]{ball:bivariate.lfs}
{Ball}, N.~M., {Loveday}, J., {Brunner}, R.~J., {Baldry}, I.~K., \&
  {Brinkmann}, J. 2006, \mnras, 373, 845

\bibitem[{{Barnes}(1988)}]{barnes:disk.halo.mergers}
{Barnes}, J.~E. 1988, \apj, 331, 699

\bibitem[{{Barnes}(1992)}]{barnes:disk.disk.mergers}
---. 1992, \apj, 393, 484

\bibitem[{{Barnes} \& {Hernquist}(1992)}]{barneshernquist92}
{Barnes}, J.~E., \& {Hernquist}, L. 1992, \araa, 30, 705

\bibitem[{{Barnes} \& {Hernquist}(1996)}]{barneshernquist96}
---. 1996, \apj, 471, 115

\bibitem[{{Barnes} \& {Hernquist}(1991)}]{barnes.hernquist.91}
{Barnes}, J.~E., \& {Hernquist}, L.~E. 1991, \apjl, 370, L65

\bibitem[{{Barton} {et~al.}(2007){Barton}, {Arnold}, {Zentner}, {Bullock}, \&
  {Wechsler}}]{barton:triggered.sf}
{Barton}, E.~J., {Arnold}, J.~A., {Zentner}, A.~R., {Bullock}, J.~S., \&
  {Wechsler}, R.~H. 2007, \apj, 671, 1538

\bibitem[{{Behroozi} {et~al.}(2010){Behroozi}, {Conroy}, \&
  {Wechsler}}]{behroozi:mgal.mhalo.uncertainties}
{Behroozi}, P.~S., {Conroy}, C., \& {Wechsler}, R.~H. 2010, \apj, in press,
  arXiv:1001.0015

\bibitem[{{Bell} \& {de Jong}(2000)}]{belldejong:disk.sfh}
{Bell}, E.~F., \& {de Jong}, R.~S. 2000, \mnras, 312, 497

\bibitem[{{Bell} \& {de Jong}(2001)}]{belldejong:tf}
---. 2001, \apj, 550, 212

\bibitem[{{Bell} {et~al.}(2003){Bell}, {McIntosh}, {Katz}, \&
  {Weinberg}}]{bell:mfs}
{Bell}, E.~F., {McIntosh}, D.~H., {Katz}, N., \& {Weinberg}, M.~D. 2003, \apjs,
  149, 289

\bibitem[{{Bell} {et~al.}(2006{\natexlab{a}}){Bell}, {Phleps}, {Somerville},
  {Wolf}, {Borch}, \& {Meisenheimer}}]{bell:merger.fraction}
{Bell}, E.~F., {Phleps}, S., {Somerville}, R.~S., {Wolf}, C., {Borch}, A., \&
  {Meisenheimer}, K. 2006{\natexlab{a}}, \apj, 652, 270

\bibitem[{{Bell} {et~al.}(2005)}]{bell:morphology.vs.sfr}
{Bell}, E.~F., {et~al.} 2005, \apj, 625, 23

\bibitem[{{Bell} {et~al.}(2006{\natexlab{b}})}]{bell:dry.mergers}
---. 2006{\natexlab{b}}, \apj, 640, 241

\bibitem[{{Bennert} {et~al.}(2008){Bennert}, {Canalizo}, {Jungwiert},
  {Stockton}, {Schweizer}, {Peng}, \& {Lacy}}]{bennert:qso.hosts}
{Bennert}, N., {Canalizo}, G., {Jungwiert}, B., {Stockton}, A., {Schweizer},
  F., {Peng}, C.~Y., \& {Lacy}, M. 2008, \apj, 677, 846

\bibitem[{{Benson}(2005)}]{benson:cosmo.orbits}
{Benson}, A.~J. 2005, \mnras, 358, 551

\bibitem[{{Benson} {et~al.}(2004){Benson}, {Lacey}, {Frenk}, {Baugh}, \&
  {Cole}}]{benson:heating.model}
{Benson}, A.~J., {Lacey}, C.~G., {Frenk}, C.~S., {Baugh}, C.~M., \& {Cole}, S.
  2004, \mnras, 351, 1215

\bibitem[{{Bertone} \&
  {Conselice}(2009)}]{bertone:merger.rate.vs.stellar.wind.model}
{Bertone}, S., \& {Conselice}, C.~J. 2009, \mnras, 396, 2345

\bibitem[{{Binney} \& {Tremaine}(1987)}]{binneytremaine}
{Binney}, J., \& {Tremaine}, S. 1987, {Galactic dynamics} (Princeton, NJ,
  Princeton University Press, 1987)

\bibitem[{{Blain} {et~al.}(1999){Blain}, {Jameson}, {Smail}, {Longair},
  {Kneib}, \& {Ivison}}]{blain:ir.lf.synthesis.model}
{Blain}, A.~W., {Jameson}, A., {Smail}, I., {Longair}, M.~S., {Kneib}, J.-P.,
  \& {Ivison}, R.~J. 1999, \mnras, 309, 715

\bibitem[{{Bluck} {et~al.}(2009){Bluck}, {Conselice}, {Bouwens}, {Daddi},
  {Dickinson}, {Papovich}, \& {Yan}}]{bluck:highz.merger.fraction}
{Bluck}, A.~F.~L., {Conselice}, C.~J., {Bouwens}, R.~J., {Daddi}, E.,
  {Dickinson}, M., {Papovich}, C., \& {Yan}, H. 2009, \mnras, 394, L51

\bibitem[{{Borch} {et~al.}(2006)}]{borch:mfs}
{Borch}, A., {et~al.} 2006, \aap, 453, 869

\bibitem[{{Borriello} \& {Salucci}(2001)}]{borriello01}
{Borriello}, A., \& {Salucci}, P. 2001, \mnras, 323, 285

\bibitem[{{Bouch{\'e}} {et~al.}(2007)}]{bouche:z2.kennicutt}
{Bouch{\'e}}, N., {et~al.} 2007, \apj, 671, 303

\bibitem[{{Bournaud} {et~al.}(2005){Bournaud}, {Jog}, \&
  {Combes}}]{bournaud:minor.mergers}
{Bournaud}, F., {Jog}, C.~J., \& {Combes}, F. 2005, \aap, 437, 69

\bibitem[{{Bower} {et~al.}(2006){Bower}, {Benson}, {Malbon}, {Helly}, {Frenk},
  {Baugh}, {Cole}, \& {Lacey}}]{bower:sam}
{Bower}, R.~G., {Benson}, A.~J., {Malbon}, R., {Helly}, J.~C., {Frenk}, C.~S.,
  {Baugh}, C.~M., {Cole}, S., \& {Lacey}, C.~G. 2006, \mnras, 370, 645

\bibitem[{{Boylan-Kolchin} {et~al.}(2008){Boylan-Kolchin}, {Ma}, \&
  {Quataert}}]{boylankolchin:merger.time.calibration}
{Boylan-Kolchin}, M., {Ma}, C.-P., \& {Quataert}, E. 2008, \mnras, 383, 93

\bibitem[{{Bridge} {et~al.}(2010){Bridge}, {Carlberg}, \&
  {Sullivan}}]{bridge:merger.fraction.new}
{Bridge}, C.~R., {Carlberg}, R.~G., \& {Sullivan}, M. 2010, \apj, 709, 1067

\bibitem[{{Bridge} {et~al.}(2007)}]{bridge:merger.fractions}
{Bridge}, C.~R., {et~al.} 2007, \apj, 659, 931

\bibitem[{{Brough} {et~al.}(2006){Brough}, {Forbes}, {Kilborn}, \&
  {Couch}}]{brough:group.dynamics}
{Brough}, S., {Forbes}, D.~A., {Kilborn}, V.~A., \& {Couch}, W. 2006, \mnras,
  370, 1223

\bibitem[{{Brown} {et~al.}(2007){Brown}, {Dey}, {Jannuzi}, {Brand}, {Benson},
  {Brodwin}, {Croton}, \& {Eisenhardt}}]{brown:mf.evolution}
{Brown}, M.~J.~I., {Dey}, A., {Jannuzi}, B.~T., {Brand}, K., {Benson}, A.~J.,
  {Brodwin}, M., {Croton}, D.~J., \& {Eisenhardt}, P.~R. 2007, \apj, 654, 858

\bibitem[{{Brown} {et~al.}(2008){Brown}, {Zheng}, {White}, {Dey}, {Jannuzi},
  {Benson}, {Brand}, {Brodwin}, \& {Croton}}]{brown:hod.evol}
{Brown}, M.~J.~I., {et~al.} 2008, \apj, 682, 937

\bibitem[{{Bundy} {et~al.}(2005){Bundy}, {Ellis}, \& {Conselice}}]{bundy:mfs}
{Bundy}, K., {Ellis}, R.~S., \& {Conselice}, C.~J. 2005, \apj, 625, 621

\bibitem[{{Bundy} {et~al.}(2009){Bundy}, {Fukugita}, {Ellis}, {Targett},
  {Belli}, \& {Kodama}}]{bundy:merger.fraction.new}
{Bundy}, K., {Fukugita}, M., {Ellis}, R.~S., {Targett}, T.~A., {Belli}, S., \&
  {Kodama}, T. 2009, \apj, 697, 1369

\bibitem[{{Bundy} {et~al.}(2006)}]{bundy:mtrans}
{Bundy}, K., {et~al.} 2006, \apj, 651, 120

\bibitem[{{Burkert} {et~al.}(2008){Burkert}, {Naab}, {Johansson}, \&
  {Jesseit}}]{burkert:anisotropy}
{Burkert}, A., {Naab}, T., {Johansson}, P.~H., \& {Jesseit}, R. 2008, \apj,
  685, 897

\bibitem[{{Calura} {et~al.}(2008){Calura}, {Jimenez}, {Panter}, {Matteucci}, \&
  {Heavens}}]{calura:sdss.gas.fracs}
{Calura}, F., {Jimenez}, R., {Panter}, B., {Matteucci}, F., \& {Heavens}, A.~F.
  2008, \apj, 682, 252

\bibitem[{{Canalizo} \&
  {Stockton}(2001)}]{canalizostockton01:postsb.qso.mergers}
{Canalizo}, G., \& {Stockton}, A. 2001, \apj, 555, 719

\bibitem[{{Carollo} {et~al.}(1998){Carollo}, {Stiavelli}, \&
  {Mack}}]{carollo98}
{Carollo}, C.~M., {Stiavelli}, M., \& {Mack}, J. 1998, \aj, 116, 68

\bibitem[{{Cassata} {et~al.}(2005)}]{cassata:merger.fraction}
{Cassata}, P., {et~al.} 2005, \mnras, 357, 903

\bibitem[{{Chabrier}(2003)}]{chabrier:imf}
{Chabrier}, G. 2003, \pasp, 115, 763

\bibitem[{{Combes} {et~al.}(1990){Combes}, {Debbasch}, {Friedli}, \&
  {Pfenniger}}]{combes:pseudobulges}
{Combes}, F., {Debbasch}, F., {Friedli}, D., \& {Pfenniger}, D. 1990, \aap,
  233, 82

\bibitem[{{Conroy} \& {Wechsler}(2009)}]{conroy:hod.vs.z}
{Conroy}, C., \& {Wechsler}, R.~H. 2009, \apj, 696, 620

\bibitem[{{Conroy} {et~al.}(2006){Conroy}, {Wechsler}, \&
  {Kravtsov}}]{conroy:monotonic.hod}
{Conroy}, C., {Wechsler}, R.~H., \& {Kravtsov}, A.~V. 2006, \apj, 647, 201

\bibitem[{{Conroy} {et~al.}(2007)}]{conroy:mhalo-mgal.evol}
{Conroy}, C., {et~al.} 2007, \apj, 654, 153

\bibitem[{{Conselice}(2009)}]{conselice:merger.timescale.obs.est}
{Conselice}, C.~J. 2009, \mnras, in press [arXiv:0906.4704]

\bibitem[{{Conselice} {et~al.}(2003){Conselice}, {Bershady}, {Dickinson}, \&
  {Papovich}}]{conselice:merger.fraction}
{Conselice}, C.~J., {Bershady}, M.~A., {Dickinson}, M., \& {Papovich}, C. 2003,
  \aj, 126, 1183

\bibitem[{{Conselice} {et~al.}(2008){Conselice}, {Rajgor}, \&
  {Myers}}]{conselice:highz.mgr.rate.vs.mass}
{Conselice}, C.~J., {Rajgor}, S., \& {Myers}, R. 2008, \mnras, 386, 909

\bibitem[{{Conselice} {et~al.}(2009){Conselice}, {Yang}, \&
  {Bluck}}]{conselice:mgr.pairs.updated}
{Conselice}, C.~J., {Yang}, C., \& {Bluck}, A.~F.~L. 2009, \mnras, 394, 1956

\bibitem[{{Cooray}(2005)}]{cooray:highz}
{Cooray}, A. 2005, \mnras, 364, 303

\bibitem[{{Cooray}(2006)}]{cooray:hod.clf}
---. 2006, \mnras, 365, 842

\bibitem[{{Cox} {et~al.}(2006{\natexlab{a}}){Cox}, {Dutta}, {Di Matteo},
  {Hernquist}, {Hopkins}, {Robertson}, \& {Springel}}]{cox:kinematics}
{Cox}, T.~J., {Dutta}, S.~N., {Di Matteo}, T., {Hernquist}, L., {Hopkins},
  P.~F., {Robertson}, B., \& {Springel}, V. 2006{\natexlab{a}}, \apj, 650, 791

\bibitem[{{Cox} {et~al.}(2006{\natexlab{b}}){Cox}, {Jonsson}, {Primack}, \&
  {Somerville}}]{cox:feedback}
{Cox}, T.~J., {Jonsson}, P., {Primack}, J.~R., \& {Somerville}, R.~S.
  2006{\natexlab{b}}, \mnras, 373, 1013

\bibitem[{{Cox} {et~al.}(2008){Cox}, {Jonsson}, {Somerville}, {Primack}, \&
  {Dekel}}]{cox:massratio.starbursts}
{Cox}, T.~J., {Jonsson}, P., {Somerville}, R.~S., {Primack}, J.~R., \& {Dekel},
  A. 2008, \mnras, 384, 386

\bibitem[{{Cresci} {et~al.}(2009)}]{cresci:dynamics.highz.disks}
{Cresci}, G., {et~al.} 2009, \apj, 697, 115

\bibitem[{{Croton} {et~al.}(2006)}]{croton:sam}
{Croton}, D.~J., {et~al.} 2006, \mnras, 365, 11

\bibitem[{{Cypriano} {et~al.}(2006){Cypriano}, {Sodr{\'e}}, {Campusano},
  {Dale}, \& {Hardy}}]{cypriano:icl.cluster.gal.stripping}
{Cypriano}, E.~S., {Sodr{\'e}}, L.~J., {Campusano}, L.~E., {Dale}, D.~A., \&
  {Hardy}, E. 2006, \aj, 131, 2417

\bibitem[{{Daddi} {et~al.}(2005)}]{daddi05:drgs}
{Daddi}, E., {et~al.} 2005, \apj, 626, 680

\bibitem[{{Damen} {et~al.}(2009){Damen}, {Labb{\'e}}, {Franx}, {van Dokkum},
  {Taylor}, \& {Gawiser}}]{damen:ssfr.highz.massive.gals}
{Damen}, M., {Labb{\'e}}, I., {Franx}, M., {van Dokkum}, P.~G., {Taylor},
  E.~N., \& {Gawiser}, E.~J. 2009, \apj, 690, 937

\bibitem[{{Darg}
  {et~al.}(2009{\natexlab{a}})}]{darg:galzoo.merger.frac.by.morph}
{Darg}, D.~W., {et~al.} 2009{\natexlab{a}}, \mnras, in press, arXiv:0903.4937

\bibitem[{{Darg} {et~al.}(2009{\natexlab{b}})}]{darg:galzoo.merger.properties}
---. 2009{\natexlab{b}}, \mnras, in press, arXiv:0903.5057

\bibitem[{{Dasyra} {et~al.}(2006)}]{dasyra:mass.ratio.conditions}
{Dasyra}, K.~M., {et~al.} 2006, \apj, 638, 745

\bibitem[{{Dasyra} {et~al.}(2007)}]{dasyra:pg.qso.dynamics}
---. 2007, \apj, 657, 102

\bibitem[{{de Lucia} \& {Blaizot}(2007)}]{delucia:sam}
{de Lucia}, G., \& {Blaizot}, J. 2007, \mnras, 375, 2

\bibitem[{{De Lucia} {et~al.}(2004){De Lucia}, {Kauffmann}, {Springel},
  {White}, {Lanzoni}, {Stoehr}, {Tormen}, \& {Yoshida}}]{delucia:subhalos}
{De Lucia}, G., {Kauffmann}, G., {Springel}, V., {White}, S.~D.~M., {Lanzoni},
  B., {Stoehr}, F., {Tormen}, G., \& {Yoshida}, N. 2004, \mnras, 348, 333

\bibitem[{{De Propris} {et~al.}(2005){De Propris}, {Liske}, {Driver}, {Allen},
  \& {Cross}}]{depropris:merger.fraction}
{De Propris}, R., {Liske}, J., {Driver}, S.~P., {Allen}, P.~D., \& {Cross},
  N.~J.~G. 2005, \aj, 130, 1516

\bibitem[{{de Ravel} {et~al.}(2009){de Ravel}, {Le F{\`e}vre}, {Tresse},
  {Bottini}, {Garilli}, {Le Brun}, {Maccagni}, {Scaramella}, {Scodeggio},
  {Vettolani}, {Zanichelli}, {Adami}, {Arnouts}, {Bardelli}, {Bolzonella},
  {Cappi}, {Charlot}, {Ciliegi}, {Contini}, {Foucaud}, {Franzetti},
  {Gavignaud}, {Guzzo}, {Ilbert}, {Iovino}, {Lamareille}, {McCracken},
  {Marano}, {Marinoni}, {Mazure}, {Meneux}, {Merighi}, {Paltani}, {Pell{\`o}},
  {Pollo}, {Pozzetti}, {Radovich}, {Vergani}, {Zamorani}, {Zucca}, {Bondi},
  {Bongiorno}, {Brinchmann}, {Cucciati}, {de La Torre}, {Gregorini}, {Memeo},
  {Perez-Montero}, {Mellier}, {Merluzzi}, \&
  {Temporin}}]{deravel:merger.fraction.to.z1}
{de Ravel}, L., {et~al.} 2009, \aap, 498, 379

\bibitem[{{Debattista} {et~al.}(2004){Debattista}, {Carollo}, {Mayer}, \&
  {Moore}}]{debattista:pseudobulges.a}
{Debattista}, V.~P., {Carollo}, C.~M., {Mayer}, L., \& {Moore}, B. 2004, \apjl,
  604, L93

\bibitem[{{Debattista} {et~al.}(2006){Debattista}, {Mayer}, {Carollo}, {Moore},
  {Wadsley}, \& {Quinn}}]{debattista:pseudobulges.b}
{Debattista}, V.~P., {Mayer}, L., {Carollo}, C.~M., {Moore}, B., {Wadsley}, J.,
  \& {Quinn}, T. 2006, \apj, 645, 209

\bibitem[{{di Matteo} {et~al.}(2007){di Matteo}, {Combes}, {Melchior}, \&
  {Semelin}}]{dimatteo:merger.induced.sb.sims}
{di Matteo}, P., {Combes}, F., {Melchior}, A.-L., \& {Semelin}, B. 2007, \aap,
  468, 61

\bibitem[{{Di Matteo} {et~al.}(2005){Di Matteo}, {Springel}, \&
  {Hernquist}}]{dimatteo:msigma}
{Di Matteo}, T., {Springel}, V., \& {Hernquist}, L. 2005, \nat, 433, 604

\bibitem[{{Doyon} {et~al.}(1994){Doyon}, {Wells}, {Wright}, {Joseph}, {Nadeau},
  \& {James}}]{Doyon94}
{Doyon}, R., {Wells}, M., {Wright}, G.~S., {Joseph}, R.~D., {Nadeau}, D., \&
  {James}, P.~A. 1994, \apjl, 437, L23

\bibitem[{{Driver} {et~al.}(2007){Driver}, {Allen}, {Liske}, \&
  {Graham}}]{driver:bulge.mfs}
{Driver}, S.~P., {Allen}, P.~D., {Liske}, J., \& {Graham}, A.~W. 2007, \apjl,
  657, L85

\bibitem[{{Eke} {et~al.}(2004)}]{eke:groups}
{Eke}, V.~R., {et~al.} 2004, \mnras, 355, 769

\bibitem[{{Elahi} {et~al.}(2008){Elahi}, {Thacker}, {Widrow}, \&
  {Scannapieco}}]{elahi:substructure.mf.vs.powerspectrum.ns}
{Elahi}, P.~J., {Thacker}, R.~J., {Widrow}, L.~M., \& {Scannapieco}, E. 2008,
  \mnras, in press, arXiv:0811.0206 [astro-ph]

\bibitem[{{Eliche-Moral} {et~al.}(2008){Eliche-Moral},
  {Gonz{\'a}lez-Garc{\'{\i}}a}, {Balcells}, {Aguerri}, {Gallego}, \&
  {Zamorano}}]{elichemoral:pseudo.from.minor}
{Eliche-Moral}, M.~C., {Gonz{\'a}lez-Garc{\'{\i}}a}, A.~C., {Balcells}, M.,
  {Aguerri}, J.~A.~L., {Gallego}, J., \& {Zamorano}, J. 2008, in Astronomical
  Society of the Pacific Conference Series, Vol. 396, Astronomical Society of
  the Pacific Conference Series, ed. J.~G. {Funes} \& E.~M. {Corsini}, 359--+

\bibitem[{{Erb}(2008)}]{erb:outflow.inflow.masses}
{Erb}, D.~K. 2008, \apj, 674, 151

\bibitem[{{Erb} {et~al.}(2006){Erb}, {Steidel}, {Shapley}, {Pettini}, {Reddy},
  \& {Adelberger}}]{erb:lbg.gasmasses}
{Erb}, D.~K., {Steidel}, C.~C., {Shapley}, A.~E., {Pettini}, M., {Reddy},
  N.~A., \& {Adelberger}, K.~L. 2006, \apj, 646, 107

\bibitem[{{Fakhouri} \& {Ma}(2008{\natexlab{a}})}]{fakhouri:merger.rate.env}
{Fakhouri}, O., \& {Ma}, C.-P. 2008{\natexlab{a}}, \mnras, in press,
  arXiv:0808.2471 [astro-ph]

\bibitem[{{Fakhouri} \& {Ma}(2008{\natexlab{b}})}]{fakhouri:halo.merger.rates}
---. 2008{\natexlab{b}}, \mnras, 386, 577

\bibitem[{{Ferrarese} \& {Merritt}(2000)}]{FM00}
{Ferrarese}, L., \& {Merritt}, D. 2000, \apjl, 539, L9

\bibitem[{{Flores} {et~al.}(2006){Flores}, {Hammer}, {Puech}, {Amram}, \&
  {Balkowski}}]{flores:tf.evolution}
{Flores}, H., {Hammer}, F., {Puech}, M., {Amram}, P., \& {Balkowski}, C. 2006,
  \aap, 455, 107

\bibitem[{{Fontana} {et~al.}(2004)}]{fontana:mfs}
{Fontana}, A., {et~al.} 2004, \aap, 424, 23

\bibitem[{{Fontana} {et~al.}(2006)}]{fontana:highz.mfs}
---. 2006, \aap, 459, 745

\bibitem[{{Forster Schreiber}
  {et~al.}(2009)}]{forsterschreiber:z2.sf.gal.spectroscopy}
{Forster Schreiber}, N.~M., {et~al.} 2009, \apj, 706, 1364

\bibitem[{{Foster} {et~al.}(2009){Foster}, {Proctor}, {Forbes}, {Spolaor},
  {Hopkins}, \& {Brodie}}]{foster:metallicity.gradients}
{Foster}, C., {Proctor}, R.~N., {Forbes}, D.~A., {Spolaor}, M., {Hopkins},
  P.~F., \& {Brodie}, J.~P. 2009, \mnras, 1562

\bibitem[{{Franceschini} {et~al.}(2006)}]{franceschini:mfs}
{Franceschini}, A., {et~al.} 2006, \aap, 453, 397

\bibitem[{{Franx} {et~al.}(2008){Franx}, {van Dokkum}, {Schreiber}, {Wuyts},
  {Labb{\'e}}, \& {Toft}}]{franx:size.evol}
{Franx}, M., {van Dokkum}, P.~G., {Schreiber}, N.~M.~F., {Wuyts}, S.,
  {Labb{\'e}}, I., \& {Toft}, S. 2008, \apj, 688, 770

\bibitem[{{Gao} {et~al.}(2004){Gao}, {White}, {Jenkins}, {Stoehr}, \&
  {Springel}}]{gao:subhalo.mf}
{Gao}, L., {White}, S.~D.~M., {Jenkins}, A., {Stoehr}, F., \& {Springel}, V.
  2004, \mnras, 355, 819

\bibitem[{{Gauthier} {et~al.}(2006){Gauthier}, {Dubinski}, \&
  {Widrow}}]{gauthier:triggered.bar.from.substructure}
{Gauthier}, J.-R., {Dubinski}, J., \& {Widrow}, L.~M. 2006, \apj, 653, 1180

\bibitem[{{Gebhardt} {et~al.}(2000)}]{Gebhardt00}
{Gebhardt}, K., {et~al.} 2000, \apjl, 539, L13

\bibitem[{{Genel} {et~al.}(2008){Genel}, {Genzel}, {Bouch{\'e}}, {Naab}, \&
  {Sternberg}}]{genel:merger.rates.perprogenitor}
{Genel}, S., {Genzel}, R., {Bouch{\'e}}, N., {Naab}, T., \& {Sternberg}, A.
  2008, \apj, in press, arXiv:0812.3154

\bibitem[{{Genzel} {et~al.}(2001){Genzel}, {Tacconi}, {Rigopoulou}, {Lutz}, \&
  {Tecza}}]{Genzel01}
{Genzel}, R., {Tacconi}, L.~J., {Rigopoulou}, D., {Lutz}, D., \& {Tecza}, M.
  2001, \apj, 563, 527

\bibitem[{{Genzel} {et~al.}(2008)}]{genzel:highz.rapid.secular}
{Genzel}, R., {et~al.} 2008, \apj, 687, 59

\bibitem[{{Gerke} {et~al.}(2007)}]{gerke:blue.frac.evol}
{Gerke}, B.~F., {et~al.} 2007, \mnras, 222

\bibitem[{{Gottl{\"o}ber} {et~al.}(2001){Gottl{\"o}ber}, {Klypin}, \&
  {Kravtsov}}]{gottlober:merger.rate.vs.env}
{Gottl{\"o}ber}, S., {Klypin}, A., \& {Kravtsov}, A.~V. 2001, \apj, 546, 223

\bibitem[{{Governato} {et~al.}(2007){Governato}, {Willman}, {Mayer}, {Brooks},
  {Stinson}, {Valenzuela}, {Wadsley}, \& {Quinn}}]{governato:disk.formation}
{Governato}, F., {Willman}, B., {Mayer}, L., {Brooks}, A., {Stinson}, G.,
  {Valenzuela}, O., {Wadsley}, J., \& {Quinn}, T. 2007, \mnras, 374, 1479

\bibitem[{{Governato} {et~al.}(2008){Governato}, {Brook}, {Brooks}, {Mayer},
  {Willman}, {Jonsson}, {Stilp}, {Pope}, {Christensen}, {Wadsley}, \&
  {Quinn}}]{governato:disk.rebuilding}
{Governato}, F., {et~al.} 2008, \mnras, in press, arXiv:0812.0379 [astro-ph]

\bibitem[{{Grazian} {et~al.}(2007)}]{grazian:drg.comparisons}
{Grazian}, A., {et~al.} 2007, \aap, 465, 393

\bibitem[{{Guo} \& {White}(2008)}]{guo:gal.growth.channels}
{Guo}, Q., \& {White}, S.~D.~M. 2008, \mnras, 384, 2

\bibitem[{{Guyon} {et~al.}(2006){Guyon}, {Sanders}, \&
  {Stockton}}]{guyon:qso.hosts.ir}
{Guyon}, O., {Sanders}, D.~B., \& {Stockton}, A. 2006, \apjs, 166, 89

\bibitem[{{Hammer} {et~al.}(2005){Hammer}, {Flores}, {Elbaz}, {Zheng}, {Liang},
  \& {Cesarsky}}]{hammer:obs.disks.w.mergers}
{Hammer}, F., {Flores}, H., {Elbaz}, D., {Zheng}, X.~Z., {Liang}, Y.~C., \&
  {Cesarsky}, C. 2005, \aap, 430, 115

\bibitem[{{Hernquist}(1989)}]{hernquist.89}
{Hernquist}, L. 1989, \nat, 340, 687

\bibitem[{{Hernquist}(1992)}]{hernquist:bulgeless.mergers}
---. 1992, \apj, 400, 460

\bibitem[{{Hernquist}(1993)}]{hernquist:bulge.mergers}
---. 1993, \apj, 409, 548

\bibitem[{{Hoffman} {et~al.}(2009){Hoffman}, {Cox}, {Dutta}, \&
  {Hernquist}}]{hoffman:dissipation.and.gal.kinematics}
{Hoffman}, L.~K., {Cox}, T.~J., {Dutta}, S.~N., \& {Hernquist}, L.~E. 2009,
  \apjl, in press, arXiv:0903.3064 [astro-ph]

\bibitem[{{Hopkins} \& {Beacom}(2006)}]{hopkinsbeacom:sfh}
{Hopkins}, A.~M., \& {Beacom}, J.~F. 2006, \apj, 651, 142

\bibitem[{{Hopkins} {et~al.}(2007{\natexlab{a}}){Hopkins}, {Bundy},
  {Hernquist}, \& {Ellis}}]{hopkins:transition.mass}
{Hopkins}, P.~F., {Bundy}, K., {Hernquist}, L., \& {Ellis}, R.~S.
  2007{\natexlab{a}}, \apj, 659, 976

\bibitem[{{Hopkins} {et~al.}(2009{\natexlab{a}}){Hopkins}, {Cox}, {Dutta},
  {Hernquist}, {Kormendy}, \& {Lauer}}]{hopkins:cusps.ell}
{Hopkins}, P.~F., {Cox}, T.~J., {Dutta}, S.~N., {Hernquist}, L., {Kormendy},
  J., \& {Lauer}, T.~R. 2009{\natexlab{a}}, \apjs, 181, 135

\bibitem[{{Hopkins} {et~al.}(2008{\natexlab{a}}){Hopkins}, {Cox}, \&
  {Hernquist}}]{hopkins:cusps.fp}
{Hopkins}, P.~F., {Cox}, T.~J., \& {Hernquist}, L. 2008{\natexlab{a}}, \apj,
  689, 17

\bibitem[{{Hopkins} {et~al.}(2008{\natexlab{b}}){Hopkins}, {Cox}, {Kere{\v s}},
  \& {Hernquist}}]{hopkins:groups.ell}
{Hopkins}, P.~F., {Cox}, T.~J., {Kere{\v s}}, D., \& {Hernquist}, L.
  2008{\natexlab{b}}, \apjs, 175, 390

\bibitem[{{Hopkins} {et~al.}(2009{\natexlab{b}}){Hopkins}, {Cox}, {Younger}, \&
  {Hernquist}}]{hopkins:disk.survival}
{Hopkins}, P.~F., {Cox}, T.~J., {Younger}, J.~D., \& {Hernquist}, L.
  2009{\natexlab{b}}, \apj, 691, 1168

\bibitem[{{Hopkins} \& {Hernquist}(2006)}]{hopkins:seyferts}
{Hopkins}, P.~F., \& {Hernquist}, L. 2006, \apjs, 166, 1

\bibitem[{{Hopkins} \& {Hernquist}(2009)}]{hopkins:seyfert.limits}
---. 2009, \apj, 694, 599

\bibitem[{{Hopkins} {et~al.}(2005{\natexlab{a}}){Hopkins}, {Hernquist}, {Cox},
  {Di Matteo}, {Martini}, {Robertson}, \&
  {Springel}}]{hopkins:lifetimes.methods}
{Hopkins}, P.~F., {Hernquist}, L., {Cox}, T.~J., {Di Matteo}, T., {Martini},
  P., {Robertson}, B., \& {Springel}, V. 2005{\natexlab{a}}, \apj, 630, 705

\bibitem[{{Hopkins} {et~al.}(2008{\natexlab{c}}){Hopkins}, {Hernquist}, {Cox},
  {Dutta}, \& {Rothberg}}]{hopkins:cusps.mergers}
{Hopkins}, P.~F., {Hernquist}, L., {Cox}, T.~J., {Dutta}, S.~N., \& {Rothberg},
  B. 2008{\natexlab{c}}, \apj, 679, 156

\bibitem[{{Hopkins} {et~al.}(2008{\natexlab{d}}){Hopkins}, {Hernquist}, {Cox},
  \& {Kere{\v s}}}]{hopkins:groups.qso}
{Hopkins}, P.~F., {Hernquist}, L., {Cox}, T.~J., \& {Kere{\v s}}, D.
  2008{\natexlab{d}}, \apjs, 175, 356

\bibitem[{{Hopkins} {et~al.}(2009{\natexlab{c}}){Hopkins}, {Hernquist}, {Cox},
  {Kere{\v s}}, \& {Wuyts}}]{hopkins:cusps.evol}
{Hopkins}, P.~F., {Hernquist}, L., {Cox}, T.~J., {Kere{\v s}}, D., \& {Wuyts},
  S. 2009{\natexlab{c}}, \apj, 691, 1424

\bibitem[{{Hopkins} {et~al.}(2007{\natexlab{b}}){Hopkins}, {Hernquist}, {Cox},
  {Robertson}, \& {Krause}}]{hopkins:bhfp.theory}
{Hopkins}, P.~F., {Hernquist}, L., {Cox}, T.~J., {Robertson}, B., \& {Krause},
  E. 2007{\natexlab{b}}, \apj, 669, 45

\bibitem[{{Hopkins} {et~al.}(2007{\natexlab{c}}){Hopkins}, {Hernquist}, {Cox},
  {Robertson}, \& {Krause}}]{hopkins:bhfp.obs}
---. 2007{\natexlab{c}}, \apj, 669, 67

\bibitem[{{Hopkins} {et~al.}(2008{\natexlab{e}}){Hopkins}, {Hernquist}, {Cox},
  {Younger}, \& {Besla}}]{hopkins:disk.heating}
{Hopkins}, P.~F., {Hernquist}, L., {Cox}, T.~J., {Younger}, J.~D., \& {Besla},
  G. 2008{\natexlab{e}}, \apj, 688, 757

\bibitem[{{Hopkins} {et~al.}(2005{\natexlab{b}}){Hopkins}, {Hernquist},
  {Martini}, {Cox}, {Robertson}, {Di Matteo}, \&
  {Springel}}]{hopkins:lifetimes.letter}
{Hopkins}, P.~F., {Hernquist}, L., {Martini}, P., {Cox}, T.~J., {Robertson},
  B., {Di Matteo}, T., \& {Springel}, V. 2005{\natexlab{b}}, \apjl, 625, L71

\bibitem[{{Hopkins} {et~al.}(2009{\natexlab{d}}){Hopkins}, {Hickox},
  {Quataert}, \& {Hernquist}}]{hopkins:seyfert.bimodality}
{Hopkins}, P.~F., {Hickox}, R., {Quataert}, E., \& {Hernquist}, L.
  2009{\natexlab{d}}, \mnras, 398, 333

\bibitem[{{Hopkins} {et~al.}(2009{\natexlab{e}}){Hopkins}, {Lauer}, {Cox},
  {Hernquist}, \& {Kormendy}}]{hopkins:cores}
{Hopkins}, P.~F., {Lauer}, T.~R., {Cox}, T.~J., {Hernquist}, L., \& {Kormendy},
  J. 2009{\natexlab{e}}, \apjs, 181, 486

\bibitem[{{Hopkins} {et~al.}(2007{\natexlab{d}}){Hopkins}, {Lidz}, {Hernquist},
  {Coil}, {Myers}, {Cox}, \& {Spergel}}]{hopkins:clustering}
{Hopkins}, P.~F., {Lidz}, A., {Hernquist}, L., {Coil}, A.~L., {Myers}, A.~D.,
  {Cox}, T.~J., \& {Spergel}, D.~N. 2007{\natexlab{d}}, \apj, 662, 110

\bibitem[{{Hopkins} {et~al.}(2007{\natexlab{e}}){Hopkins}, {Richards}, \&
  {Hernquist}}]{hopkins:bol.qlf}
{Hopkins}, P.~F., {Richards}, G.~T., \& {Hernquist}, L. 2007{\natexlab{e}},
  \apj, 654, 731

\bibitem[{{Hopkins}
  {et~al.}(2009{\natexlab{f}})}]{hopkins:merger.rates.methods}
{Hopkins}, P.~F., {et~al.} 2009{\natexlab{f}}, \mnras, in preparation

\bibitem[{{Hopkins} {et~al.}(2009{\natexlab{g}}){Hopkins}, {Somerville}, {Cox},
  {Hernquist}, {Jogee}, {Kere{\v s}}, {Ma}, {Robertson}, \&
  {Stewart}}]{hopkins:disk.survival.cosmo}
---. 2009{\natexlab{g}}, \mnras, 397, 802

\bibitem[{{Ilbert} {et~al.}(2009)}]{ilbert:cosmos.morph.mfs}
{Ilbert}, O., {et~al.} 2009, \apj, in press, arXiv:0903.0102

\bibitem[{{James} {et~al.}(1999){James}, {Bate}, {Wells}, {Wright}, \&
  {Doyon}}]{James99}
{James}, P., {Bate}, C., {Wells}, M., {Wright}, G., \& {Doyon}, R. 1999,
  \mnras, 309, 585

\bibitem[{{Jesseit} {et~al.}(2009){Jesseit}, {Cappellari}, {Naab}, {Emsellem},
  \& {Burkert}}]{jesseit:merger.rem.spin.vs.gas}
{Jesseit}, R., {Cappellari}, M., {Naab}, T., {Emsellem}, E., \& {Burkert}, A.
  2009, \mnras, 397, 1202

\bibitem[{{Jesseit} {et~al.}(2007){Jesseit}, {Naab}, {Peletier}, \&
  {Burkert}}]{jesseit:kinematics}
{Jesseit}, R., {Naab}, T., {Peletier}, R.~F., \& {Burkert}, A. 2007, \mnras,
  376, 997

\bibitem[{{Jiang} {et~al.}(2008){Jiang}, {Jing}, {Faltenbacher}, {Lin}, \&
  {Li}}]{jiang:dynfric.calibration}
{Jiang}, C.~Y., {Jing}, Y.~P., {Faltenbacher}, A., {Lin}, W.~P., \& {Li}, C.
  2008, \apj, 675, 1095

\bibitem[{{Jogee} {et~al.}(2005){Jogee}, {Scoville}, \&
  {Kenney}}]{jogee:H2.masses}
{Jogee}, S., {Scoville}, N., \& {Kenney}, J.~D.~P. 2005, \apj, 630, 837

\bibitem[{{Jogee} {et~al.}(2008)}]{jogee:merger.density.08.conf.proc}
{Jogee}, S., {et~al.} 2008, in Astronomical Society of the Pacific Conference
  Series, Vol. 396, Astronomical Society of the Pacific Conference Series, ed.
  J.~G. {Funes} \& E.~M. {Corsini}, 337--+

\bibitem[{{Jogee} {et~al.}(2009)}]{jogee:merger.density.08}
{Jogee}, S., {et~al.} 2009, \apj, 697, 1971

\bibitem[{{Joseph} \& {Wright}(1985)}]{joseph85}
{Joseph}, R.~D., \& {Wright}, G.~S. 1985, \mnras, 214, 87

\bibitem[{{Kajisawa} {et~al.}(2009)}]{kajisawa:stellar.mf.to.z3}
{Kajisawa}, M., {et~al.} 2009, \apj, 702, 1393

\bibitem[{{Kannappan}(2004)}]{kannappan:gfs}
{Kannappan}, S.~J. 2004, \apjl, 611, L89

\bibitem[{{Kartaltepe} {et~al.}(2007)}]{kartaltepe:pair.fractions}
{Kartaltepe}, J.~S., {et~al.} 2007, \apjs, 172, 320

\bibitem[{{Kazantzidis} {et~al.}(2008){Kazantzidis}, {Bullock}, {Zentner},
  {Kravtsov}, \& {Moustakas}}]{kazantzidis:mw.merger.hist.sim}
{Kazantzidis}, S., {Bullock}, J.~S., {Zentner}, A.~R., {Kravtsov}, A.~V., \&
  {Moustakas}, L.~A. 2008, \apj, 688, 254

\bibitem[{{Kazantzidis} {et~al.}(2009){Kazantzidis}, {Zentner}, {Kravtsov},
  {Bullock}, \& {Debattista}}]{kazantzidis:thin.disk.thickening}
{Kazantzidis}, S., {Zentner}, A.~R., {Kravtsov}, A.~V., {Bullock}, J.~S., \&
  {Debattista}, V.~P. 2009, \apj, in press, arXiv:0902.1983 [astro-ph]

\bibitem[{{Khochfar} \& {Burkert}(2006)}]{khochfar:cosmo.orbits}
{Khochfar}, S., \& {Burkert}, A. 2006, \aap, 445, 403

\bibitem[{{Khochfar} \& {Silk}(2006)}]{khochfar:size.evolution.model}
{Khochfar}, S., \& {Silk}, J. 2006, \apjl, 648, L21

\bibitem[{{Khochfar} \& {Silk}(2008)}]{khochfarsilk:new.sam.dry.mergers}
---. 2008, ArXiv e-prints

\bibitem[{{Kitzbichler} \&
  {White}(2008)}]{kitzbichler:mgr.rate.pair.calibration}
{Kitzbichler}, M.~G., \& {White}, S.~D.~M. 2008, \mnras, 391, 1489

\bibitem[{{Komatsu} {et~al.}(2009)}]{komatsu:wmap5}
{Komatsu}, E., {et~al.} 2009, \apjs, 180, 330

\bibitem[{{Kormendy} {et~al.}(2009){Kormendy}, {Fisher}, {Cornell}, \&
  {Bender}}]{jk:profiles}
{Kormendy}, J., {Fisher}, D.~B., {Cornell}, M.~E., \& {Bender}, R. 2009, \apjs,
  182, 216

\bibitem[{{Kormendy} \&
  {Kennicutt}(2004)}]{kormendy.kennicutt:pseudobulge.review}
{Kormendy}, J., \& {Kennicutt}, Jr., R.~C. 2004, \araa, 42, 603

\bibitem[{{Kravtsov} {et~al.}(2004){Kravtsov}, {Berlind}, {Wechsler}, {Klypin},
  {Gottl{\"o}ber}, {Allgood}, \& {Primack}}]{kravtsov:subhalo.mfs}
{Kravtsov}, A.~V., {Berlind}, A.~A., {Wechsler}, R.~H., {Klypin}, A.~A.,
  {Gottl{\"o}ber}, S., {Allgood}, B., \& {Primack}, J.~R. 2004, \apj, 609, 35

\bibitem[{{Kriek} {et~al.}(2006)}]{kriek:drg.seds}
{Kriek}, M., {et~al.} 2006, \apjl, 649, L71

\bibitem[{{Krivitsky} \& {Kontorovich}(1997)}]{krivitsky.kontorovich}
{Krivitsky}, D.~S., \& {Kontorovich}, V.~M. 1997, \aap, 327, 921

\bibitem[{{Kuijken} \& {Merrifield}(1995)}]{kuijken:pseudobulges.obs}
{Kuijken}, K., \& {Merrifield}, M.~R. 1995, \apjl, 443, L13

\bibitem[{{Labb{\'e}} {et~al.}(2005)}]{labbe05:drgs}
{Labb{\'e}}, I., {et~al.} 2005, \apjl, 624, L81

\bibitem[{{Lagan{\'a}} {et~al.}(2008){Lagan{\'a}}, {Lima Neto},
  {Andrade-Santos}, \& {Cypriano}}]{lagana:intracluster.light}
{Lagan{\'a}}, T.~F., {Lima Neto}, G.~B., {Andrade-Santos}, F., \& {Cypriano},
  E.~S. 2008, \aap, 485, 633

\bibitem[{{Lake} \& {Dressler}(1986)}]{LakeDressler86}
{Lake}, G., \& {Dressler}, A. 1986, \apj, 310, 605

\bibitem[{{Lauer} {et~al.}(2007)}]{lauer:bimodal.profiles}
{Lauer}, T.~R., {et~al.} 2007, \apj, 664, 226

\bibitem[{{Lee} {et~al.}(2009){Lee}, {Giavalisco}, {Conroy}, {Wechsler},
  {Ferguson}, {Somerville}, {Dickinson}, \& {Urry}}]{lee:2009.uv.lum.vs.mhalo}
{Lee}, K., {Giavalisco}, M., {Conroy}, C., {Wechsler}, R.~H., {Ferguson},
  H.~C., {Somerville}, R.~S., {Dickinson}, M.~E., \& {Urry}, C.~M. 2009, \apj,
  695, 368

\bibitem[{{Lin} {et~al.}(2004)}]{lin:merger.fraction}
{Lin}, L., {et~al.} 2004, \apjl, 617, L9

\bibitem[{{Lin} {et~al.}(2008)}]{lin:mergers.by.type}
---. 2008, \apj, 681, 232

\bibitem[{{Lin} \& {Mohr}(2004)}]{lin:intracluster.light.measurements}
{Lin}, Y.-T., \& {Mohr}, J.~J. 2004, \apj, 617, 879

\bibitem[{{L{\'o}pez-Sanjuan} {et~al.}(2009{\natexlab{a}}){L{\'o}pez-Sanjuan},
  {Balcells}, {P{\'e}rez-Gonz{\'a}lez}, {Barro}, {Garc{\'{\i}}a-Dab{\'o}},
  {Gallego}, \& {Zamorano}}]{lopezsanjuan:merger.fraction.to.z1}
{L{\'o}pez-Sanjuan}, C., {Balcells}, M., {P{\'e}rez-Gonz{\'a}lez}, P.~G.,
  {Barro}, G., {Garc{\'{\i}}a-Dab{\'o}}, C.~E., {Gallego}, J., \& {Zamorano},
  J. 2009{\natexlab{a}}, \aap, in press [arXiv:0905.2765]

\bibitem[{{L{\'o}pez-Sanjuan}
  {et~al.}(2009{\natexlab{b}})}]{lopezsanjuan:mgr.rate.pairs}
{L{\'o}pez-Sanjuan}, C., {et~al.} 2009{\natexlab{b}}, \apj, 694, 643

\bibitem[{{Lotz} {et~al.}(2008{\natexlab{a}}){Lotz}, {Jonsson}, {Cox}, \&
  {Primack}}]{lotz:merger.selection}
{Lotz}, J.~M., {Jonsson}, P., {Cox}, T.~J., \& {Primack}, J.~R.
  2008{\natexlab{a}}, \mnras, 391, 1137

\bibitem[{{Lotz} {et~al.}(2009{\natexlab{a}}){Lotz}, {Jonsson}, {Cox}, \&
  {Primack}}]{lotz:gasfraction.vs.mergertime}
---. 2009{\natexlab{a}}, \mnras, in press, arXiv:0912.1593

\bibitem[{{Lotz} {et~al.}(2009{\natexlab{b}}){Lotz}, {Jonsson}, {Cox}, \&
  {Primack}}]{lotz:mgr.timescale.vs.massratio.morphology}
---. 2009{\natexlab{b}}, \mnras, in press, arXiv:0912.1590

\bibitem[{{Lotz} {et~al.}(2006){Lotz}, {Madau}, {Giavalisco}, {Primack}, \&
  {Ferguson}}]{lotz:morphology.evol}
{Lotz}, J.~M., {Madau}, P., {Giavalisco}, M., {Primack}, J., \& {Ferguson},
  H.~C. 2006, \apj, 636, 592

\bibitem[{{Lotz} {et~al.}(2004){Lotz}, {Primack}, \& {Madau}}]{lotz:gini-m20}
{Lotz}, J.~M., {Primack}, J., \& {Madau}, P. 2004, \aj, 128, 163

\bibitem[{{Lotz} {et~al.}(2008{\natexlab{b}})}]{lotz:merger.fraction}
{Lotz}, J.~M., {et~al.} 2008{\natexlab{b}}, \apj, 672, 177

\bibitem[{{Magorrian} {et~al.}(1998)}]{magorrian}
{Magorrian}, J., {et~al.} 1998, \aj, 115, 2285

\bibitem[{{Makino} \& {Hut}(1997)}]{makino:merger.cross.sections}
{Makino}, J., \& {Hut}, P. 1997, \apj, 481, 83

\bibitem[{{Malin} \& {Carter}(1980)}]{malin80}
{Malin}, D.~F., \& {Carter}, D. 1980, \nat, 285, 643

\bibitem[{{Malin} \& {Carter}(1983)}]{malin83}
---. 1983, \apj, 274, 534

\bibitem[{{Maller} {et~al.}(2006){Maller}, {Katz}, {Kere{\v s}}, {Dav{\'e}}, \&
  {Weinberg}}]{maller:sph.merger.rates}
{Maller}, A.~H., {Katz}, N., {Kere{\v s}}, D., {Dav{\'e}}, R., \& {Weinberg},
  D.~H. 2006, \apj, 647, 763

\bibitem[{{Mamon}(2006)}]{mamon:groups.review}
{Mamon}, G.~A. 2006, in Groups of Galaxies in the Nearby Universe, ed.
  I.~{Saviane}, V.~{Ivanov}, \& J.~{Borissova}

\bibitem[{{Mandelbaum} {et~al.}(2006){Mandelbaum}, {Seljak}, {Kauffmann},
  {Hirata}, \& {Brinkmann}}]{mandelbaum:mhalo}
{Mandelbaum}, R., {Seljak}, U., {Kauffmann}, G., {Hirata}, C.~M., \&
  {Brinkmann}, J. 2006, \mnras, 368, 715

\bibitem[{{Mannucci} {et~al.}(2009){Mannucci}, {Cresci}, {Maiolino}, {Marconi},
  {Pastorini}, {Pozzetti}, {Gnerucci}, {Risaliti}, {Schneider}, {Lehnert}, \&
  {Salvati}}]{mannucci:z3.gal.gfs.tf}
{Mannucci}, F., {et~al.} 2009, \mnras, 398, 1915

\bibitem[{{Marchesini} {et~al.}(2009){Marchesini}, {van Dokkum}, {F{\"o}rster
  Schreiber}, {Franx}, {Labb{\'e}}, \& {Wuyts}}]{marchesini:highz.stellar.mfs}
{Marchesini}, D., {van Dokkum}, P.~G., {F{\"o}rster Schreiber}, N.~M., {Franx},
  M., {Labb{\'e}}, I., \& {Wuyts}, S. 2009, \apj, 701, 1765

\bibitem[{{Martin} {et~al.}(2007){Martin}, {Small}, {Schiminovich}, {Wyder},
  {P{\'e}rez-Gonz{\'a}lez}, {Johnson}, {Wolf}, {Barlow}, {Forster}, {Friedman},
  {Morrissey}, {Neff}, {Seibert}, {Welsh}, {Bianchi}, {Donas}, {Heckman},
  {Lee}, {Madore}, {Milliard}, {Rich}, {Szalay}, {Yi}, {Meisenheimer}, \&
  {Rieke}}]{martin:uv.lum.history}
{Martin}, D.~C., {et~al.} 2007, \apjs, 173, 415

\bibitem[{{McGaugh}(2005)}]{mcgaugh:tf}
{McGaugh}, S.~S. 2005, \apj, 632, 859

\bibitem[{{Mihos} \& {Hernquist}(1994{\natexlab{a}})}]{mihos:gradients}
{Mihos}, J.~C., \& {Hernquist}, L. 1994{\natexlab{a}}, \apj, 427, 112

\bibitem[{{Mihos} \& {Hernquist}(1994{\natexlab{b}})}]{mihos:starbursts.94}
---. 1994{\natexlab{b}}, \apjl, 431, L9

\bibitem[{{Mihos} \& {Hernquist}(1996)}]{mihos:starbursts.96}
---. 1996, \apj, 464, 641

\bibitem[{{Moster} {et~al.}(2009){Moster}, {Somerville}, {Maulbetsch}, {van den
  Bosch}, {Maccio'}, {Naab}, \& {Oser}}]{moster:stellar.vs.halo.mass.to.z1}
{Moster}, B.~P., {Somerville}, R.~S., {Maulbetsch}, C., {van den Bosch}, F.~C.,
  {Maccio'}, A.~V., {Naab}, T., \& {Oser}, L. 2009, \apj, in press,
  arXiv:0903.4682 [astro-ph]

\bibitem[{{Naab} \& {Burkert}(2003)}]{naab:minor.mergers}
{Naab}, T., \& {Burkert}, A. 2003, \apj, 597, 893

\bibitem[{{Naab} {et~al.}(2006){Naab}, {Jesseit}, \& {Burkert}}]{naab:gas}
{Naab}, T., {Jesseit}, R., \& {Burkert}, A. 2006, \mnras, 372, 839

\bibitem[{{Naab} {et~al.}(2009){Naab}, {Johansson}, \&
  {Ostriker}}]{naab:size.evol.from.minor.mergers}
{Naab}, T., {Johansson}, P.~H., \& {Ostriker}, J.~P. 2009, \apjl, 699, L178

\bibitem[{{Naab} {et~al.}(2007){Naab}, {Johansson}, {Ostriker}, \&
  {Efstathiou}}]{naab:etg.formation}
{Naab}, T., {Johansson}, P.~H., {Ostriker}, J.~P., \& {Efstathiou}, G. 2007,
  \apj, 658, 710

\bibitem[{{Naab} \& {Trujillo}(2006)}]{naab:profiles}
{Naab}, T., \& {Trujillo}, I. 2006, \mnras, 369, 625

\bibitem[{{Neistein} {et~al.}(2006){Neistein}, {van den Bosch}, \&
  {Dekel}}]{neistein:natural.downsizing}
{Neistein}, E., {van den Bosch}, F.~C., \& {Dekel}, A. 2006, \mnras, 372, 933

\bibitem[{{Noeske} {et~al.}(2007{\natexlab{a}})}]{noeske:sfh}
{Noeske}, K.~G., {et~al.} 2007{\natexlab{a}}, \apjl, 660, L47

\bibitem[{{Noeske} {et~al.}(2007{\natexlab{b}})}]{noeske:2007.sfh.part1}
---. 2007{\natexlab{b}}, \apjl, 660, L43

\bibitem[{{Nurmi} {et~al.}(2006){Nurmi}, {Hein{\"a}m{\"a}ki}, {Saar},
  {Einasto}, {Holopainen}, {Mart{\'\i}nez}, \& {Einasto}}]{nurmi:subhalo.mf}
{Nurmi}, P., {Hein{\"a}m{\"a}ki}, P., {Saar}, E., {Einasto}, M., {Holopainen},
  J., {Mart{\'\i}nez}, V.~J., \& {Einasto}, J. 2006, \aap, in press
  [astro-ph/0611941]

\bibitem[{{Okamoto} {et~al.}(2005){Okamoto}, {Eke}, {Frenk}, \&
  {Jenkins}}]{okamoto:feedback.vs.disk.morphology}
{Okamoto}, T., {Eke}, V.~R., {Frenk}, C.~S., \& {Jenkins}, A. 2005, \mnras,
  363, 1299

\bibitem[{{O'Neill} \& {Dubinski}(2003)}]{oniell:bar.obs}
{O'Neill}, J.~K., \& {Dubinski}, J. 2003, \mnras, 346, 251

\bibitem[{{Pannella} {et~al.}(2006){Pannella}, {Hopp}, {Saglia}, {Bender},
  {Drory}, {Salvato}, {Gabasch}, \& {Feulner}}]{pannella:mfs}
{Pannella}, M., {Hopp}, U., {Saglia}, R.~P., {Bender}, R., {Drory}, N.,
  {Salvato}, M., {Gabasch}, A., \& {Feulner}, G. 2006, \apjl, 639, L1

\bibitem[{{Papovich} {et~al.}(2006)}]{papovich:ssfr}
{Papovich}, C., {et~al.} 2006, \apj, 640, 92

\bibitem[{{Parry} {et~al.}(2009){Parry}, {Eke}, \&
  {Frenk}}]{parry:sam.merger.vs.morph}
{Parry}, O.~H., {Eke}, V.~R., \& {Frenk}, C.~S. 2009, \mnras, 396, 1972

\bibitem[{{Patton} \& {Atfield}(2008)}]{patton:mgr.rate.vs.rmag}
{Patton}, D.~R., \& {Atfield}, J.~E. 2008, \apj, 685, 235

\bibitem[{{Patton} {et~al.}(2002)}]{patton:merger.fraction}
{Patton}, D.~R., {et~al.} 2002, \apj, 565, 208

\bibitem[{{P{\'e}rez-Gonz{\'a}lez}
  {et~al.}(2008{\natexlab{a}}){P{\'e}rez-Gonz{\'a}lez}, {Trujillo}, {Barro},
  {Gallego}, {Zamorano}, \& {Conselice}}]{perezgonzalez:hod.ell.evol}
{P{\'e}rez-Gonz{\'a}lez}, P.~G., {Trujillo}, I., {Barro}, G., {Gallego}, J.,
  {Zamorano}, J., \& {Conselice}, C.~J. 2008{\natexlab{a}}, \apj, 687, 50

\bibitem[{{P{\'e}rez-Gonz{\'a}lez}
  {et~al.}(2008{\natexlab{b}})}]{perezgonzalez:mf.compilation}
{P{\'e}rez-Gonz{\'a}lez}, P.~G., {et~al.} 2008{\natexlab{b}}, \apj, 675, 234

\bibitem[{{Persic} \& {Salucci}(1988)}]{persic88}
{Persic}, M., \& {Salucci}, P. 1988, \mnras, 234, 131

\bibitem[{{Persic} {et~al.}(1996){Persic}, {Salucci}, \& {Stel}}]{persic96}
{Persic}, M., {Salucci}, P., \& {Stel}, F. 1996, \mnras, 281, 27

\bibitem[{{Pfenniger}(1984)}]{pfenniger:bar.dynamics}
{Pfenniger}, D. 1984, \aap, 134, 373

\bibitem[{{Puech} {et~al.}(2007{\natexlab{a}}){Puech}, {Hammer}, {Flores},
  {Neichel}, {Yang}, \& {Rodrigues}}]{puech:minor.merger.at.z06}
{Puech}, M., {Hammer}, F., {Flores}, H., {Neichel}, B., {Yang}, Y., \&
  {Rodrigues}, M. 2007{\natexlab{a}}, \aap, 476, L21

\bibitem[{{Puech} {et~al.}(2007{\natexlab{b}}){Puech}, {Hammer}, {Lehnert}, \&
  {Flores}}]{puech:highz.vsigma.disks}
{Puech}, M., {Hammer}, F., {Lehnert}, M.~D., \& {Flores}, H.
  2007{\natexlab{b}}, \aap, 466, 83

\bibitem[{{Puech} {et~al.}(2008)}]{puech:tf.evol}
{Puech}, M., {et~al.} 2008, \aap, 484, 173

\bibitem[{{Purcell} {et~al.}(2009){Purcell}, {Kazantzidis}, \&
  {Bullock}}]{purcell:minor.merger.thindisk.destruction}
{Purcell}, C.~W., {Kazantzidis}, S., \& {Bullock}, J.~S. 2009, \apjl, 694, L98

\bibitem[{{Quinn} \& {Goodman}(1986)}]{quinn86:dynfric.on.sats}
{Quinn}, P.~J., \& {Goodman}, J. 1986, \apj, 309, 472

\bibitem[{{Raha} {et~al.}(1991){Raha}, {Sellwood}, {James}, \&
  {Kahn}}]{raha:bar.instabilities}
{Raha}, N., {Sellwood}, J.~A., {James}, R.~A., \& {Kahn}, F.~D. 1991, \nat,
  352, 411

\bibitem[{{Robertson} {et~al.}(2006){Robertson}, {Bullock}, {Cox}, {Di Matteo},
  {Hernquist}, {Springel}, \& {Yoshida}}]{robertson:disk.formation}
{Robertson}, B., {Bullock}, J.~S., {Cox}, T.~J., {Di Matteo}, T., {Hernquist},
  L., {Springel}, V., \& {Yoshida}, N. 2006, \apj, 645, 986

\bibitem[{{Robertson} \&
  {Bullock}(2008)}]{robertsonbullock:highz.disk.vs.model}
{Robertson}, B.~E., \& {Bullock}, J.~S. 2008, \apjl, 685, L27

\bibitem[{{Rothberg} \& {Joseph}(2004)}]{rj:profiles}
{Rothberg}, B., \& {Joseph}, R.~D. 2004, \aj, 128, 2098

\bibitem[{{Rothberg} \&
  {Joseph}(2006{\natexlab{a}})}]{rothberg.joseph:kinematics}
---. 2006{\natexlab{a}}, \aj, 131, 185

\bibitem[{{Rothberg} \&
  {Joseph}(2006{\natexlab{b}})}]{rothberg.joseph:rotation}
---. 2006{\natexlab{b}}, \aj, 132, 976

\bibitem[{{Ruhland} {et~al.}(2009){Ruhland}, {Bell}, {Haeussler}, {Taylor},
  {Barden}, \& {McIntosh}}]{ruhland:dry.mergers.and.color.mag.relation}
{Ruhland}, C., {Bell}, E.~F., {Haeussler}, B., {Taylor}, E.~N., {Barden}, M.,
  \& {McIntosh}, D.~H. 2009, \apj, in press, arXiv:0901.4340 [astro-ph]

\bibitem[{{Salucci} {et~al.}(1999){Salucci}, {Szuszkiewicz}, {Monaco}, \&
  {Danese}}]{salucci:bhmf}
{Salucci}, P., {Szuszkiewicz}, E., {Monaco}, P., \& {Danese}, L. 1999, \mnras,
  307, 637

\bibitem[{{Sanders} \& {Mirabel}(1996)}]{sanders96:ulirgs.mergers}
{Sanders}, D.~B., \& {Mirabel}, I.~F. 1996, \araa, 34, 749

\bibitem[{{Sanders} {et~al.}(1988){Sanders}, {Soifer}, {Elias}, {Madore},
  {Matthews}, {Neugebauer}, \& {Scoville}}]{sanders88:quasars}
{Sanders}, D.~B., {Soifer}, B.~T., {Elias}, J.~H., {Madore}, B.~F., {Matthews},
  K., {Neugebauer}, G., \& {Scoville}, N.~Z. 1988, \apj, 325, 74

\bibitem[{{Scannapieco} {et~al.}(2008){Scannapieco}, {Tissera}, {White}, \&
  {Springel}}]{scannapieco:fb.disk.sims}
{Scannapieco}, C., {Tissera}, P.~B., {White}, S.~D.~M., \& {Springel}, V. 2008,
  \mnras, 389, 1137

\bibitem[{{Schweizer}(1980)}]{schweizer80}
{Schweizer}, F. 1980, \apj, 237, 303

\bibitem[{{Schweizer}(1982)}]{schweizer82}
---. 1982, \apj, 252, 455

\bibitem[{{Schweizer}(1992)}]{schweizer92}
{Schweizer}, F. 1992, in Physics of Nearby Galaxies: Nature or Nurture?, ed.
  T.~X. {Thuan}, C.~{Balkowski}, \& J.~{Tran Thanh van}, 283--+

\bibitem[{{Schweizer}(1996)}]{schweizer96}
---. 1996, \aj, 111, 109

\bibitem[{{Schweizer} \& {Seitzer}(1992)}]{schweizerseitzer92}
{Schweizer}, F., \& {Seitzer}, P. 1992, \aj, 104, 1039

\bibitem[{{Shankar} {et~al.}(2009){Shankar}, {Weinberg}, \&
  {Miralda-Escud{\'e}}}]{shankar:bol.qlf}
{Shankar}, F., {Weinberg}, D.~H., \& {Miralda-Escud{\'e}}, J. 2009, \apj, 690,
  20

\bibitem[{{Shapiro} {et~al.}(2008)}]{shapiro:highz.kinematics}
{Shapiro}, K.~L., {et~al.} 2008, \apj, 682, 231

\bibitem[{{Shapley} {et~al.}(2005){Shapley}, {Coil}, {Ma}, \&
  {Bundy}}]{shapley:z1.abundances}
{Shapley}, A.~E., {Coil}, A.~L., {Ma}, C.-P., \& {Bundy}, K. 2005, \apj, 635,
  1006

\bibitem[{{Sheth} {et~al.}(2001){Sheth}, {Mo}, \& {Tormen}}]{shethtormen}
{Sheth}, R.~K., {Mo}, H.~J., \& {Tormen}, G. 2001, \mnras, 323, 1

\bibitem[{{Shier} \& {Fischer}(1998)}]{ShierFischer98}
{Shier}, L.~M., \& {Fischer}, J. 1998, \apj, 497, 163

\bibitem[{{Soifer} {et~al.}(1984{\natexlab{a}})}]{soifer84a}
{Soifer}, B.~T., {et~al.} 1984{\natexlab{a}}, \apjl, 278, L71

\bibitem[{{Soifer} {et~al.}(1984{\natexlab{b}})}]{soifer84b}
---. 1984{\natexlab{b}}, \apjl, 283, L1

\bibitem[{{Soltan}(1982)}]{soltan82}
{Soltan}, A. 1982, \mnras, 200, 115

\bibitem[{{Somerville} {et~al.}(2008){Somerville}, {Hopkins}, {Cox},
  {Robertson}, \& {Hernquist}}]{somerville:new.sam}
{Somerville}, R.~S., {Hopkins}, P.~F., {Cox}, T.~J., {Robertson}, B.~E., \&
  {Hernquist}, L. 2008, \mnras, 391, 481

\bibitem[{{Spergel} {et~al.}(2003)}]{spergel:wmap1}
{Spergel}, D.~N., {et~al.} 2003, \apjs, 148, 175

\bibitem[{{Spergel} {et~al.}(2007)}]{spergel:wmap3}
---. 2007, \apjs, 170, 377

\bibitem[{{Springel} {et~al.}(2005{\natexlab{a}}){Springel}, {Di Matteo}, \&
  {Hernquist}}]{springel:red.galaxies}
{Springel}, V., {Di Matteo}, T., \& {Hernquist}, L. 2005{\natexlab{a}}, \apjl,
  620, L79

\bibitem[{{Springel} {et~al.}(2005{\natexlab{b}}){Springel}, {Di Matteo}, \&
  {Hernquist}}]{springel:models}
---. 2005{\natexlab{b}}, \mnras, 361, 776

\bibitem[{{Springel} {et~al.}(2006){Springel}, {Frenk}, \&
  {White}}]{springel:millenium.review}
{Springel}, V., {Frenk}, C.~S., \& {White}, S.~D.~M. 2006, \nat, 440, 1137

\bibitem[{{Springel} \& {Hernquist}(2005)}]{springel:spiral.in.merger}
{Springel}, V., \& {Hernquist}, L. 2005, \apjl, 622, L9

\bibitem[{{Springel} {et~al.}(2005{\natexlab{c}})}]{springel:millenium}
{Springel}, V., {et~al.} 2005{\natexlab{c}}, \nat, 435, 629

\bibitem[{{Stewart}(2009)}]{stewart:massratio.defn.conf.proc}
{Stewart}, K.~R. 2009, To appear in proceedings of "Galaxy Evolution: Emerging
  Insights and Future Challenges", arXiv:0902.2214

\bibitem[{{Stewart} {et~al.}(2009{\natexlab{a}}){Stewart}, {Bullock}, {Barton},
  \& {Wechsler}}]{stewart:merger.rates}
{Stewart}, K.~R., {Bullock}, J.~S., {Barton}, E.~J., \& {Wechsler}, R.~H.
  2009{\natexlab{a}}, \apj, 702, 1005

\bibitem[{{Stewart} {et~al.}(2009{\natexlab{b}}){Stewart}, {Bullock},
  {Wechsler}, \& {Maller}}]{stewart:disk.survival.vs.mergerrates}
{Stewart}, K.~R., {Bullock}, J.~S., {Wechsler}, R.~H., \& {Maller}, A.~H.
  2009{\natexlab{b}}, \apj, 702, 307

\bibitem[{{Stewart} {et~al.}(2008){Stewart}, {Bullock}, {Wechsler}, {Maller},
  \& {Zentner}}]{stewart:mw.minor.accretion}
{Stewart}, K.~R., {Bullock}, J.~S., {Wechsler}, R.~H., {Maller}, A.~H., \&
  {Zentner}, A.~R. 2008, \apj, 683, 597

\bibitem[{{Tacconi} {et~al.}(2002){Tacconi}, {Genzel}, {Lutz}, {Rigopoulou},
  {Baker}, {Iserlohe}, \& {Tecza}}]{tacconi:ulirgs.sb.profiles}
{Tacconi}, L.~J., {Genzel}, R., {Lutz}, D., {Rigopoulou}, D., {Baker}, A.~J.,
  {Iserlohe}, C., \& {Tecza}, M. 2002, \apj, 580, 73

\bibitem[{{Tacconi} {et~al.}(2008)}]{tacconi:smg.mgr.lifetime.to.quiescent}
{Tacconi}, L.~J., {et~al.} 2008, \apj, 680, 246

\bibitem[{{Tacconi} {et~al.}(2010)}]{tacconi:high.molecular.gf.highz}
---. 2010, \nat, in press, arXiv:1002.2149

\bibitem[{{Tinker} {et~al.}(2005){Tinker}, {Weinberg}, {Zheng}, \&
  {Zehavi}}]{tinker:hod}
{Tinker}, J.~L., {Weinberg}, D.~H., {Zheng}, Z., \& {Zehavi}, I. 2005, \apj,
  631, 41

\bibitem[{{Toft} {et~al.}(2007)}]{toft:z2.sizes.vs.sfr}
{Toft}, S., {et~al.} 2007, \apj, 671, 285

\bibitem[{{Toomre}(1977)}]{toomre77}
{Toomre}, A. 1977, in Evolution of Galaxies and Stellar Populations (New Haven:
  Yale University Observatory), ed. B.~M. {Tinsley} \& R.~B. {Larson}, 401

\bibitem[{{Trujillo} {et~al.}(2007){Trujillo}, {Conselice}, {Bundy}, {Cooper},
  {Eisenhardt}, \& {Ellis}}]{trujillo:ell.size.evol.update}
{Trujillo}, I., {Conselice}, C.~J., {Bundy}, K., {Cooper}, M.~C., {Eisenhardt},
  P., \& {Ellis}, R.~S. 2007, \mnras, 382, 109

\bibitem[{{Trujillo} \& {Pohlen}(2005)}]{trujillo:truncation.scale.evol}
{Trujillo}, I., \& {Pohlen}, M. 2005, \apjl, 630, L17

\bibitem[{{Vale} \& {Ostriker}(2006)}]{valeostriker:monotonic.hod}
{Vale}, A., \& {Ostriker}, J.~P. 2006, \mnras, 371, 1173

\bibitem[{{van den Bosch} {et~al.}(2005){van den Bosch}, {Tormen}, \&
  {Giocoli}}]{vandenbosch:subhalo.mf}
{van den Bosch}, F.~C., {Tormen}, G., \& {Giocoli}, C. 2005, \mnras, 359, 1029

\bibitem[{{van den Bosch} {et~al.}(2007)}]{vandenbosch:concordance.hod}
{van den Bosch}, F.~C., {et~al.} 2007, \mnras, 376, 841

\bibitem[{{van Dokkum}(2005)}]{vandokkum:dry.mergers}
{van Dokkum}, P.~G. 2005, \aj, 130, 2647

\bibitem[{{van Dokkum} {et~al.}(2006)}]{vandokkum06:drgs}
{van Dokkum}, P.~G., {et~al.} 2006, \apjl, 638, L59

\bibitem[{{Wang} {et~al.}(2006){Wang}, {Li}, {Kauffmann}, \& {de
  Lucia}}]{wang:sdss.hod}
{Wang}, L., {Li}, C., {Kauffmann}, G., \& {de Lucia}, G. 2006, \mnras, 371, 537

\bibitem[{{Weil} \& {Hernquist}(1994)}]{weil94:multiple.merger.kinematics}
{Weil}, M.~L., \& {Hernquist}, L. 1994, \apjl, 431, L79

\bibitem[{{Weil} \& {Hernquist}(1996)}]{weil96:multiple.merger.scalings}
---. 1996, \apj, 460, 101

\bibitem[{{Weinberg} {et~al.}(2008){Weinberg}, {Colombi}, {Dav{\'e}}, \&
  {Katz}}]{weinberg:baryons.and.substructure}
{Weinberg}, D.~H., {Colombi}, S., {Dav{\'e}}, R., \& {Katz}, N. 2008, \apj,
  678, 6

\bibitem[{{Weinmann} {et~al.}(2006{\natexlab{a}}){Weinmann}, {van den Bosch},
  {Yang}, \& {Mo}}]{weinmann:obs.hod}
{Weinmann}, S.~M., {van den Bosch}, F.~C., {Yang}, X., \& {Mo}, H.~J.
  2006{\natexlab{a}}, \mnras, 366, 2

\bibitem[{{Weinmann} {et~al.}(2006{\natexlab{b}}){Weinmann}, {van den Bosch},
  {Yang}, {Mo}, {Croton}, \& {Moore}}]{weinmann:group.cat.vs.sam}
{Weinmann}, S.~M., {van den Bosch}, F.~C., {Yang}, X., {Mo}, H.~J., {Croton},
  D.~J., \& {Moore}, B. 2006{\natexlab{b}}, \mnras, 372, 1161

\bibitem[{{Weinzirl} {et~al.}(2009){Weinzirl}, {Jogee}, {Khochfar}, {Burkert},
  \& {Kormendy}}]{weinzirl:b.t.dist}
{Weinzirl}, T., {Jogee}, S., {Khochfar}, S., {Burkert}, A., \& {Kormendy}, J.
  2009, \apj, 696, 411

\bibitem[{{Wetzel}(2010)}]{wetzel:sat.orbit.vs.halomz}
{Wetzel}, A.~R. 2010, \mnras, in press, arXiv:1001.4792

\bibitem[{{Wetzel} {et~al.}(2009{\natexlab{a}}){Wetzel}, {Cohn}, \&
  {White}}]{wetzel:mgr.rate.subhalos}
{Wetzel}, A.~R., {Cohn}, J.~D., \& {White}, M. 2009{\natexlab{a}}, \mnras, 395,
  1376

\bibitem[{{Wetzel} {et~al.}(2009{\natexlab{b}}){Wetzel}, {Cohn}, \&
  {White}}]{wetzel:merger.bias}
---. 2009{\natexlab{b}}, \mnras, 281

\bibitem[{{White}(1976)}]{white:cross.section}
{White}, S.~D.~M. 1976, \mnras, 174, 467

\bibitem[{{White} \& {Rees}(1978)}]{whiterees78}
{White}, S.~D.~M., \& {Rees}, M.~J. 1978, \mnras, 183, 341

\bibitem[{{Wolf} {et~al.}(2005)}]{wolf:merger.mf}
{Wolf}, C., {et~al.} 2005, \apj, 630, 771

\bibitem[{{Woods} \& {Geller}(2007)}]{woods:minor.mergers}
{Woods}, D.~F., \& {Geller}, M.~J. 2007, \aj, 134, 527

\bibitem[{{Woods} {et~al.}(2006){Woods}, {Geller}, \&
  {Barton}}]{woods:tidal.triggering}
{Woods}, D.~F., {Geller}, M.~J., \& {Barton}, E.~J. 2006, \aj, 132, 197

\bibitem[{{Xu} {et~al.}(2004){Xu}, {Sun}, \& {He}}]{xu:merger.mf}
{Xu}, C.~K., {Sun}, Y.~C., \& {He}, X.~T. 2004, \apjl, 603, L73

\bibitem[{{Yan} {et~al.}(2003){Yan}, {Madgwick}, \&
  {White}}]{yan:clf.evolution}
{Yan}, R., {Madgwick}, D.~S., \& {White}, M. 2003, \apj, 598, 848

\bibitem[{{Yang} {et~al.}(2005){Yang}, {Mo}, {Jing}, \& {van den
  Bosch}}]{yang:obs.clf}
{Yang}, X., {Mo}, H.~J., {Jing}, Y.~P., \& {van den Bosch}, F.~C. 2005, \mnras,
  358, 217

\bibitem[{{Younger} {et~al.}(2008{\natexlab{a}}){Younger}, {Hopkins}, {Cox}, \&
  {Hernquist}}]{younger:minor.mergers}
{Younger}, J.~D., {Hopkins}, P.~F., {Cox}, T.~J., \& {Hernquist}, L.
  2008{\natexlab{a}}, \apj, 686, 815

\bibitem[{{Younger} {et~al.}(2008{\natexlab{b}})}]{younger:smg.sizes}
{Younger}, J.~D., {et~al.} 2008{\natexlab{b}}, \apj, 688, 59

\bibitem[{{Yu} \& {Tremaine}(2002)}]{yutremaine:bhmf}
{Yu}, Q., \& {Tremaine}, S. 2002, \mnras, 335, 965

\bibitem[{{Zentner} {et~al.}(2005){Zentner}, {Berlind}, {Bullock}, {Kravtsov},
  \& {Wechsler}}]{zentner:substructure.sam.hod}
{Zentner}, A.~R., {Berlind}, A.~A., {Bullock}, J.~S., {Kravtsov}, A.~V., \&
  {Wechsler}, R.~H. 2005, \apj, 624, 505

\bibitem[{{Zheng} {et~al.}(2005){Zheng}, {Hammer}, {Flores}, {Ass{\'e}mat}, \&
  {Rawat}}]{zheng:morphological.sfr.evol}
{Zheng}, X.~Z., {Hammer}, F., {Flores}, H., {Ass{\'e}mat}, F., \& {Rawat}, A.
  2005, \aap, 435, 507

\bibitem[{{Zheng} {et~al.}(2007){Zheng}, {Coil}, \&
  {Zehavi}}]{zheng:hod.evolution}
{Zheng}, Z., {Coil}, A.~L., \& {Zehavi}, I. 2007, \apj, 667, 760

\end{thebibliography}

\end{document}